\documentclass[prd,twocolumn,twoside,preprintnumbers,superscriptaddress,nofootinbib]{revtex4}

\usepackage{amsmath,slashed}
\usepackage{graphicx,graphics,color}
\usepackage{dcolumn}
\usepackage[hyperfootnotes=false]{hyperref}
\usepackage{xspace}
\usepackage{enumerate}
\usepackage{varwidth}

\usepackage{physics}
\usepackage{amssymb}
\usepackage{mathtools}
\usepackage{dsfont}
\usepackage{multirow}

\hypersetup{
    colorlinks=true,
    linkcolor=blue,
    citecolor=blue
}

\newcommand{\GeV}{\,\text{GeV}}

\newcommand{\TeV}{\,\text{TeV}}

\renewcommand{\Im}{\text{Im}\,}
\renewcommand{\Re}{\text{Re}\,}

\allowdisplaybreaks[1]

\begin{document}

\title{Light new physics and the $\boldsymbol{\tau}$ lepton dipole moments}

\author{Martin Hoferichter}
\affiliation{Albert Einstein Center for Fundamental Physics, Institute for Theoretical Physics, University of Bern, Sidlerstrasse 5, 3012 Bern, Switzerland}

\author{Gabriele Levati}
\affiliation{Albert Einstein Center for Fundamental Physics, Institute for Theoretical Physics, University of Bern, Sidlerstrasse 5, 3012 Bern, Switzerland}

\begin{abstract} 
Testing New-Physics (NP) scenarios that couple predominantly to the third generation is notoriously difficult experimentally, as exemplified by comparing limits for the $\tau$ lepton dipole moments to those of electrons and muons. In this case, extracting limits from processes such as $e^+e^-\to\tau^+\tau^-$ often relies on effective-field-theory (EFT) arguments, which allow for model-independent statements, but only apply if the NP scale is sufficiently large compared to the center-of-mass energy. In this work we offer a comprehensive analysis of light NP contributions to the $\tau$ dipole moments, providing a detailed account of the interpretation of asymmetry measurements in $e^+e^-\to\tau^+\tau^-$ that are tailored towards the extraction of dipole moments, for the test cases of new light spin-$0$ and spin-$1$ bosons. Moreover, we study the decoupling to the EFT limit in these scenarios and discuss the complementarity to constraints from other related processes, such as production in $e^+e^-$ reactions. While covering a wide range of light NP scenarios, as specific case study we present a detailed discussion of a tauphilic gauge vector boson at Belle II.
\end{abstract}

\maketitle

\section{Introduction}

Precision observables involving leptons are among some of the most interesting low-energy probes of the Standard Model (SM) of particle physics. This broad class of observables includes the electric (EDM, $d_\ell$) and the anomalous magnetic dipole moments (AMM, $a_\ell$) of leptons. The large majority of these quantities have been measured at extremely high precision levels at experimental facilities, thus making them particularly relevant as tests of the SM.
This is the case, for instance, for the AMM of the electron and the muon, as well as for the experimental bounds on the corresponding EDMs:
\begin{align}
\label{dipole_moments_exp}
a_\mu^{\text{exp}} &= 116\, 592\, 071.5(14.5) \times
10^{-11} &\text{\cite{Muong-2:2025xyk}}\,, \nonumber \\
a_e^{\text{exp}} &= 115 \, 965\,  218 \, 059\,(13) \times
10^{-14}  &\text{\cite{Fan:2022eto}}\,,\nonumber\\
d_e^\text{exp} &< 4.1 \times 10^{-30} \, e\, \text{cm}\,&\text{\cite{Roussy:2022cmp}}\,,\nonumber\\
d_\mu^\text{exp} &< 2 \times 10^{-19} \, e\, \text{cm}\,&\text{\cite{Muong-2:2008ebm}}\,.
\end{align}
Thanks to the precision of their measurements and the relative theoretical predictions, the dipole moments of the electron and the muon play a pivotal role in constraining possible New-Physics (NP) scenarios.

More in detail,
$d_\mu$ clearly represents the least explored NP probe among those listed in Eq.~\eqref{dipole_moments_exp}, and various efforts are ongoing to improve its precision in the future~\cite{Abe:2019thb,Aiba:2021bxe,Adelmann:2025nev,Crivellin:2018qmi}.
On the other hand, even limits for the equivalent electron EDM $d_e$ constrained in paramagnetic molecules, albeit more sensitive by ten orders of magnitude, are still around five orders of magnitude away from the SM prediction~\cite{Ema:2022yra} (and, accordingly, for $d_{\mu,\tau}$~\cite{Yamaguchi:2020eub} the difference is much larger). A positive detection of a nonvanishing lepton EDM at current experimental facilities would therefore constitute a clear NP signal.
For $a_\ell$,
experimental and theoretical efforts need to proceed at the same pace in order to maximize the sensitivity to NP.
As far as $a_e$ is concerned, the current limiting factor in the reach is a persistent tension between measurements of the fine-structure constant in Cs~\cite{Parker:2018vye} and Rb~\cite{Morel:2020dww} atom-interferometry experiments. Theoretical uncertainties~\cite{Aoyama:2019ryr,Volkov:2019phy,Volkov:2024yzc,Aoyama:2024aly,DiLuzio:2024sps,Hoferichter:2025fea} arise instead a factor of four below the uncertainty of the direct measurement~\cite{Fan:2022eto}.
Contrary to the case of the electron, the global experimental average of $a_\mu^\text{exp}$~\cite{Muong-2:2025xyk,Muong-2:2023cdq,Muong-2:2024hpx,Muong-2:2021ojo,Muong-2:2021vma,Muong-2:2021ovs,Muong-2:2021xzz,Muong-2:2006rrc} currently exhibits a precision that is a factor of four better than that of the corresponding theoretical prediction~\cite{Aliberti:2025beg,Aoyama:2012wk,Volkov:2019phy,Volkov:2024yzc,Aoyama:2024aly,Parker:2018vye,Morel:2020dww,Fan:2022eto,Czarnecki:2002nt,Gnendiger:2013pva,Ludtke:2024ase,Hoferichter:2025yih,RBC:2018dos,Giusti:2019xct,Borsanyi:2020mff,Lehner:2020crt,Wang:2022lkq,Aubin:2022hgm,Ce:2022kxy,ExtendedTwistedMass:2022jpw,RBC:2023pvn,Kuberski:2024bcj,Boccaletti:2024guq,Spiegel:2024dec,RBC:2024fic,Djukanovic:2024cmq,ExtendedTwistedMass:2024nyi,MILC:2024ryz,FermilabLatticeHPQCD:2024ppc,Keshavarzi:2019abf,DiLuzio:2024sps,Kurz:2014wya,Colangelo:2015ama,Masjuan:2017tvw,Colangelo:2017qdm,Colangelo:2017fiz,Hoferichter:2018dmo,Hoferichter:2018kwz,Eichmann:2019tjk,Bijnens:2019ghy,Leutgeb:2019gbz,Cappiello:2019hwh,Masjuan:2020jsf,Bijnens:2020xnl,Bijnens:2021jqo,Danilkin:2021icn,Stamen:2022uqh,Leutgeb:2022lqw,Hoferichter:2023tgp,Hoferichter:2024fsj,Estrada:2024cfy,Deineka:2024mzt,Eichmann:2024glq,Bijnens:2024jgh,Hoferichter:2024vbu,Hoferichter:2024bae,Holz:2024lom,Holz:2024diw,Cappiello:2025fyf,Colangelo:2014qya,Blum:2019ugy,Chao:2021tvp,Chao:2022xzg,Blum:2023vlm,Fodor:2024jyn}, and a broad research program is aimed at realizing the NP sensitivity  set by the experimental measurement~\cite{Colangelo:2022jxc,Aliberti:2025beg,Hertzog:2025ssc}.

Tests in $a_\ell$ and $d_\ell$ display an interesting complementarity in probing the $CP$ and flavor structure of NP~\cite{Giudice:2012ms}. Accessing these quantities for the $\tau$ lepton as well would therefore be of primary importance in charting the NP landscape: besides discerning chirally enhanced scenarios~\cite{Giudice:2012ms,Crivellin:2021rbq,Athron:2025ets}, $\tau$ dipole moments would also allow one to probe NP scenarios whose predictions feature larger couplings to the third generation of fermions~\cite{Barbieri:1995uv, Barbieri:1996ae, Barbieri:2012uh, Matsedonskyi:2014iha,Panico:2016ull,Glioti:2024hye,Froggatt:1978nt,Berezhiani:1983hm,Bordone:2017bld,Fuentes-Martin:2022xnb, Davighi:2023iks,Nakai:2025dmp}.
Unfortunately, the short lifetime of the $\tau$ lepton renders its dipole moments much more challenging to access in experiment at a competitive level of precision.
New experiments and techniques are being conceived to constrain $a_\tau$ and $d_\tau$ more precisely, see, e.g., Refs.~\cite{Kim:1982ry,Samuel:1990su,Eidelman:2016aih,Koksal:2017nmy,Fomin:2018ybj,Fu:2019utm,Gutierrez-Rodriguez:2019umw,Beresford:2019gww,Dyndal:2020yen,Haisch:2023upo,Shao:2023bga,Beresford:2024dsc,Dittmaier:2025ikh,Buttazzo:2026amk}, among which are, for instance, recent measurements in peripheral Pb--Pb collisions at LHC~\cite{ATLAS:2022ryk,CMS:2022arf,CMS:2024qjo}. Nonetheless, it seems challenging to scale these techniques to a sensitivity much beyond the Schwinger term (except for future lepton colliders~\cite{Buttazzo:2026amk}), whereas testing realistic NP scenarios requires a precision of at least $10^{-5}$ in $a_\tau$~\cite{Crivellin:2021spu}.

To this end, other proposals~\cite{Bernabeu:2004ww,Bernabeu:2006wf,Bernabeu:2007rr,Bernabeu:2008ii} have been put forward to considerably improve the constraints on the $\tau$ dipole moments at high-luminosity $B$ factories. These works showed that it is, in principle, possible to access $a_\tau$ and $d_\tau$ by measuring properly defined asymmetries in the polarized cross section of the process $e^+ e^- \to \tau^+ \tau^-$, and recent works have discussed the feasibility of such a program~\cite{Crivellin:2021spu, USBelleIIGroup:2022qro, Aihara:2024zds, Banerjee:2020rww, Ulrich:2025fij, Gogniat:2025eom}.
However, it is important to stress that what is measured via these methods is not $a_\tau$ or $d_\tau$ directly, but the corresponding electromagnetic form factors $F_2(q^2)$ and $F_3(q^2)$ at a nonzero momentum transfer $q^2\ne 0$. Such a momentum dependence has to be properly taken into account in order to infer the consequences for $a_\tau$ and $d_\tau$.
In particular, this observation assumes a role of primary importance when it comes to placing bounds on possible NP candidates via the impact they can have on the $\tau$ dipole moments at the loop level. Indeed, once a form factor $F_i(q^2)$ is experimentally bound, one can employ the difference between its SM prediction at $q^2 \ne 0$ and the corresponding measurement to put bounds on any NP scenario that induces a nonzero contribution to such observables.
If the mass scale of NP fields ($m^2_\text{NP}$) is considerably larger than the center-of-mass (CM) energy of the experiment, NP states effectively decouple and the difference between the measured form factor and its SM prediction provides a direct constraint on the $\tau$ dipole moments. This is the scenario that has so far received the largest attention, both within an effective-field-theory (EFT) approach and under the assumption of specific NP models, with a sensitivity to heavy NP scenarios reaching levels up to  $\simeq 10^{-20} \, e\, \text{cm}$ for $d_\tau$ and $\simeq 10^{-6}$ for $a_\tau$~\cite{Crivellin:2021spu, USBelleIIGroup:2022qro}.

However, if there exist NP fields whose masses are lighter than, or comparable to, the available CM energy, model-dependent NP contributions to the form factors $F_i(q^2)$ arise. These have to be properly subtracted before meaningful constraints can be derived for the impact of light NP on the dipole moments of the $\tau$ lepton.
In light of this observation and of the ever-growing attention that light NP scenarios have been receiving in recent years, we find it timely to discuss such model-dependent contributions to electromagnetic form factors for a wide variety of well-motivated classes of NP models, including axions and axionlike particles, generic light scalars and pseudoscalars, and light vector bosons.
We have first explored this possibility in Ref.~\cite{Hoferichter:2025ijh}, where we focused on the possibility to constrain these NP effects at Belle II. In particular, we argued that the generation of nonzero imaginary parts for $F^\text{NP}_2(q^2)$ and $F^\text{NP}_3(q^2)$ in the presence of light NP candidates can be accessed via asymmetries that do not require polarized electron beams, thus marking a novel opportunity for NP searches at $B$ factories.

The first purpose of this paper is to provide the details of the calculations leading to the constraints presented in Ref.~\cite{Hoferichter:2025ijh}. In addition,
whenever dynamically propagating fields are considered in the assessment of the magnitude of their impact on indirect observables, such as dipole moments, the same couplings can be often tested directly in tree-level scattering processes, and the interplay between direct and indirect searches is far from obvious. For that reason, a large part of this paper is devoted to exploring
the complementarity between direct and indirect search strategies for light NP candidates.

The paper is organized as follows. In Sec.~\ref{sec:light_NP}, we first introduce the main light NP scenarios we consider in our analysis. We then provide explicit expressions for their impact on the momentum-dependent form factors $F_2$ and $F_3 $, including the relevant kinematic limits.
In Sec.~\ref{sec:Asymmetries} we discuss in detail some technicalities related to the definition and implementation of those asymmetries that are necessary to access $F_2$ and $F_3$ in $e^+ e^- \to \tau^+ \tau^-$ collisions.
In Sec.~\ref{sec:other_processes}, we explore complementary strategies that can be employed to probe the relevant light NP scenarios, with a particular focus on collider probes at $B$ factories.
In particular, we apply our results to a specific case study, that of tauphilic gauge vector bosons at Belle II, which have recently received some attention due to the possibility to explain the observed rate for $B \to K^{(*)} + E_{\text{miss}}$~\cite{DiLuzio:2025qkc}, to illustrate the interplay between direct and indirect searches. Our conclusions are presented in Sec.~\ref{sec:conclusions}, while further details are relegated to the appendices.

\section{Light new physics scenarios}
\label{sec:light_NP}

Our main objective is to investigate the impact of light new states on the $\tau$ EDM and AMM. A nonzero contribution to these quantities for a generic lepton species $\ell$ can be induced by the virtual exchange of a light NP degree of freedom, which has an impact on the form factors $F_i(q^2)$ parameterizing the interaction with the electromagnetic current.
These form factors are  defined via the general parametrization of the $\gamma \ell \ell$ vertex
\begin{align}
\langle \ell(p')|j_{\text{em}}^\mu|\ell(p)\rangle &=
 \,e \,\bar{u}(p')\, \big[\gamma^\mu F_1 + (i F_2 + F_3 \gamma_5)\, \frac{\sigma^{\mu\nu}q_\nu}{2m_\ell}\notag\\
& + \left(q^2 \gamma^\mu - q^\mu \slashed{q}\right)\gamma_5 F_A \big]\, u(p)\,,
\end{align}
where $q = p'-p$ is the momentum carried by the photon, and a dependence of the form factors on its square is understood, $F_i = F_i(q^2)$. $F_1$ describes the vectorial component of the electromagnetic vertex, while $F_A(0)$ is the so-called anapole moment.
Finally, in the limit $q^2\to 0$ the form factors $F_2$ and $F_3$ are in direct relation to the AMM $a_\ell$ and the EDM $d_\ell$
\begin{equation}
a_\ell = \Re F_2(0) \,, \qquad d_\ell = \frac{e}{2m_\ell} \Re F_3(0)\,.
\end{equation}
As far as $\tau$ leptons are concerned, no direct access on $a_\tau$ or $d_\tau$ can be obtained, and the quantities that can be measured experimentally are $F_2(q^2)$ and $F_3(q^2)$, where $q^2$ corresponds to the typical energy scale of the experiment under consideration.
Information on $a_\tau$ and $d_\tau$ can then be obtained only provided that loop-level momentum-dependent corrections are properly subtracted~\cite{Crivellin:2021spu}.
In the case of heavy NP, $q^2 \ll m^2_\text{NP}$, heavy states effectively decouple from the theory, leaving only a constant, momentum-independent imprint on $a_\tau$ and $d_\tau$, and only the loop-induced, momentum-dependent contributions  from the virtual exchange of SM states need to be subtracted.

If NP is light, i.e., $m^2_\text{NP}\simeq q^2, m_\tau^2$, momentum-dependent contributions are generated as well by the virtual exchange of such NP states. Projecting $F_{2,3}(q^2)$ onto $a_\tau$ and $d_\tau$ therefore requires handling also such contributions.
In particular, for $q^2 > (m_i+m_j)^2$, where $m_{i,j}$ are the masses of two of the particle species circulating in the loop, form factors develop an imaginary part.
The physically interesting quantities are the real parts of the form factors, which can be related to the $\tau$ dipole moments by extrapolating to their $q^2\to 0$ limit. Imaginary parts can be measured, but they cannot be directly related to the dipole moments; however, in the case of light NP candidates, they are sensitive to the same coupling constants responsible for the generation of the physically relevant magnetic moments, leading to an opportunity for indirect access as highlighted in Ref.~\cite{Hoferichter:2025ijh}.\footnote{After Ref.~\cite{Hoferichter:2025ijh}, a similar strategy was suggested in Ref.~\cite{Huang:2025ghw}.}

In light of these observations, in this section we will discuss the one-loop corrections to the form factors $F_2(q^2)$ and $F_3(q^2)$ for different well-motivated light NP candidates, among them general spin-$0$ and spin-$1$ particles. We provide the complete expression for such contributions, highlighting in each case the most relevant limits.
Throughout our computations we have worked in naive dimensional regularization (NDR) for $\gamma_5$.
The conventions we adopted match those of \texttt{Package-X}~\cite{Patel:2015tea,Patel:2016fam}, which we employed in order to get analytical expressions for our results.

\subsection{Light scalars and pseudoscalars}

In this section we consider the contributions to $F_2(q^2)$ and $F_3(q^2)$ as induced by the exchange of a light spin-$0$ particle. These are naturally seen to arise in a wide variety of NP models involving the spontaneous breaking of some accidental global symmetry in the UV at some energy scale $\Lambda$. These light new states are then to be identified as the pseudo-Nambu--Goldstone bosons (pNGBs) emerging from such a spontaneous symmetry breaking pattern. It is precisely the presence of a symmetry in the UV that protects the mass of such states from radiative corrections and allows them to remain light with respect to the typical symmetry breaking scale, a condition that we can write as $M_\phi =m_{\text{NP}}\ll \Lambda_\text{NP}= \Lambda$. Typical examples of such a class of candidates are for instance axions~\cite{Peccei:1977hh, Peccei:1977ur, Wilczek:1977pj} and their direct generalizations, so-called axionlike particles (ALPs) ~\cite{Jaeckel:2010ni, Preskill:1982cy, Abbott:1982af, Dine:1982ah, Davidson:1981zd, Wilczek:1982rv, Graham:2015cka}.
In the spirit of the most general parameterization possible, we write down the most relevant dimension-5 $U(1)_\text{em}$ invariant interactions of such new states with SM fields as follows~\cite{DiLuzio:2020oah,DiLuzio:2023lmd,DiLuzio:2023cuk}:
    \begin{align}
    \label{eq:ALP_Lag}
            \mathcal{L}^\text{int}_{\phi} &= \phi \frac{m_\tau}{\Lambda}\,\bar{\tau}\left(c_S^\tau + i \,c_P^\tau \,\gamma_5\right) \tau\,\nonumber  \\
            &  \quad \,+ c_{\gamma\gamma} \frac{\alpha_\text{em}}{4\pi}\frac{\phi}{\Lambda}F_{\mu\nu}F^{\mu\nu}+  \, \tilde{c}_{\gamma\gamma} \frac{\alpha_\text{em}}{4\pi}\frac{\phi}{\Lambda}F_{\mu\nu}\tilde{F}^{\mu\nu}\,,
    \end{align}
where $F_{\mu\nu}$ is the electromagnetic field-strength tensor and $\tilde{F}_{\mu\nu}=\frac{1}{2}\epsilon_{\mu\nu\rho\sigma}F^{\rho\sigma}$ its dual.
The normalization of the scalar and pseudoscalar couplings with a power of the lepton mass is arbitrary, but reasonable: pseudoscalar interactions can be understood as emerging from the integration by part of derivative couplings of the field $\phi$ to SM fermions. The scalar couplings instead are justified in the light of the simplest possible UV origin of such terms from a direct coupling of $\phi$ with the SM Yukawa interactions. The alignment between the couplings of the operator $\phi\bar{\ell}He$ and that of the Yukawa operator $\bar{\ell}He$ is in principle not guaranteed, yet we find it useful to have an expression that treats scalar and pseudoscalar couplings on the same footing. 
Weak interactions can be neglected as long as we are interested in experiments, such as Belle II, operating at a CM energy that is considerably lighter than the electroweak scale, $s_B \ll M_W^2$~\cite{Gogniat:2026zvf}.

We stress two important points about the Lagrangian in Eq.~\eqref{eq:ALP_Lag}: first, it has to be noticed that the operators $\phi\bar{\tau}\tau$ and $\phi FF$ possess opposite $CP$ transformation properties with respect to the operators $\phi\bar{\tau}\gamma_5\tau$ and $\phi F\tilde{F}$. As a consequence, their simultaneous presence with nonzero Wilson coefficients necessarily implies that $CP$ symmetry is violated. This in turn means that a nonzero contribution to the $\tau$ EDM $d_\tau$ is generated, and is proportional to the Jarlskog invariants $c_S^\tau \tilde{c}_{\gamma\gamma}$ and $c_P^\tau c_{\gamma\gamma}$.
Second, the coefficient $c_{\gamma\gamma}$ ($\tilde{c}_{\gamma\gamma}$) consists of two contributions: a UV-dependent term $c_{\gamma\gamma}^0$ ($\tilde{c}_{\gamma\gamma}^0$), possibly sourced by heavy fermion loops, and a loop-level contribution from $\tau$ loops~\cite{Alda:2024cxn}. These drastically differ according to the nature of the spin-$0$ particle one is considering. We can distinguish three cases:
\begin{enumerate}
\item $\phi$ is an ALP. In this case, $\tilde{c}_{\gamma\gamma}$ has an anomalous component. In the case of on-shell photons it reads:
\begin{equation}
    \tilde{c}_{\gamma\gamma} = \tilde{c}^0_{\gamma\gamma} - c_\tau\, B_1\left(\frac{4m_\tau^2}{m_a^2}\right)\,.
    \label{eq:cgamma_eff}
\end{equation}
\item $\phi$ is a generic pseudoscalar particle. In this case, no anomaly is present and the corresponding Wilson coefficient reads
\begin{equation}
    \tilde{c}_{\gamma\gamma} = \tilde{c}^0_{\gamma\gamma} - c_\tau\, \left[B_1\left(\frac{4m_\tau^2}{m_a^2}\right)-1\right]\,.
\end{equation}
\item $\phi$ is a generic scalar particle. No anomaly is present and the corresponding Wilson coefficient reads
\begin{equation}
    c_{\gamma\gamma} = c^0_{\gamma\gamma} - c_\tau\, A_1\left(\frac{4m_\tau^2}{m_a^2}\right)\,.
\end{equation}
\end{enumerate}
The loop functions are defined by
\begin{align}
B_1(x) &= 1- x \big[f(x)\big]^2\,, \notag\\
A_1(x) &= -x\left[1-(x-1)\big[f(x)\big]^2\right]\,,\notag\\
f(x) &= \begin{cases} \arcsin\frac{1}{\sqrt{x}} & x \geq 1\,, \\ \frac{\pi}{2}+\frac{i}{2}\log \frac{1+\sqrt{1-x}}{1-\sqrt{1-x}} & x<1\,. \end{cases}
\end{align}

In our computations we will consider $c_{\gamma \gamma }$ and $\tilde{c}_{\gamma \gamma }$ as local Wilson coefficients and we will disregard altogether any possible momentum dependence of such quantities. However, these have to be properly accounted for if a complete two-loop analysis is to be performed.
We can conveniently distinguish two kinds of contributions: those due to the simultaneous presence of two Yukawa-like $\phi\tau\tau$ renormalizable interactions, or those due a Yukawa-like $\phi\tau\tau$ renormalizable interaction and a nonrenormalizable $\phi \gamma \gamma$ vertex. In the following we will provide both of them and discuss the behavior of the two classes of contributions.

\subsubsection{Yukawa couplings only}

\begin{figure}[t]
    \centering
    \includegraphics[width=\linewidth]{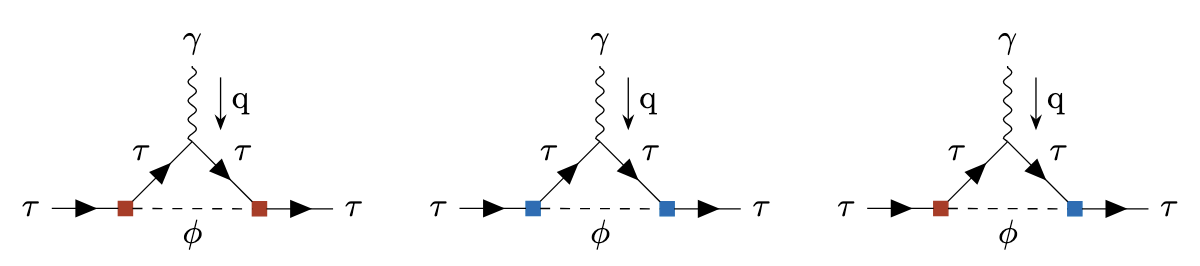}
    \caption{Feynman diagrams contributing to the $\tau$ magnetic (first two diagrams) and electric (last diagram) dipole moments. Blue dots represent an insertion of the operator $\phi \bar{\tau}\tau$, while red dots represent insertions of the operator $\phi \bar{\tau}\gamma_5\tau$.}
    \label{fig:YukDiagrams}
\end{figure}

The contributions to $F_2$
and $F_3$ generated by the virtual exchange of a light spin-$0$ particle in  Fig.~\ref{fig:YukDiagrams} read:
\begin{align}
F_2^\text{y}(q^2)&= \frac{m_\tau^2}{16\pi^2}\bigg(\frac{c_P^\tau m_\tau}{\Lambda}\bigg)^2\Big[C_{11}(\theta)+ 2C_{12}(\theta)+C_{11}(\theta) \Big]  \nonumber\\
&+\frac{m_\tau^2}{16\pi^2}\bigg(\frac{c_S^\tau m_\tau}{\Lambda}\bigg)^2\Big[C_{11}(\theta)+2C_{12}(\theta)+C_{11}(\theta)\notag\\
&  \qquad+2C_1(\theta) + 2 C_2(\theta)\Big]\,, \notag\\
F_3^\text{y}(q^2)&= \frac{m_\tau^2}{8\pi^2}\frac{m_\tau^2 \,c_S^\tau c_P^\tau}{\Lambda^2}\Big[C_1(\theta) + C_2(\theta)\Big]\,,
\end{align}
where the arguments are summarized as $\theta = \{m_\tau^2, q^2, m_\tau^2, m_\phi, m_\tau, m_\tau\}$ and the $C_i$ denote standard Passarino--Veltman functions, see App.~\ref{app:conventions} and Ref.~\cite{Denner:1991kt} for conventions. We find agreement with the results reported in the literature, in the relevant limits~\cite{Cornella:2019uxs}. By evaluating the same loop diagrams for $s=q^2\to 0$ we then obtain the corresponding contribution to the $\tau$ AMM and EDM, which we report in App.~\ref{sec:Explicit_Expressions}.
In the limit $m_\tau \gg M_\phi$ these expressions reduce to 
\begin{align}
 a_\tau^\text{y}(M_\phi \ll m_\tau) &= \frac{3}{16\pi^2} \bigg(\frac{c_S^\tau m_\tau}{\Lambda}\bigg)^2-\frac{1}{16\pi^2}\bigg(\frac{c_P^\tau m_\tau}{\Lambda}\bigg)^2\,, \nonumber\\
 d_\tau^\text{y}(M_\phi \ll m_\tau) &= \frac{1}{8\pi^2}\frac{e}{m_\tau}\frac{m_\tau^2 c_P^\tau c_S^\tau}{\Lambda^2}\,,
\end{align}
while in the opposite limit, $m_\tau \ll M_\phi$, one has
\begin{align}
 a_\tau^\text{y}(M_\phi \gg m_\tau) &= \frac{1}{4\pi^2}\frac{m_\tau^2}{M_\phi^2}\bigg[\bigg(\frac{c_P^\tau m_\tau}{\Lambda}\bigg)^2\left(\frac{11}{12}+  \log\frac{m_\tau}{M_\phi}\right)\nonumber\\
 & \qquad -\bigg(\frac{c_S^\tau m_\tau}{\Lambda}\bigg)^2\left(\frac{7}{12}+ \log \frac{m_\tau}{M_\phi}\right)\bigg]\,, \nonumber \\
 d_\tau^\text{y}(M_\phi \gg m_\tau) &= -\frac{1}{4\pi^2}\frac{m_\tau^2 c_P^\tau c_S^\tau}{\Lambda^2}\frac{m_\tau^2}{M_\phi^2}\frac{e}{m_\tau}\left[\frac{3}{4} + \log \frac{m_\tau}{M_\phi}\right]\,.
\end{align}
Contributions from scalar and pseudoscalar interactions have opposite signs, so that in their simultaneous presence accidental cancellations can take place, depending on the magnitude of the corresponding Wilson coefficients.

In the high-energy limit, $q^2 \gg m_\tau^2, M_\phi^2$, one finds the following expansion for the form factors:
\begin{align}
F_2^\text{y}(q^2\gg m_\tau^2, M_\phi^2) &\simeq \frac{1}{16\pi^2}\frac{m_\tau^2}{s} \log \frac{m_\tau^2}{-s}\notag\\
& \quad\times\bigg[2\, \bigg(\frac{c_{P}^\tau m_\tau}{\Lambda}\bigg)^2 - 3 \,\bigg(\frac{c_{S}^\tau m_\tau}{\Lambda}\bigg)^2\bigg] \,,\nonumber\\
F_3^\text{y}(q^2\gg m_\tau^2, M_\phi^2) &\simeq \frac{c_{P}^\tau c_{S}^\tau}{4\pi^2}\frac{m_\tau^2}{s} \left(\frac{m_\tau}{\Lambda}\right)^2 \log \frac{-s}{m_\tau^2}\,,
\end{align}
whereas for $M_\phi^2 \gg q^2 \gg m_\tau^2$ one has
\begin{align}
F_2^\text{y}(M_\phi^2 \gg q^2 \gg m_\tau^2) &\simeq \frac{1}{48\pi^2}\frac{m_\tau^2}{M_\phi^2}\notag\\
&\times\bigg[\bigg(\frac{c_{S}^\tau m_\tau}{\Lambda}\bigg)^2\bigg(5+6 \log\frac{M_\phi^2}{-s}\bigg) \notag\\
&\quad -\bigg(\frac{c_{P}^\tau m_\tau}{\Lambda}\bigg)^2\bigg(1+6 \log\frac{M_\phi^2}{-s}\bigg)\bigg]\,, \nonumber \\
F_3^\text{y}(M_\phi^2 \gg q^2 \gg m_\tau^2) &\simeq \frac{c_{P}^\tau c_{S}^\tau}{8\pi^2}\frac{m_\tau^2}{M_\phi^2} \frac{m_\tau^2}{\Lambda^2}\bigg[1+2\log \frac{M_\phi^2}{-s}\bigg]\,.
\end{align}
Another interesting kinematic limit is the one in which the CM energy of the collision is tuned around $s = 4 m_\tau^2$, which could be reached, for instance, by applying radiative-return techniques. In such a limit one finds the expansion ($\beta_\tau = \sqrt{1-4m_\tau^2/s}$):
\begin{align}
F_2^{\text{y}}(s \to  4 m_\tau^2) &= \frac{3i}{8\pi} \frac{m_\tau M_\phi^2}{(s\beta_\tau)^{3/2}}\bigg[\bigg(\frac{c_{P}^\tau m_\tau}{\Lambda}\bigg)^2+\bigg(\frac{c_{S}^\tau m_\tau}{\Lambda}\bigg)^2\bigg] \nonumber \\
& - \frac{1}{8\pi^2m_\tau^2 s\beta_\tau}\bigg[\bigg(\frac{c_{P}^\tau m_\tau}{\Lambda}\bigg)^2+\bigg(\frac{c_{S}^\tau m_\tau}{\Lambda}\bigg)^2\bigg]\notag\\
&\times\bigg[3-2M_\phi^2 m_\tau^2 + M_\phi^4 \log \frac{m_\tau}{M_\phi}\notag\\
&\qquad- m_\tau^2(2m_\tau^2 + M_\phi^2)\mathcal{B}_0^{\tau\tau\phi}\bigg]  + \mathcal{O}\big(\beta_\tau^{-1/2}\big)\,,\nonumber \\
F_3^{\text{y}}(s \to 4 m_\tau^2) &=\frac{i }{4\pi}\frac{m_\tau^2}{\Lambda^2}\frac{m_\tau}{ \sqrt{s\beta_\tau}}c_{P}^\tau c_{S}^\tau+ \mathcal{O}(\beta_\tau^0)\,,
\end{align}
which shows, in particular, a significant enhancement of the imaginary part of the form factor near threshold.

\begin{figure*}[t]
    \centering
    \includegraphics[width=0.45\linewidth]{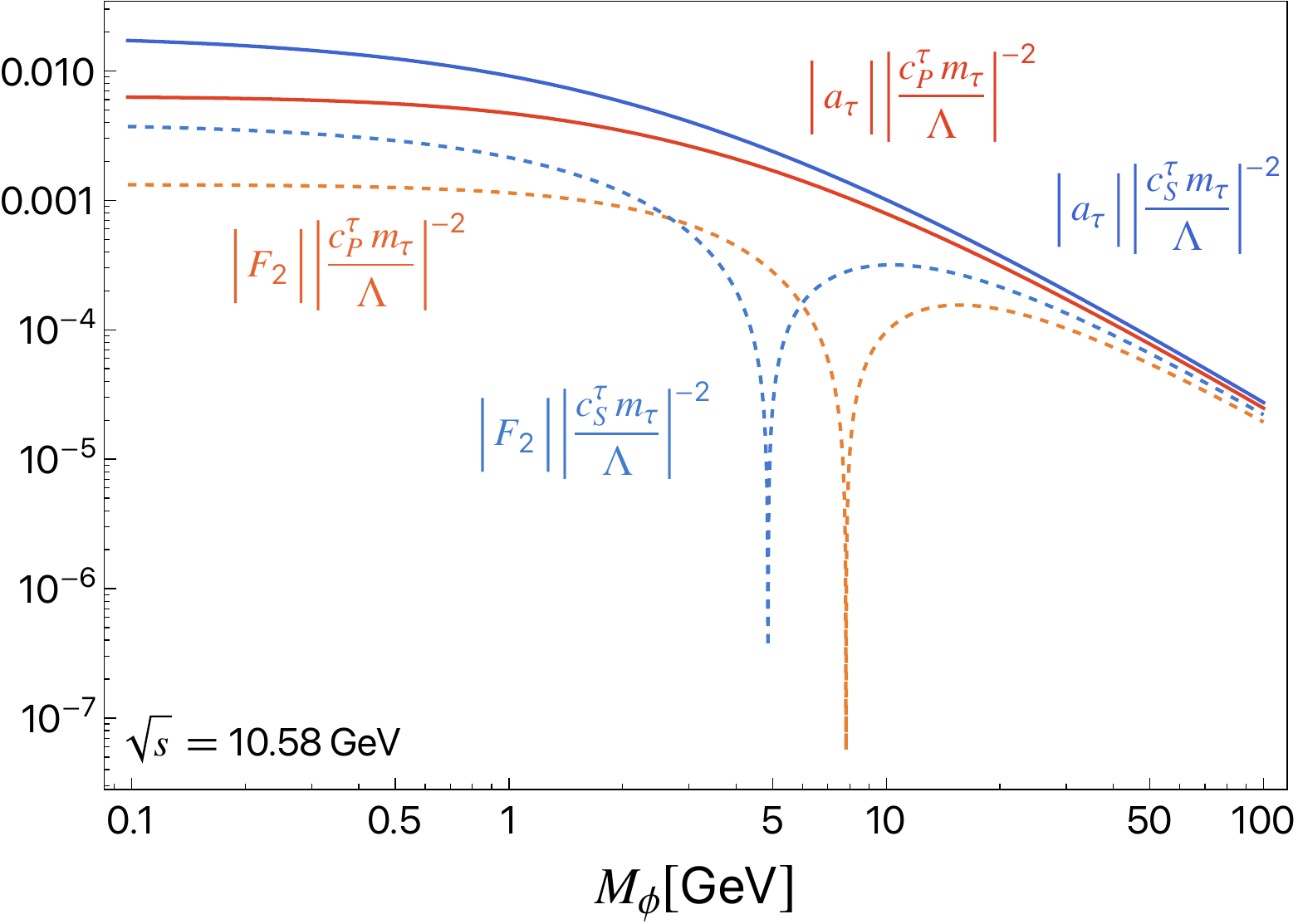}
    \includegraphics[width=0.45\linewidth]{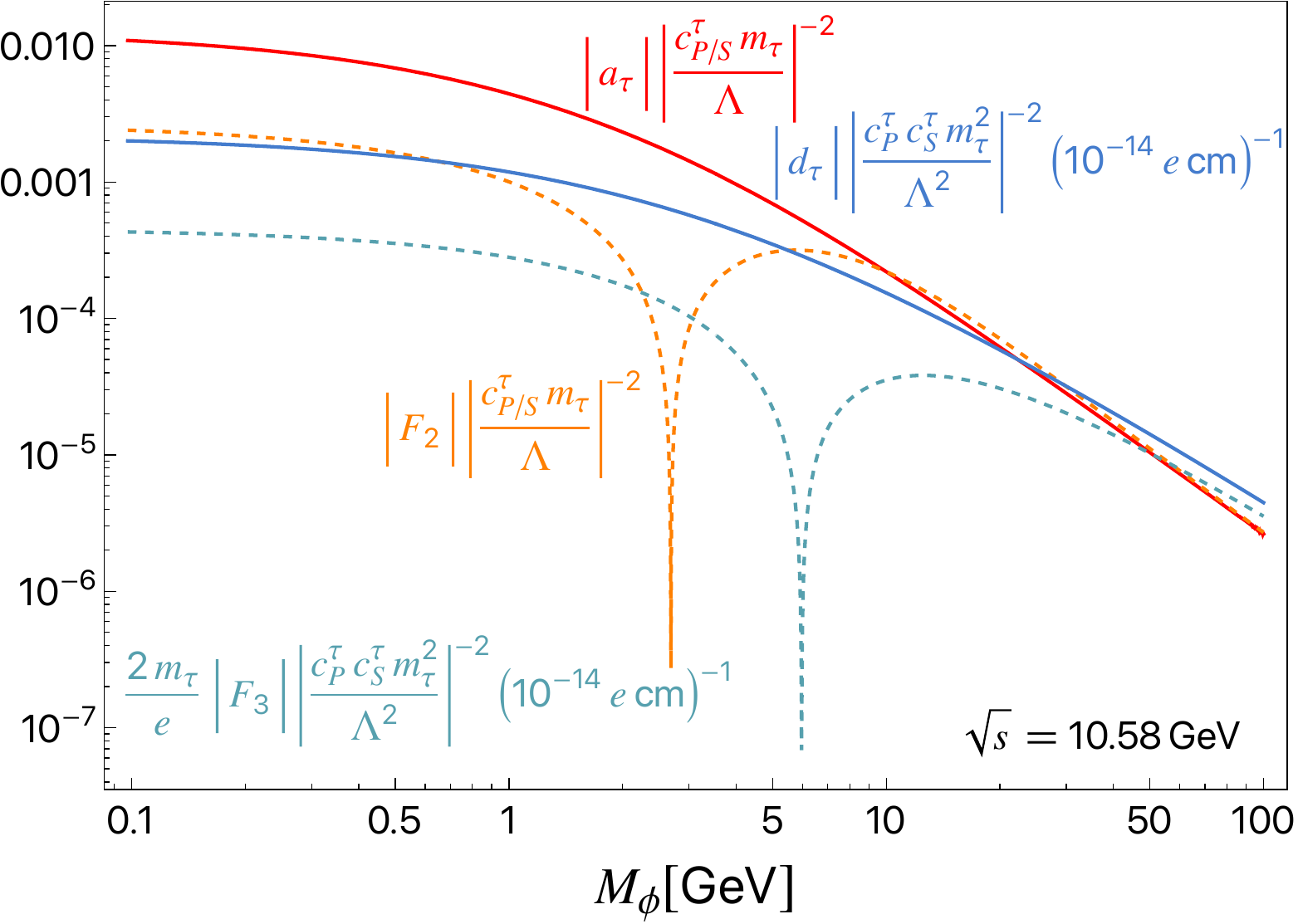}
    \caption{\textit{Left:} Contributions to $a_\tau$ induced by a $CP$-violating scalar with only Yukawa interactions to $\tau$ leptons. Its contributions to the electromagnetic form factor $F_2$ are reported as well. In the limit of a very large scalar mass, form factors tend to the value of the corresponding  dipole moments.
    \textit{Right:} In the simultaneous presence of $c_P^\tau$ and $c_S^\tau$ a contribution to $d_\tau$ is generated as well. In the figure we report the corresponding predictions for the choice $c_P^\tau = c_S^\tau$.}
    \label{fig:CPVALP_tau_yuk}
\end{figure*}

The results we have obtained in this section agree with the literature~\cite{Cornella:2019uxs, Marciano:2016yhf, DiLuzio:2020oah} in the relevant limits.
As can be appreciated in Fig.~\ref{fig:CPVALP_tau_yuk}, the dependence of form factors on the momentum transfer (given at the Belle II operational energy $\sqrt{s_B}=10.58\GeV$)
always leads to an effective reduction. The general behavior of the ratio between the form factor and the corresponding dipole moment is exemplified for the case of $F^\text{y}_2/a^\text{y}_\tau$ in Fig.~\ref{fig:Energy_Comparison_CPVALP_tau_yuk}.

\subsubsection{Including nonrenormalizable couplings}

\begin{figure}[t]
    \centering
    \includegraphics[width=1\linewidth]{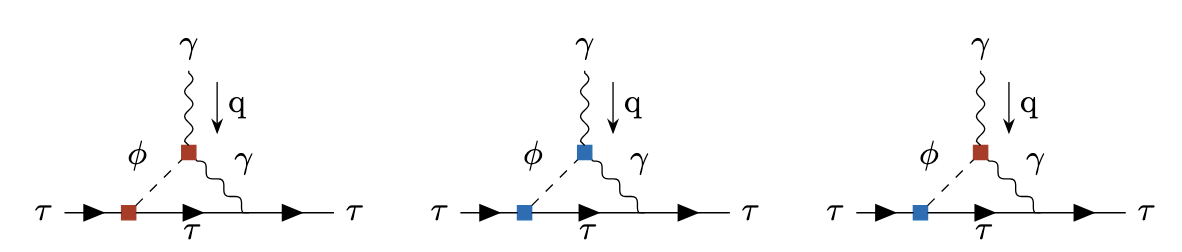}
    \caption{Feynman diagrams contributing to the ALP-mediated effects to $a_\tau$ and $d_\tau$ as induced by the exchange of a virtual scalar. Blue and red dots denote insertions of effective operators with opposite $CP$-transformation properties. The contributions in which the two EFT vertices are swapped need to be included as well.}
\label{fig:MixDiagrams}
\end{figure}

\begin{figure*}[t]
    \centering
    \includegraphics[width=0.45\linewidth]{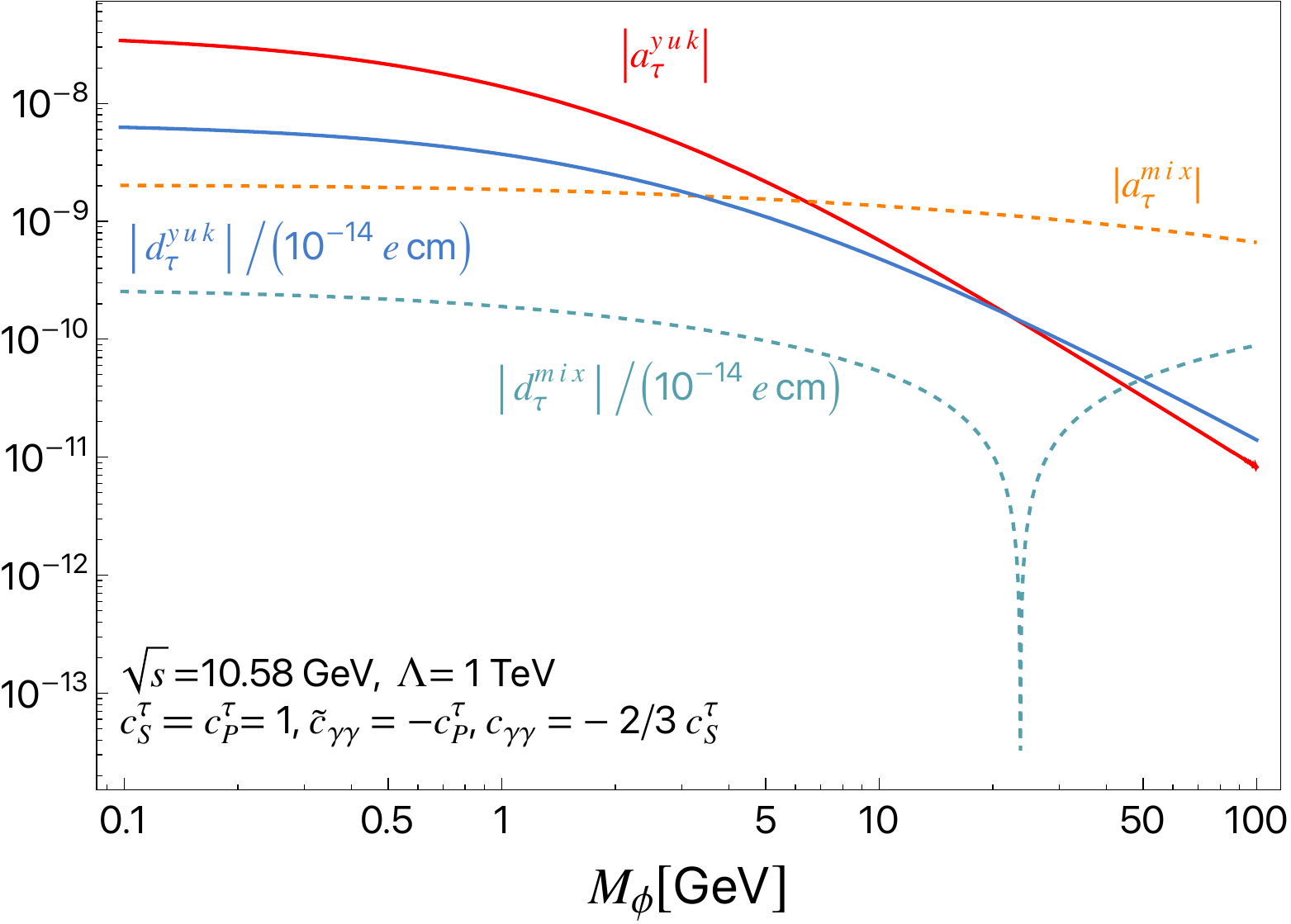}
    \includegraphics[width=0.45\linewidth]{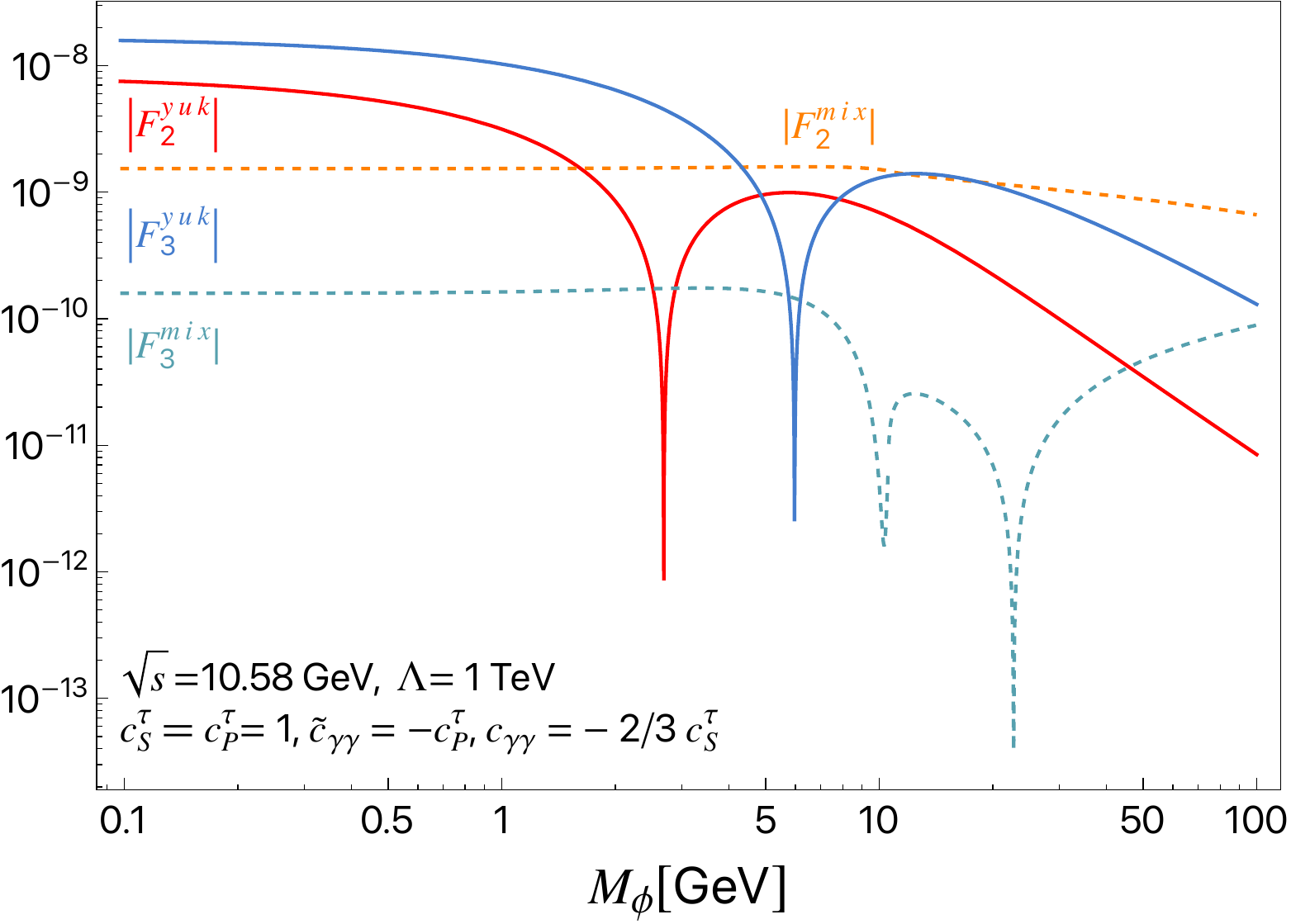}
    \caption{\textit{Left}: Comparison between the contributions to AMM and EDM as induced by pure Yukawa couplings and by the presence of mixed couplings. \textit{Right}: The same but for the form factors $F_2$ and $F_3$.}
    \label{fig:Mixed_vs_Yukawa}
\end{figure*}

The second class of contributions we consider are generated by the simultaneous presence of a renormalizable Yukawa interaction and of a nonrenormalizable coupling of the scalar field $\phi$ with photons, see Fig.~\ref{fig:MixDiagrams}.
We obtain:
\begin{widetext}
\begin{align}
F_2^\text{m}(q^2)&= \frac{\tilde{c}_{\gamma\gamma}c_P^\tau}{8\pi}\frac{m_\tau^2}{\Lambda^2}(d-3)\Big[2m_\tau^2\big(C_{11}(\theta_1)+ C_{11}(\theta_0)\big) + 2 (d-2)\big(C_{00}(\theta_1)+ C_{00}(\theta_0)\big)\Big]  \nonumber \\
& +\frac{c_{\gamma\gamma}c_S^\tau}{8\pi}\frac{m_\tau^2}{\Lambda^2}\Big[8 \big(C_{00}(\theta_1)+C_{00}(\theta_0)\big)- 4 q^2\big(C_{12}(\theta_1)+C_{12}(\theta_0)\big)\Big]\,,\nonumber \\
F_3^\text{m}(q^2)&=  \frac{c_{\gamma\gamma} c_P^\tau}{4\pi^2 }\frac{m_\tau^2}{\Lambda^2}\Big[2\, B_0(m_\tau^2, m_\tau, 0) +q^2\big(C_1(\theta_1)+C_2(\theta_0)\big) + m_\phi^2 \big(C_0(\theta_1)+ C_0(\theta_0)+ C_1(\theta_0)+ C_2(\theta_1)\big)\Big] \nonumber \\
&+\frac{\tilde{c}_{\gamma\gamma} c_S^\tau}{2\pi^2}\frac{m_\tau^2}{\Lambda^2} (d-3)(d-2)\big(C_{00}(\theta_0)+C_{00}(\theta_1) \big)\,,
\end{align}
\end{widetext}
where $\theta_0 = \{m_\tau^2, m_\tau^2, q^2, m_\phi, m_\tau,0\}$ and $\theta_1 = \{m_\tau^2, m_\tau^2, q^2, 0, m_\tau,m_\phi\}$. These expressions agree with those in the literature, for the cases available~\cite{Cornella:2019uxs}.
By evaluating the same loop diagrams for $q^2\to 0$ we then obtain the corresponding contribution to the $\tau$ AMM and EDM, which we report in App.~\ref{sec:Explicit_Expressions}.
In the limit $m_\tau \gg M_\phi$, identifying $1/\epsilon + \log \mu^2 \to \log \Lambda^2$ and focusing on the leading logarithmic dependence on $\Lambda$, these expressions reduce to
\begin{align}
 a_\tau^\text{m} (m_\tau\gg M_\phi)
 &\simeq \frac{\alpha_\text{em}}{4\pi}\frac{(\tilde{c}_{\gamma\gamma}\, c_P^\tau+ c_{\gamma\gamma}\, c_S^\tau )\,m_\tau^2}{2\pi^2\Lambda^2} \log \frac{\Lambda^2}{m_\tau^2}\,, \nonumber \\
 d_\tau^\text{m} (m_\tau\gg M_\phi)
 &\simeq \frac{\alpha_\text{em}}{4\pi}\frac{(c_{\gamma\gamma}\, c_P^\tau- \tilde{c}_{\gamma\gamma}\, c_S^\tau )\,m_\tau^2}{4\pi^2\Lambda^2}\frac{e}{2m_\tau} \log \frac{\Lambda^2}{m_\tau^2}\,.
\end{align}
The leading logarithmic terms are unaltered in the opposite limit, $m_\tau \ll M_\phi$, provided that $\log \Lambda^2/m_\tau^2 \leftrightarrow \log \Lambda^2/M_\phi^2$; they do feature, however, different finite terms, which we omit here.
In the high-energy limit, $q^2 \gg m_\tau^2, M_\phi^2$, one finds the following expansions: 
\begin{align}
F_2^\text{m}(q^2\gg m_\tau^2, M_\phi^2) &\simeq \frac{\alpha_\text{em}}{4\pi}\frac{m_\tau^2}{\Lambda^2}\frac{\tilde{c}_{\gamma\gamma} c_P^\tau + c_{\gamma\gamma}c_S^\tau}{2\pi^2} \, \log\frac{-\Lambda^2}{s}\,, \nonumber\\
F_3^\text{m}(q^2\gg m_\tau^2, M_\phi^2) &\simeq \frac{\alpha_\text{em}}{4\pi}\frac{m_\tau^2}{\Lambda^2}\frac{\tilde{c}_{\gamma\gamma} c_P^\tau - c_{\gamma\gamma}c_S^\tau}{2\pi^2} \, \log\frac{-\Lambda^2}{s}\,.
\end{align}
Also in this case an interesting limit to be considered is $s\to 4m_\tau^2 $, where we find again a power-law enhancement, which is, however, independent of the large logarithm characterizing the mixed contribution:
\begin{align}
F_2^\text{m}(s \to 4 m_\tau^2)&= \frac{1}{2 \pi^2 \,\beta_\tau}\frac{m_\tau}{s\,\Lambda}\left[\tilde{c}_{\gamma\gamma}\frac{c_P^\tau m_\tau}{\Lambda}-c_{\gamma\gamma}\frac{c_S^\tau m_\tau}{\Lambda}\right]  \nonumber \\&\times \bigg[M_\phi^2 - 4 m_\tau^2+ (6m_\tau^2-M_\phi^2)\mathcal{B}_0^{\tau\tau\phi}\notag\\
&- \frac{(M_\phi^2-4m_\tau^2)^2}{2m_\tau^2} \log \frac{m_\tau^2}{M_\phi^2-4m_\tau^2}\bigg]\notag\\
&+ \mathcal{O}(\beta_\tau^0)\,,
\end{align}
while $F_3$ does not display any kind of enhancement in such a regime.

All of the expressions above display a pole $1/\epsilon$, where $2\epsilon = 4-d$. Such a divergence has to be properly reabsorbed by considering an effective dimension-5 operator in the ALP EFT, having the following general structure:
\begin{equation}
\label{eq:UVpole_Ren}
\mathcal{L}^{d=6}_{\text{dip}} \supset \frac{v}{\sqrt{2}\Lambda^2}\,\bar{\tau} \sigma^{\mu\nu}\left(\Re C_{\tau\gamma} + i \text{ Im }C_{\tau\gamma}\right)\tau F_{\mu\nu}\,,
\end{equation}
where $\Re C_{\tau\gamma}$ and $\Im C_{\tau\gamma}$ are the real and imaginary parts of the SMEFT contribution to the $\tau$ dipole moment operator. Their magnitude can be estimated only once a specific UV completion for the scalar state $\phi$ is considered~\cite{Bauer:2017ris}. In light of this observation, it makes little sense to include also the finite parts from the previous expressions unless a UV completion is specified; for estimating the impact of these loop factors it is better justified to consider only the universal logarithmic dependence.

In Fig.~\ref{fig:Mixed_vs_Yukawa} we report a comparison between the pure Yukawa contributions and the mixed ones at Belle II operational energies, $\sqrt{s_B}= 10.58\GeV$. The pure Yukawa contributions dominate the overall value of the dipole moments and the corresponding form factors for small scalar mass values, whereas the mixed contribution becomes dominant at larger energies. This behavior is not accidental and is directly related to the origin of the two contributions, which becomes apparent in the high-energy regime.
Indeed, mixed contributions feature an insertion of the nonrenormalizable operator $\phi FF$ (or $\phi F\tilde{F}$). The corresponding contributions can be understood as running effects, which show therefore a logarithmic energy dependence. Instead, the Yukawa-like contributions stem from renormalizable operators whose energy dependence has a power-law behavior. Hence, in the large energy limit the power-law suppression of the Yukawa-induced contributions becomes more relevant than the one-loop suppression accompanying the $\phi FF$ and $\phi F\tilde{F}$ operators, resulting in a larger overall contribution from the latter.
More explicitly, we can observe that schematically
\begin{equation}
\frac{\Re F_{2,3}^\text{m}}{\Re F_{2,3}^\text{y}} \overset{s\to \infty}{\propto} \frac{\log \frac{m_\tau^2}{s}}{\log \frac{\Lambda^2}{s}}\frac{s}{m_\tau^2}\,,
\end{equation}
which indeed features the aforementioned hierarchy between the two contributions at high energies \cite{Alda:2024cxn}.

\subsection{Light vector bosons}

In this section we consider the contributions to $F_2(q^2)$ and $F_3(q^2)$ as induced by the exchange of a light spin-$1$ particle. These can be thought of as the gauge bosons associated to some extra gauge symmetry existing beyond the standard ones, or they can be regarded as being composite states emerging from some dark confining sector. They can acquire a mass term either via a dark Higgs mechanism, in the first case, or as the consequence of the confinement dynamics in the second one, in analogy to spin-$1$ QCD resonances.

In an EFT approach, we can parameterize the most relevant $ U(1)_\text{em}$ invariant interactions of such new states with SM fields as follows~\cite{Kribs:2022gri}:
\begin{align}
    \label{eq:XBos}
            \mathcal{L}^\text{int}_{X} &= i \, g_D\, X_\mu \,\bar{\tau}\, \gamma^\mu \left(c_V + \,c_A \,\gamma_5\right) \tau\,\nonumber \\
            & \quad +g_D \, e^2\, C_{A} \,\epsilon^{\mu\nu\alpha\beta}X_\mu A_\nu \partial_\alpha A_\beta \,.
\end{align}
The first terms describe the minimal interactions of a massive spin-$1$ particle with $\tau$ leptons. The second interaction, with Wilson coefficient $C_A$, represents a Chern--Simons coupling to the electromagnetic field. In the case of an anomalous gauge boson such a term can emerge from integrating out some heavy fermion fields, called anomalons, which are required to exist at high energies to cancel the anomalous contribution from the coupling of the gauge boson with SM fields. These terms crucially depend on the scheme chosen for describing the gauge anomalies and cannot be considered on their own. A fully gauge-invariant result can be obtained only provided that the one-loop amplitude involving SM leptons is considered and its divergences are treated within the same scheme~\cite{Anastasopoulos:2006cz, Dror:2017nsg, Michaels:2020fzj}, see App.~\ref{sec:Tauphilic_VBs} for further discussion. Alternatively, if the vector is a St\"uckelberg field, no gauge symmetry exists and such a class of operators has no reason to be neglected~\cite{Kribs:2022gri}.
A relevant phenomenological feature common to both scenarios is the power-law enhancement $s/M_V^2$ for small vector boson masses, which is realized by those amplitudes that involve either axial or Chern--Simons couplings of the new vector state $X$. This enhancement is nothing but a consequence of the coupling of a new vector state to a nonconserved current, and signals the nonrenormalizability of the corresponding theory, which manifests itself in energy-enhanced processes~\cite{Dror:2017ehi, Dror:2018wfl}.

\begin{figure}[t]
    \centering
    \includegraphics[width=0.4\linewidth]{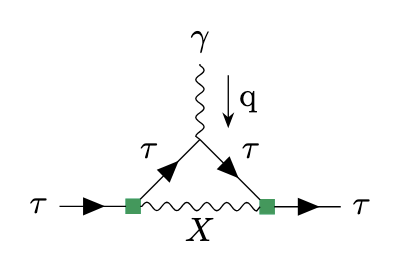}
    \caption{Feynman diagram contributing to the $\tau$ magnetic dipole moment.}
    \label{fig:MinCouDiagrams}
\end{figure}

The Lagrangian in Eq.~\eqref{eq:XBos} respects $CP$ symmetry and hence cannot generate any contribution to $d_\tau$. As done for light scalars and pseudoscalars, we report in the following sections the contributions to $a_\tau$ as induced by the exchange of a virtual spin-$1$ particle. In doing so, we will work with the standard Proca field propagator,
\begin{equation}
\Delta_{\mu\nu}(q^2)=\frac{-i}{q^2-M_V^2} \left(g_{\mu\nu}- \frac{q_\mu q_\nu}{M_V^2}\right)\,,
\end{equation}
which corresponds to the choice of working in unitary gauge if the spin-$1$ particle is a gauge boson.

\subsubsection{Minimal couplings only}

The first contribution we consider is the one induced exclusively by the minimal couplings in Eq.~\eqref{eq:XBos} and is depicted in Fig.~\ref{fig:MinCouDiagrams}.
We obtain the following result:
\begin{align}
\label{eq:F2MinCou}
F_2^{\text{mc}} &= - \frac{g_D^2 |c_V|^2}{8\pi^2} m_\tau^2 (d-2) \Big[C_{22}(\theta)+2C_{12}(\theta)+C_{22}(\theta)\nonumber \\
& \qquad \qquad \qquad \qquad\quad+2C_{1}(\theta)+2C_{2}(\theta)\Big] \nonumber \\
&- \frac{g_D^2 |c_A|^2}{2\pi^2} m_\tau^2 \frac{m_\tau^2}{m_V^2} \Big[C_{22}(\theta)+2C_{12}(\theta)+C_{22}(\theta)\Big]\nonumber \\
&- \frac{g_D^2 |c_A|^2}{2\pi^2}m_\tau^2 \Big[(d-2)  \big(C_{11}(\theta)+2C_{12}(\theta)+C_{22}(\theta)\big) \nonumber\\
& \qquad \qquad \qquad(d-1)\big(C_{1}(\theta)+C_{21}(\theta)\big)+ 8C_0(\theta)\Big]\,.
\end{align}
The contribution to the $\tau$ AMM is obtained by evaluating the same loop diagrams in the limit $q^2\to 0$; we report the corresponding expressions in App.~\ref{sec:Explicit_Expressions}.
In the limit $M_V \ll m_\tau$, the latter simplifies to 
\begin{align}
a_\tau^\text{mc}(m_\tau \gg M_V) &=\frac{g_D^2}{8\pi^2}|c_V|^2\notag\\
&- \frac{g_D^2}{4 \pi^2}|c_A|^2\bigg[\frac{m_\tau^2}{M_V^2} - \frac{5}{2} + \log \frac{m_\tau^2}{M_V^2}\bigg]\,,
\end{align}
while in the opposite limit, $M_V \gg m_\tau$, one finds
\begin{equation}
a^\text{mc}_\tau(m_\tau \ll M_V) =\frac{g_D^2}{12\pi^2} \frac{m_\tau^2}{M_V^2}\,\big(|c_V|^2-5|c_A|^2\big)\,,
\end{equation}
in accordance with the results available in the literature~\cite{Leveille:1977rc,Heeck:2016xkh, Crivellin:2022gfu}.
In the high-energy limit, the form factor contains logarithmic terms of the form
\begin{align}
F_2(q^2 \gg M_V^2, m_\tau^2) &\supset  -\frac{g_D^2|c_V|^{2}}{4 \pi^2}\frac{m_\tau^2}{s}\log \frac{-s}{m_\tau^2}  \nonumber \\
&+\frac{g_D^2|c_A|^{2}}{4 \pi^2}\frac{m_\tau^2}{s} \bigg[ 2 \frac{m_\tau^2}{M_V^2}\log \frac{-s}{m_\tau^2}\notag\\
&\qquad+2\log^2 \frac{-s}{M_V^2}- 9 \log\frac{-s}{m_\tau^2} \bigg]\,,
\end{align}
where we have not reported further terms involving (poly-)logarithms of functions of $M_V$ and $m_\tau$ that do not show a logarithmic dependence on $s$.

\begin{figure}[t]
    \centering
    \includegraphics[width=0.9\linewidth]{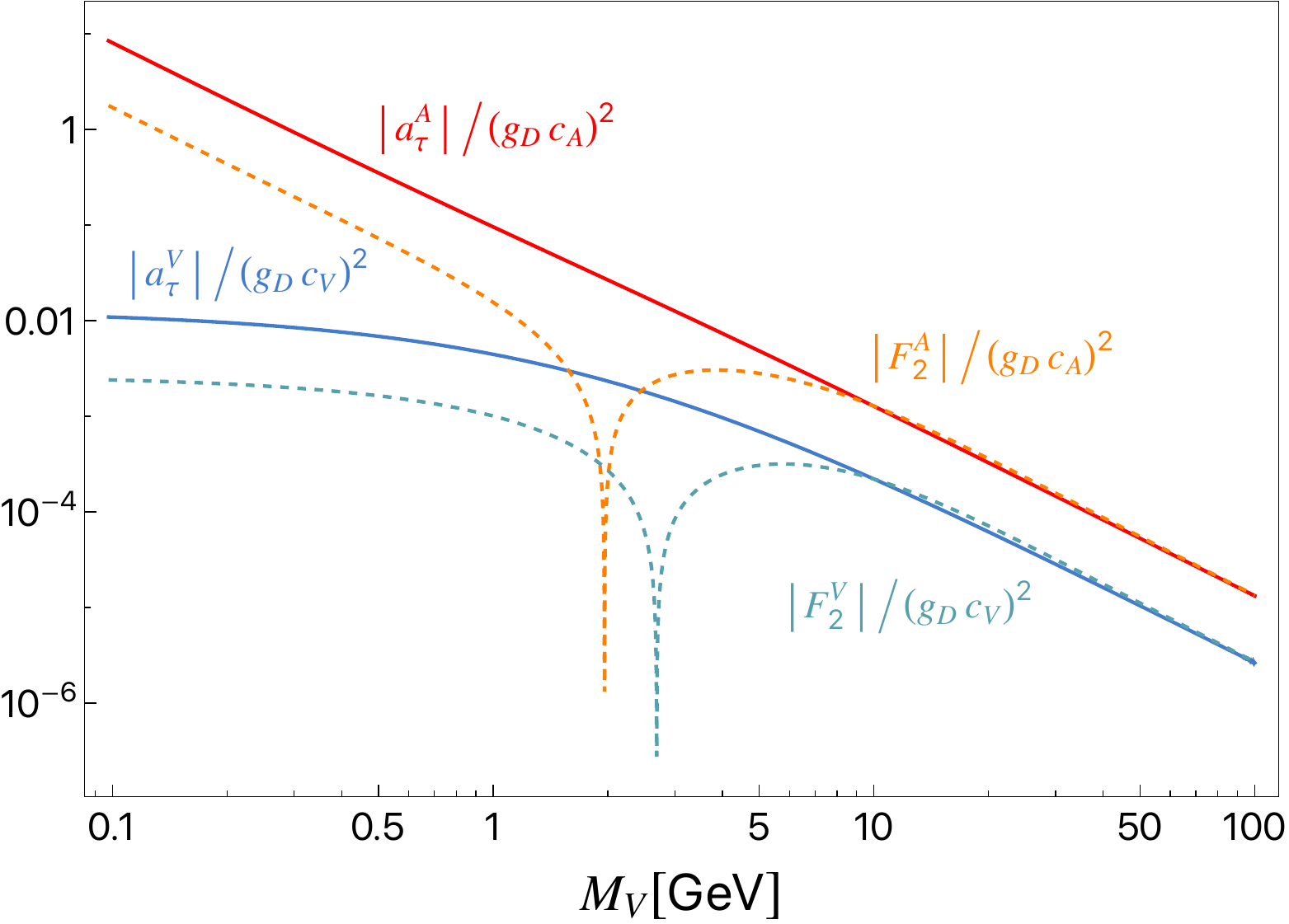}
    \caption{Total contribution to $a_\tau$ from the virtual exchange of a light vector boson. }
    \label{fig:VB_tot}
\end{figure}

\begin{figure}[t]
    \centering
    \includegraphics[width=0.33\linewidth]{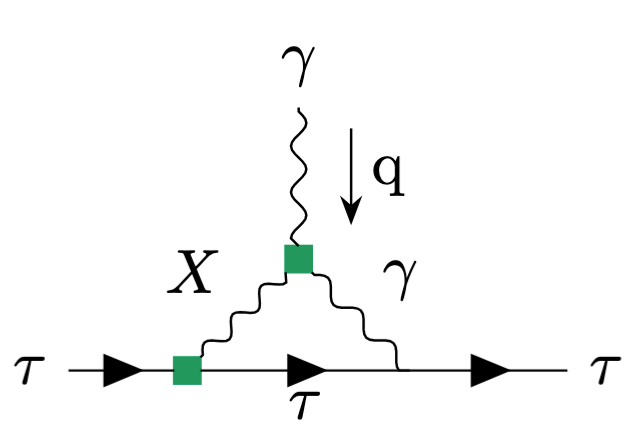}
    \caption{Feynman diagram contributing to the vector-mediated effects to $a_\tau$.}
    \label{fig:CSdiagram}
\end{figure}

An interesting point to be stressed is the large enhancement related to the axial coupling of the light vector boson to leptons~\cite{Dror:2017nsg}.
In the limit $M_V^2\gg s\gg m_\tau^2$ we rather find:
\begin{align}
F_2(M_V^2 \gg q^2 \gg m_\tau^2) &= -\frac{g_D^2}{12\pi^2}\frac{m_\tau^2}{M_V^2}\bigg[5 |c_A|^2- |c_V|^2 \nonumber \\
& -\frac{s}{M_V^2}\frac{1}{24}\bigg\{3|c_A|^2 \Big(\log \frac{M_V^2}{-s}-1\Big)\notag\\
&+|c_V|^2\Big(5-\log \frac{M_V^2}{-s}\Big)\bigg\}\bigg]\,,
\end{align}
while approaching the $\tau^+\tau^-$ threshold $s=q^2 \to 4m_\tau^2$ we find again an interesting power-law enhancement for $F_2$, which can be exploited by properly tuning the energy entering the collision under consideration:
\begin{align}
F_2^{\text{mc}}(s \to 4m_\tau^2)&=
\frac{3ig_D^2}{4\pi} \frac{m_\tau}{(s\beta_\tau)^{3/2}}\notag\\
&\quad\times\Big[2m_\tau^2 |c_A|^2+ M_V^2 (|c_A|^2+|c_V|^2)\Big] \nonumber \\
&+ \frac{g_D^2}{4 \pi^2 m_\tau^2 M_V^2\,s\,\beta_\tau}\notag\\
&\quad\times\Big[2M_V^2|c_A|^2+ M_V^2(|c_A|^2+ |c_V|^2)\Big]\notag\\
&\quad\times\bigg[m_\tau^2\Big((2m_\tau^2+ M_V^2)\mathcal{B}_0^{\tau\tau V}- 2M_V^2\Big)\notag\\
&\qquad + M_V^4 \log \frac{m_\tau}{M_V}\bigg]
+ \mathcal{O}(\beta_\tau^{-1/2})\,.
\end{align}
Our results are displayed in Fig.~\ref{fig:VB_tot}.

\subsubsection{Including Chern--Simons couplings}

\begin{figure}[t]
    \centering
    \includegraphics[width=0.9\linewidth]{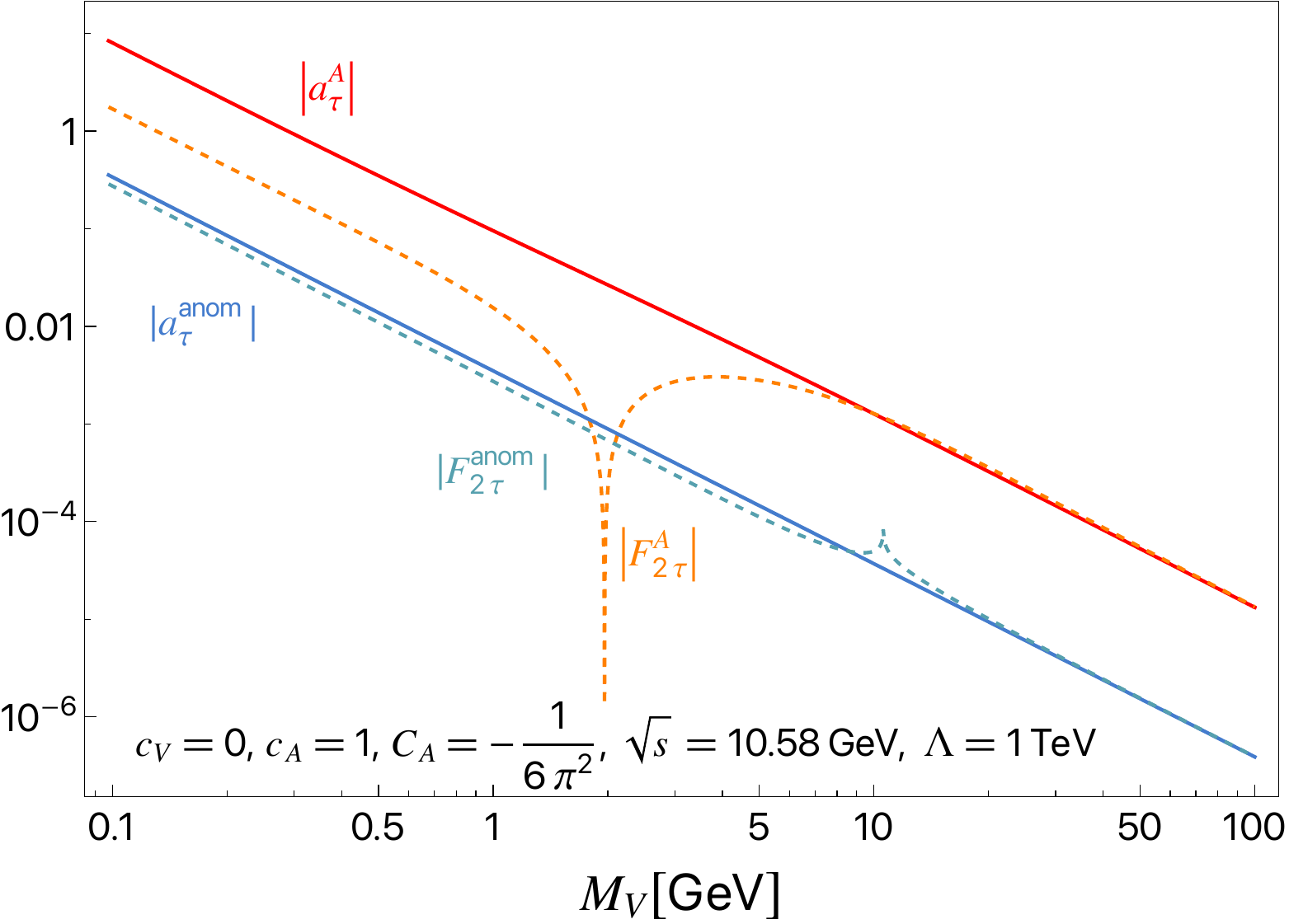}
    \caption{Total contribution to $a_\tau$ from the virtual exchange of a light vector boson.}
    \label{fig:VB_anom}
\end{figure}

The second class of contributions we consider are generated by the simultaneous presence of an interaction of the vector field $X$ and a Chern--Simons three-vector coupling with photons, see Fig.~\ref{fig:CSdiagram}.
We obtain:
\begin{align}
F^\text{an}_2 &=  \alpha_{\text{em}}g_D^2c_A C_A\frac{m_\tau^2}{M_V^2} \bigg[-d \,M_V^2 \Big( C_{22}(\theta_2)+ 2 C_{12}(\theta_2)\nonumber \\
& \qquad  + C_{11}(\theta_2)-3 C_2(\theta_2)- C_1(\theta_2)- 12 C_{00}(\theta_2) \nonumber \\
& \qquad  - 4 C_{002}(\theta_2)- 4 C_{001}(\theta_2)\Big) \notag\\
&+ 2m_\tau^2\Big(6 C_{22}(\theta_2)+ 12C_{12}(\theta_2) + 6C_{11}(\theta_2)\notag\\
&\qquad+2C_{222}(\theta_2)+2C_{111}(\theta_2)
+6C_{122}(\theta_2)\notag\\
&\qquad+6C_{1}(\theta_2)+6C_{2}(\theta_2)+2C_{0}(\theta_2)\Big)\notag\\
&- 2q^2 \Big(2C_{12}(\theta_2)+2C_{112}(\theta_2)+2C_{122}(\theta_2)\Big)\nonumber\\
&+ 2 M_V^2 \Big(C_{22}(\theta_2)+ 2 C_{12}(\theta_2)+C_{11}(\theta_2)\nonumber \\
& \qquad  -3 C_{0}(\theta_2)-6C_{2}(\theta_2)-4C_{1}(\theta_2)\Big) \nonumber \\
& -8\Big(C_{001}(\theta_2)+C_{002}(\theta_2)+2C_{00}(\theta_2)\Big)  + \theta_2 \leftrightarrow \theta_3 \bigg]\,,
\end{align}
where $\theta_2 = \{m_\tau^2, q^2, m_\tau^2, m_\tau, 0, M_V\}$ and $\theta_3 = \{m_\tau^2, q^2, m_\tau^2, m_\tau, M_V,0\}$. The corresponding effect on $a_\tau$ can be obtained by considering the same loops in the $q^2 \to 0$ limit, see App.~\ref{sec:Explicit_Expressions}.
In the limit $M_V \ll m_\tau$, the expression for $a_\tau$ simplifies to
\begin{equation}
a^\text{an}_\tau(m_\tau \gg M_V) =  -\frac{C_A\,c_A}{2\pi^2}\frac{m_\tau^2}{M_V^2} e^2 g_D^2 \left(1+  \log \frac{\Lambda^2}{m_\tau^2}\right)\,,
\end{equation}
whereas in the opposite limit, $M_V \gg m_\tau$, one finds
\begin{align}
a^\text{an}_\tau(m_\tau \ll M_V) &= -\frac{C_A\,c_A}{72 \pi^2}\frac{m_\tau^2}{M_V^2} e^2 g_D^2\notag\\
&\times\left(65+ 36 \log \frac{\Lambda^2}{m_\tau^2} + 6 \log \frac{M_V^2}{m_\tau^2}\right)\,.
\end{align}
Regarding the form factor, we find for its high-energy limit
\begin{equation}
F^\text{an}_2(q^2 \gg m_\tau^2, M_V^2) = -\frac{C_A\,c_A}{2\pi^2}\frac{m_\tau^2}{M_V^2}e^2 g_D^2\left[2+ \log\frac{-\Lambda^2}{s}\right]\,.
\end{equation}
Close to the $\tau^+\tau^-$ threshold, it behaves as
\begin{align}
\label{eq:F2an_thr}
F_2^\text{an}(s \to 4m_\tau^2)&=-C_A e^2 g_D^2 c_A \frac{4m_\tau^2-M_V^2}{16M_V^2\, s\,\pi^2 \beta_\tau}  \notag\\
&\times\bigg[M_V^2  - 4 m_\tau^2 +  (6m_\tau^2 - M_V^2)\mathcal{B}_0^{\tau \tau V} \nonumber \\
&\qquad - \frac{(M_V^2-4m_\tau^2)^2}{2m_\tau^2} \log\frac{m_\tau^2}{M_V^2-4m_\tau^2}\bigg]\notag\\
&+ \mathcal{O}(\beta_\tau^0)\,.
\end{align}
Our results are displayed in Fig.~\ref{fig:VB_anom}.
Also in this case a power-law enhancement for light vector boson masses is present, which can be understood as being induced by the longitudinal component of the light vector boson. This is another example of the low-energy realization of the Goldstone boson equivalence theorem (GBET), according to which, up to corrections of order $M_V^2/s$, the amplitudes involving the longitudinal component of a massive vector are identical to the ones involving the Goldstone boson that has been incorporated within the massive spin-$1$ field in unitary gauge.
In addition to this, it has to be noticed that the contribution from Chern--Simons couplings has a logarithmic sensitivity to the scale of NP. This signals the effective nature of such an interaction, which can be easily understood again in the light of the GBET: by expressing the longitudinal component of the vector field $X_\mu$ as $X^\mu_L = \partial^\mu \phi_X/M_V$ and by integrating by parts the Chern--Simons interaction, it is possible to reduce it to an effective dimension-5 coupling that is analogous to the $\phi F\tilde{F}$ interaction we have already analyzed for a pseudoscalar~\cite{Dror:2017nsg,Michaels:2020fzj, Kribs:2022gri}. More explicitly, denoting by $\Omega_\mu = \epsilon_{\mu\nu\alpha\beta}A^{\nu}F^{\alpha \beta} = \frac{1}{2}\epsilon_{\mu\nu\alpha\beta}A^{\nu}\partial^\alpha A^\beta$, one has
\begin{align}
\Omega_\mu X^\mu_L \rightarrow \Omega_\mu \frac{\partial^\mu \phi_X}{M_V} &= - \frac{\phi_X}{M_V}\partial^\mu \Omega_\mu=  -\frac{\phi_X}{M_V}F\tilde{F} \notag\\
&\propto -\frac{\,\phi_X}{ g_D\,\Lambda} F\tilde{F}\,,
\end{align}
where in the last step we have assumed that $M_V \propto g_D\Lambda$, as it would be the case in a Higgs mechanism providing a mass to the vector boson.
In the case of a gauge vector boson acquiring its mass via a Higgs mechanism, these contributions have been taken into account comprehensively in the evaluation of the impact of a light gauge vector boson on $a_\mu$ in Ref.~\cite{Anastasopoulos:2022ywj}, where the full evaluation up to two-loop order has been carried out.
We also remark that, once the full fermion content of the theory is spelled out, as it is the case for UV complete theories, the closed formula provided in Refs.~\cite{Jegerlehner:2017gek,Ludtke:2024ase} can be appropriately adapted to the computation of the full, finite contribution to $a_\tau$ and $F_2$.

\section{Constraints from asymmetries in \texorpdfstring{$\boldsymbol{e^+e^-\to\tau^+\tau^-}$}{}}
\label{sec:Asymmetries}

\begin{figure*}[t]
    \centering
    \includegraphics[width=0.45\linewidth]{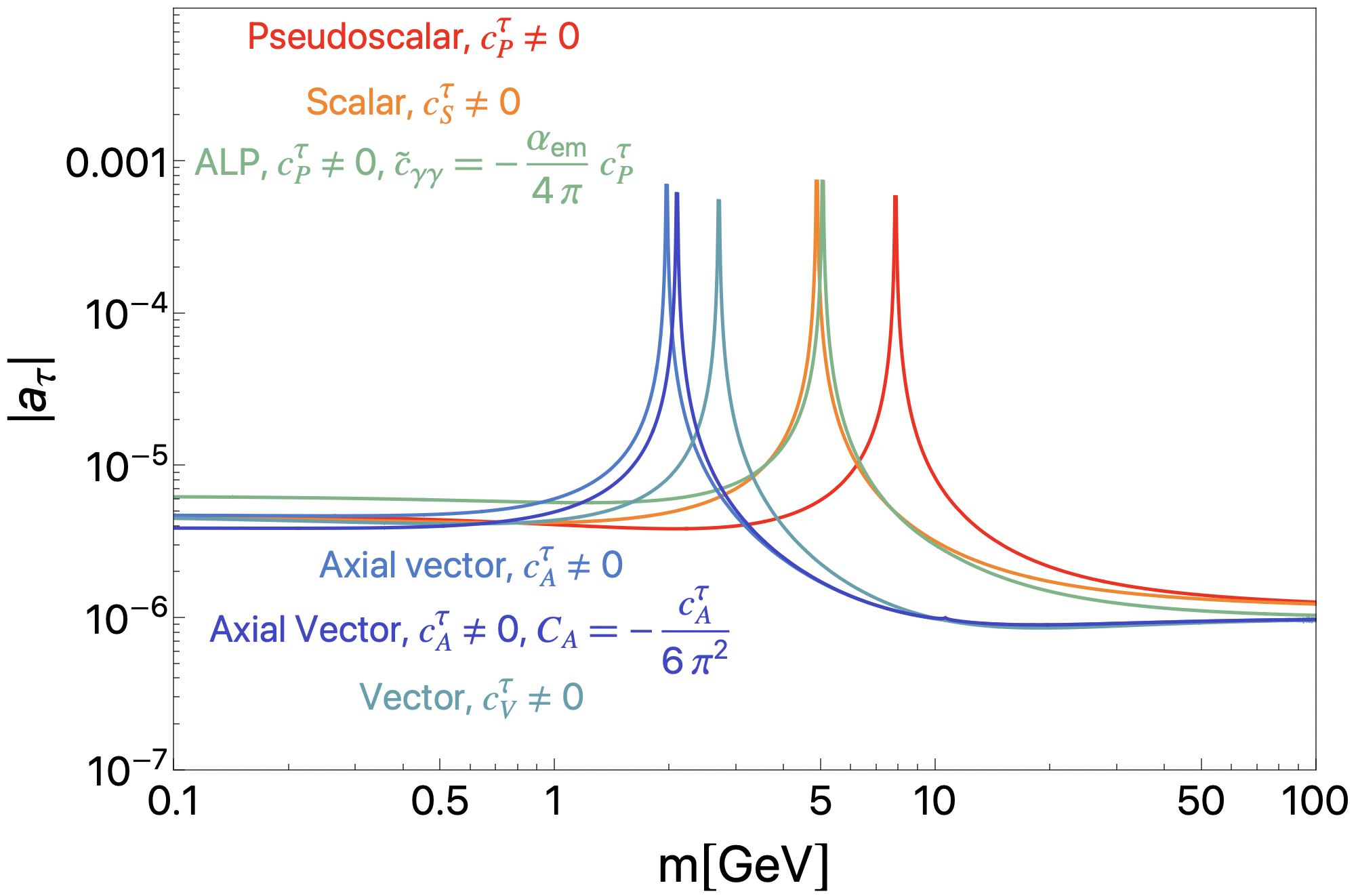}
    \includegraphics[width=0.45\linewidth]{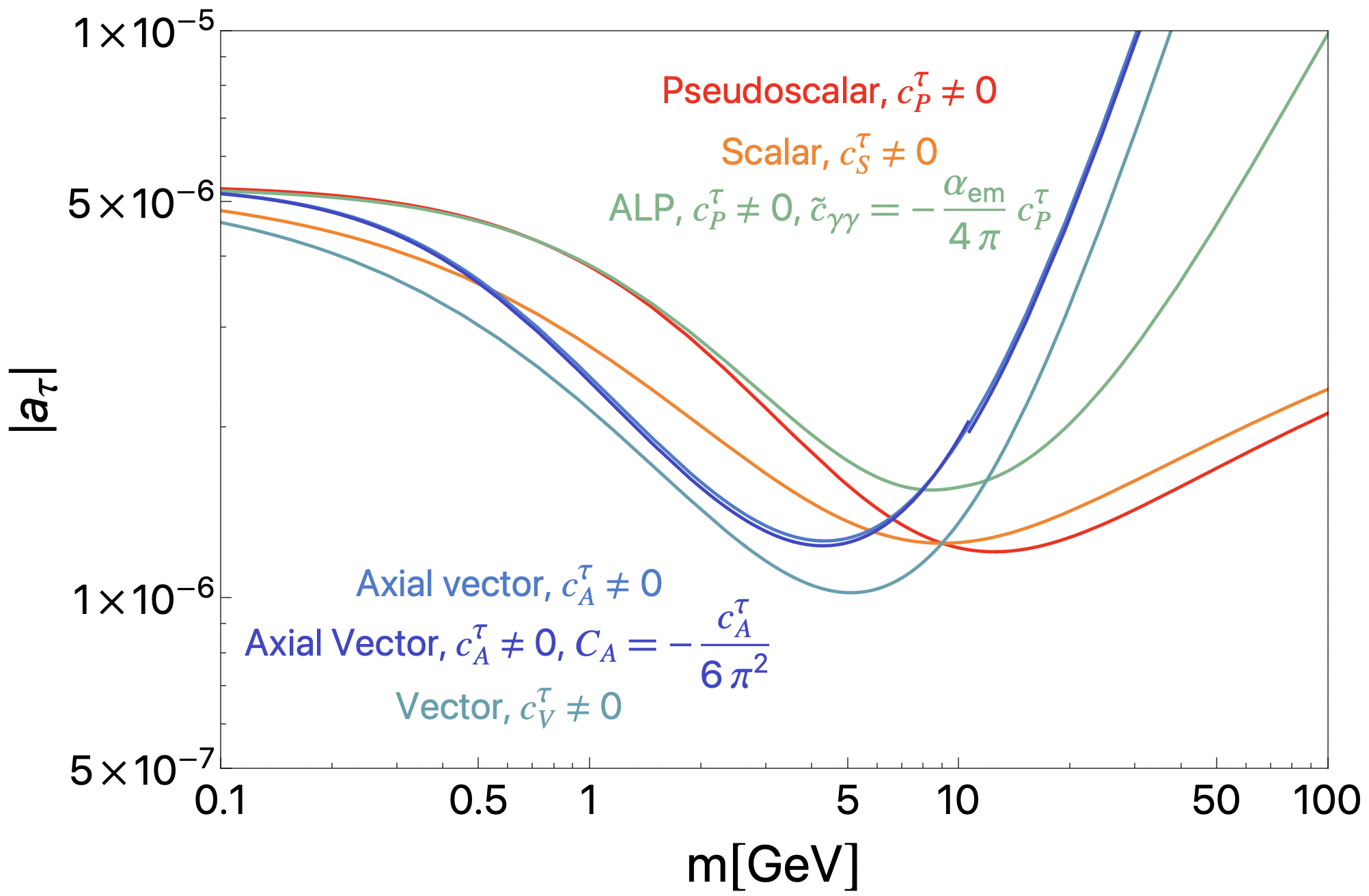}
    \caption{Comparison of the sensitivity of the Belle II experiments to light NP affecting $a_\tau$ for $\Lambda = 1\TeV$. The choice of $c_{\gamma\gamma} = -\alpha_\text{em}/(4\pi) \,c_P^\tau$ corresponds to the minimal coupling present in the case of a derivatively coupled ALP, being unavoidably generated in passing from the derivative to the nonderivative basis. The choice of $C_A = -y_A/(6\pi^2)$ is dictated by the requirement of gauge anomaly cancellation. \textit{Left (Right)}: Bounds obtained assuming a sensitivity on $\Re F_2^\text{eff} (\Im F_2^\text{eff})= 10^{-6}$ as a function of the mass of the NP mediator. Figure from Ref.~\cite{Hoferichter:2025ijh}.}
    \label{fig:Money_Plots}
\end{figure*}

In the previous sections we discussed the impact of light NP on the AMM and the EDM of the $\tau$ lepton.
In Ref.~\cite{Hoferichter:2025ijh} we proposed to assess the sensitivity of the Belle II experiment to such light NP candidates by investigating their impact on the electromagnetic form factors $F_2$ and $F_3$.
It is our purpose in this section to provide further details on these findings.

The fundamental idea underlying our work consists of observing that direct information on $F_{2,3}(q^2)$ can be accessed by considering properly defined asymmetries in the scattering process $e^+ e^- \to \tau^+ \tau^-$.
Such a possibility has been discussed in detail in the literature~\cite{Bernabeu:2007rr, Bernabeu:2008ii, Crivellin:2021spu, Gogniat:2025eom}, where the relation between experimentally measurable asymmetries and form factors $F_2$ and $F_3$ was spelled out.
The latter can be written as 
\begin{equation}
F_{\{2,3\}}(s) = F^\text{SM}_{\{2,3\}}(s) + F^\text{NP}_{\{2,3\}}(s)\,,
\end{equation}
so that a potential NP effect can be extracted from the experimentally measured $F_i(s)$ by properly subtracting the SM contribution.
In particular, in the case of heavy NP, the quantities in which one is interested are the real parts of the form factors $F_2$ and $F_3$, since
\begin{align}
\label{ReF2F3_SM_NP}
\Re  F_2(s) &= \Re  F^\text{SM}_2(s) + a_\tau^\text{NP}\,, \nonumber \\
\Re  F_3(s) &= \Re  F^\text{SM}_3(s) + \frac{2m_\tau}{e}d_\tau^\text{NP} \simeq \frac{2m_\tau}{e}d_\tau^\text{NP}\,.
\end{align}
This result is valid as long as effects of order $s/m_\text{NP}^2$ can be neglected, as it is the case for heavy NP candidates; moreover, in the last equality we have implicitly neglected SM contributions to the $\tau$ EDM~\cite{Yamaguchi:2020eub}.
In case polarized electron beams are available, $\Re F_2$ and $\Re F_3$ can be experimentally measured by considering appropriate asymmetries built out of the polarized cross sections
\begin{align}
d \sigma_\text{pol}^S &= d \sigma^{S\lambda}|_{\lambda=1}-d \sigma^{S\lambda}|_{\lambda=-1}\,, 
\nonumber \\
\sigma_\text{pol,FB}^S&= \int_0^1 dz \frac{d \sigma_\text{pol}^S}{d\Omega}-\int_{-1}^0 dz \frac{d \sigma_\text{pol}^S}{d\Omega}\,,
\end{align}
where $\lambda$ denotes the polarization of the incoming electron beam and $z = \cos \theta$. Referring with $\phi_\pm, \theta^{*}_\pm$ to the azimuthal and polar angles of the produced hadron $h^\pm$ in the semileptonic $\tau^\pm$ decay in its rest frame, we can then construct the following quantities:
\begin{align}
\sigma_{L, F_3}^\pm &= \int_\pi^{2\pi} d\phi_\pm \frac{d\sigma_\text{pol}^S}{d\phi^\pm}\,, &\sigma_{R, F_3}^\pm &= \int_0^{\pi} d\phi_\pm \frac{d\sigma_\text{pol}^S}{d\phi^\pm}\,, \nonumber \\
\sigma_{L, \text{pol}}^\pm &= \int_{-\pi/2}^{\pi/2} d\phi_\pm \frac{d\sigma_\text{pol}^S}{d\phi^\pm}\,, &\sigma_{R, \text{pol}}^\pm &= \int_{\pi/2}^{3/2\pi} d\phi_\pm \frac{d\sigma_\text{pol}^S}{d\phi^\pm}\,, \nonumber \\
\sigma_{\text{FB}, R}^\pm &= \int_0^{1} dz^*_\pm \frac{d\sigma_\text{pol}^S}{dz^*_\pm}\,, &\sigma_{\text{FB}, L}^\pm &= \int_{-1}^{0} dz^*_\pm \frac{d\sigma_\text{pol}^S}{dz^*_\pm}\,.
\end{align}
These, in turn, can be employed to construct the asymmetries~\cite{Bernabeu:2008ii, Crivellin:2021spu,Gogniat:2025eom}:
\begin{align}
\label{asymmetries}
A_T^\pm &= \frac{\sigma_{R, \text{pol}}^\pm-\sigma_{L, \text{pol}}^\pm}{\sigma_{R, \text{pol}}^\pm+\sigma_{L, \text{pol}}^\pm}\,, \qquad
A_L^\pm = \frac{\sigma_{\text{FB},R}^\pm-\sigma_{\text{FB}, L}^\pm}{\sigma_{\text{FB},R}^\pm+\sigma_{\text{FB}, L}^\pm}\,,\notag\\
A_{N, F_3}^\pm &= \frac{\sigma_{L, F_3}^\pm-\sigma_{R, F_3}^\pm}{\sigma_{L, F_3}^\pm+\sigma_{R, F_3}^\pm}\,,
\end{align}
which are experimentally accessible and can be related to quantities of theoretical interest via the relations
\begin{align}
\label{eq:F2F3eff_anomalies}
\Re  F_2^\text{eff} &\equiv \mp \frac{4s\beta_e\sigma_\text{tot}}{\pi^2\alpha^2\beta_\tau^3\gamma_\tau \alpha_\pm} \left(A_T^\pm - \frac{\pi}{2\gamma_\tau}A_L^\pm\right)\notag\\
&=\Re (F_2F_1^*)+ |F_2|^2\,,\notag\\
\Re  F_3^\text{eff} &\equiv \frac{4s\beta_e\sigma_\text{tot}}{\pi^2\alpha^2\beta_\tau^2\gamma_\tau \alpha_\pm}A_{N, F_3}^\pm\notag\\
&=\Re (F_3F_1^*)+ \Re (F_3F_2^*)\,,
\end{align}
where we have defined 
\begin{equation}
 \beta_\ell = \sqrt{1-\frac{4m_\ell^2}{s}}\,, \quad \gamma_\ell = \frac{\sqrt{s}}{2m_\ell}\,,
\end{equation}
and $\alpha_\pm$ depends on the spin of $h^\pm$~\cite{Gogniat:2025eom}.
Accordingly, a measurement of the asymmetries~\eqref{asymmetries} gives access to the sought interference terms in Eq.~\eqref{eq:F2F3eff_anomalies}, from which constraints on $a_\tau$ and $d_\tau$ can be inferred via Eq.~\eqref{ReF2F3_SM_NP}. The feasibility of this program, of course, depends crucially on the availability of a polarized electron beam, as could become possible with the polarization upgrade of the SuperKEKB $e^+e^-$ collider, as well as on control over radiative corrections~\cite{Gogniat:2025eom} to remove the SM background.

In the case of light NP, an additional complication arises, since NP states contribute dynamically to the measured form factors:
\begin{align}
\Re  F_2(s) &= \Re  F^\text{SM}_2(s) + \Re F_2^\text{NP}(s)\,,\nonumber\\
\Re  F_3(s) &= \Re  F^\text{SM}_3(s) + \Re  F_3^\text{NP}(s) \simeq \Re  F_3^\text{NP}(s)\,.
\end{align}
However, the NP contributions to the form factors are in a one-to-one correspondence to the induced contribution to AMM and EDM, as they depend on the same combination of couplings. The same techniques discussed above can therefore be employed in order to gain some information on $\Re  F_{\{2,3\}}$, and hence on the impact of NP states on $a_\tau$ and $d_\tau$, leading to the main result from Ref.~\cite{Hoferichter:2025ijh} reproduced in Fig.~\ref{fig:Money_Plots}, which shows the limits for the various NP scenarios that follow from $\Re F_2^\text{eff}$ when reinterpreted in terms of $a_\tau$. While the actual limits become model dependent, for heavy mediators they converge to the EFT expectation, and apart from accidental cancellations the general sensitivity stays below $10^{-5}$ (when assuming $\Re F_2^\text{eff}$ to be measured at $10^{-6}$ precision). The EFT decoupling proceeds much faster for spin-$1$ states than for spin-$0$ states, and the general EFT arguments from Ref.~\cite{Hoferichter:2025ijh} for this behavior can be verified with the explicit expressions provided in Sec.~\ref{sec:light_NP} and App.~\ref{sec:Explicit_Expressions}.

\begin{figure*}[t]
    \centering
    \includegraphics[width=0.45\linewidth]{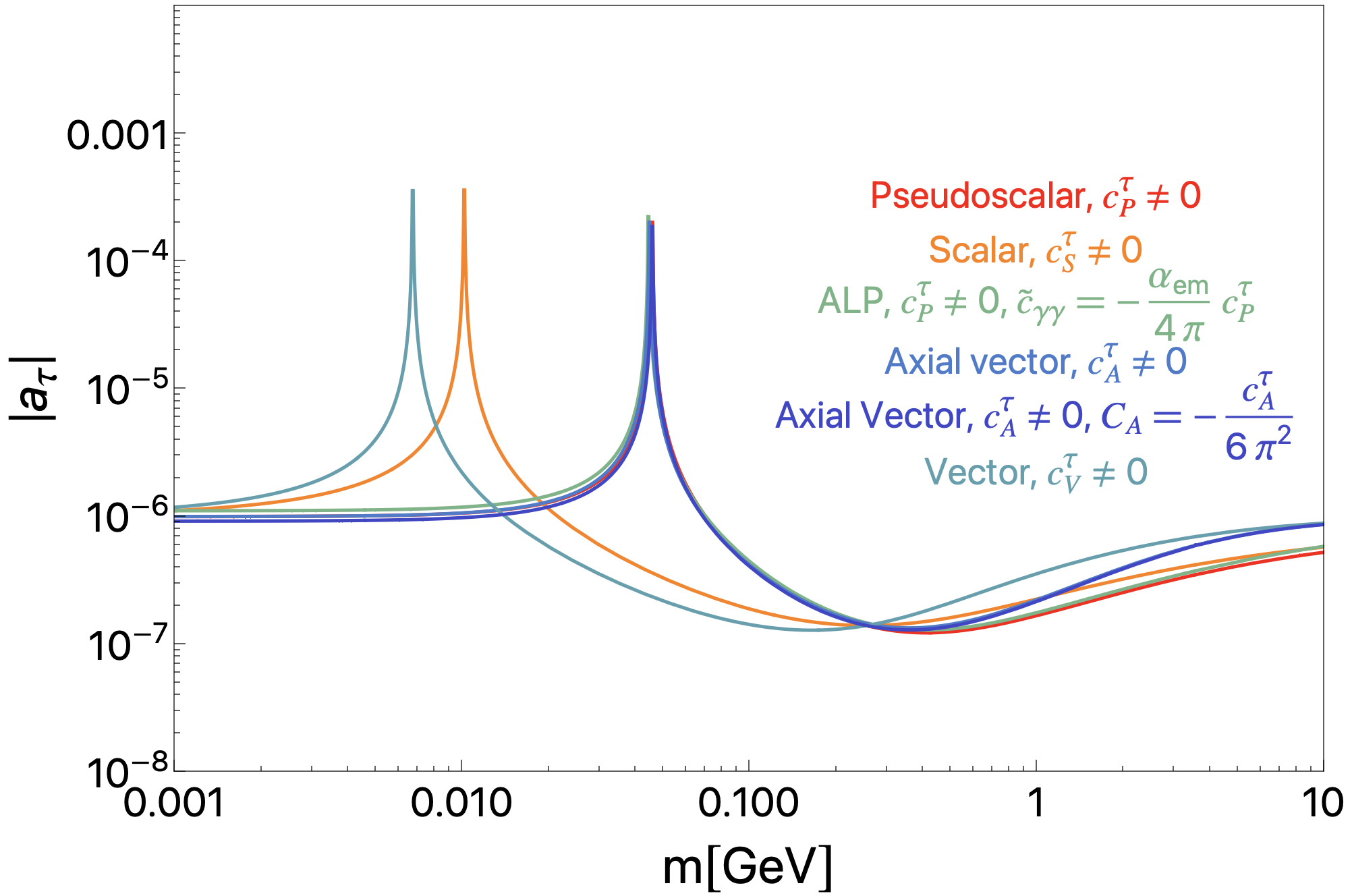}
    \includegraphics[width=0.45\linewidth]{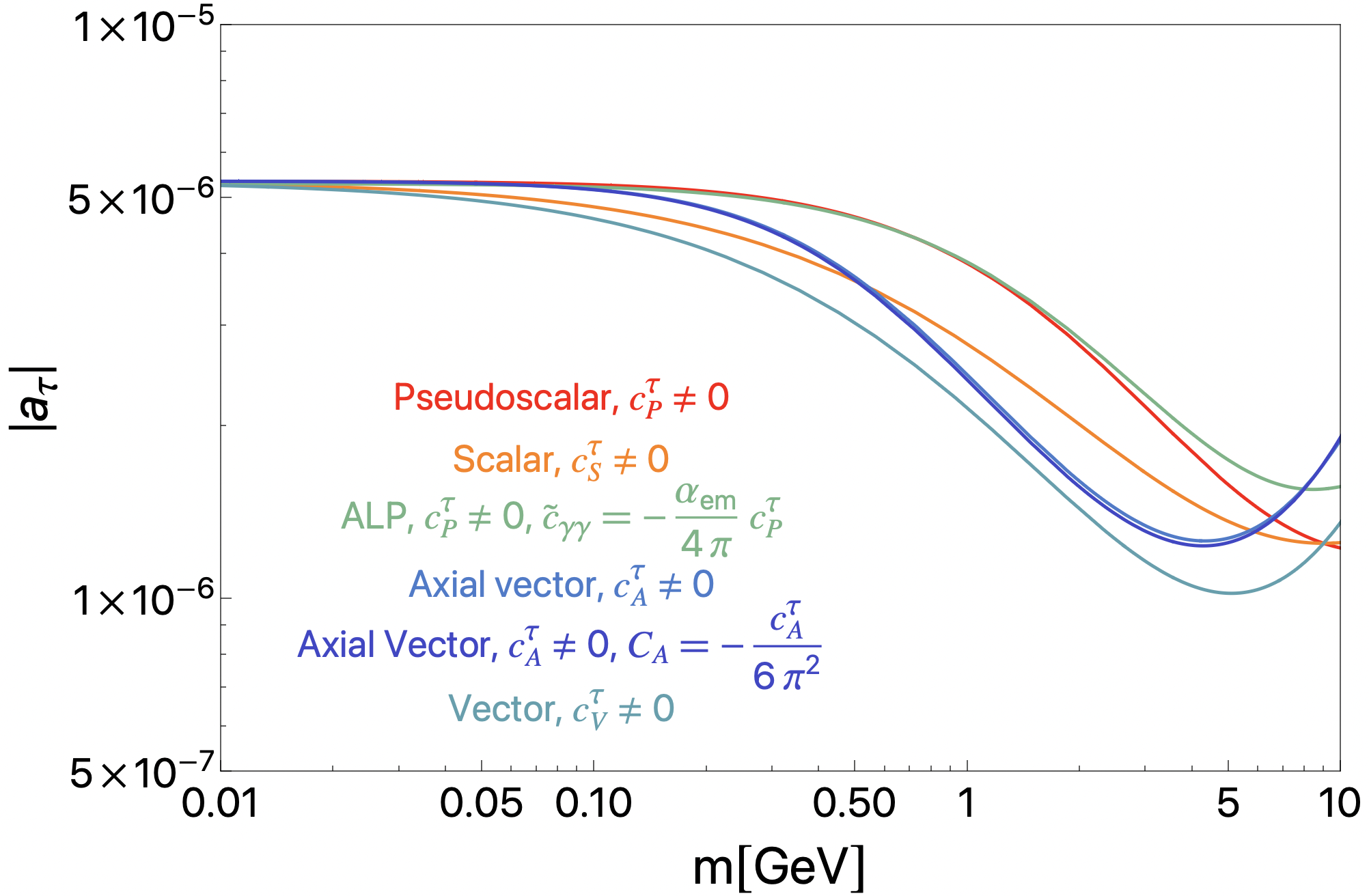}
    \caption{Comparison of the sensitivity of the Belle II experiment to light NP affecting $a_\tau$ just above the $\tau^+\tau^-$ threshold, $s_{\tau\tau} = 4(1.78\GeV)^2$ (same notation as in Fig.~\ref{fig:Money_Plots}). Figure from Ref.~\cite{Hoferichter:2025ijh}.}
    \label{fig:DiTau_threshold}
\end{figure*}

Figure~\ref{fig:Money_Plots} also includes the results obtained from a second strategy unique to light NP~\cite{Hoferichter:2025ijh}. For sufficiently large momentum transfer, $q^2 > (m_i+m_j)^2$, where $i,j$ denote two of the particles involved in the loop, the virtual exchange of new light states can lead to an imaginary part in the form factors in addition to the SM one, $\Im  F_{\{2,3\}}^\text{NP}$. As this contribution stems from the same diagrams that contribute to $\Re  F_{\{2,3\}}^\text{NP}$, the imaginary part of such form factors is sensitive to the same combination of NP couplings as their real counterpart. Moreover, the imaginary part of form factors can be accessed also in the absence of polarized electron beams~\cite{Bernabeu:2008ii, Crivellin:2021spu, Gogniat:2025eom}, by considering the quantity
\begin{equation}
A_N^\pm = \frac{\sigma_L^\pm - \sigma_R^\pm}{\sigma_L^\pm + \sigma_R^\pm}\,,
\end{equation}
where
\begin{align}
&\sigma_L^\pm = \int_{\pi}^{2\pi}d\phi_\pm \frac{d\sigma_\text{FB}}{d\phi_\pm}\,,\quad \sigma_R^\pm = \int_{0}^{\pi}d\phi_\pm \frac{d\sigma_\text{FB}}{d\phi_\pm}\,, \nonumber \\
&\sigma_\text{FB} = \int_0^1 dz\frac{d\sigma^S}{d\Omega}-\int_{-1}^0 dz\frac{d\sigma^S}{d\Omega}\,,
\end{align}
from which another (effective) observable can be constructed
\begin{equation}
\Im F_2^\text{eff} \equiv \pm \frac{3s\sigma_\text{tot}}{\pi\alpha^2\beta_e\beta_\tau^3\gamma_\tau\alpha_\pm}A_N^\pm
=\Im (F_2F_1^*)\,.
\end{equation}
When the mass of the mediator is increased, the sensitivity disappears  in line with the EFT decoupling, but for moderate masses the limits are competitive with those extracted via the real part, with largest sensitivity around the CM energy of the collider. In addition, one could try to profit from the threshold enhancement near $s=4m_\tau^2$; these results are reproduced in Fig.~\ref{fig:DiTau_threshold}, following from the expansions around threshold provided in Sec.~\ref{sec:light_NP}.

These results show how the asymmetry measurements at Belle II could be interpreted in the case of light NP, and which additional opportunities arise for such scenarios. However, we need to emphasize that such contributions to $e^+e^-\to\tau^+\tau^-$ are not the only possible effects. That is, one has to subtract from the overall result all of those unavoidable nuisance contributions other than the virtual corrections to the electromagnetic $\tau$ vertex. These include, in general, box diagrams connecting the initial and the final state via the exchange of a virtual NP candidate, or vertex corrections to the initial-state electron electromagnetic vertex. Such effects can be neglected in the presence of a certain hierarchy between the coupling of NP states and SM fields. This could be the case, for instance, for ALPs that couple derivatively to SM fermions, or for light vector bosons gauging specific combinations of lepton numbers, e.g., $L_{\tau}-L_{\mu}$. It is important to stress, however, that this is not guaranteed to happen, and that one should always consider both the specific NP model and the experimental technique that is being employed to extract bounds on $F_2$ and $F_3$. We leave the study of such corrections as an avenue for future work.

\section{Constraints from other processes}
\label{sec:other_processes}

\begin{figure*}[]
    \centering
    \includegraphics[width=0.45\linewidth]{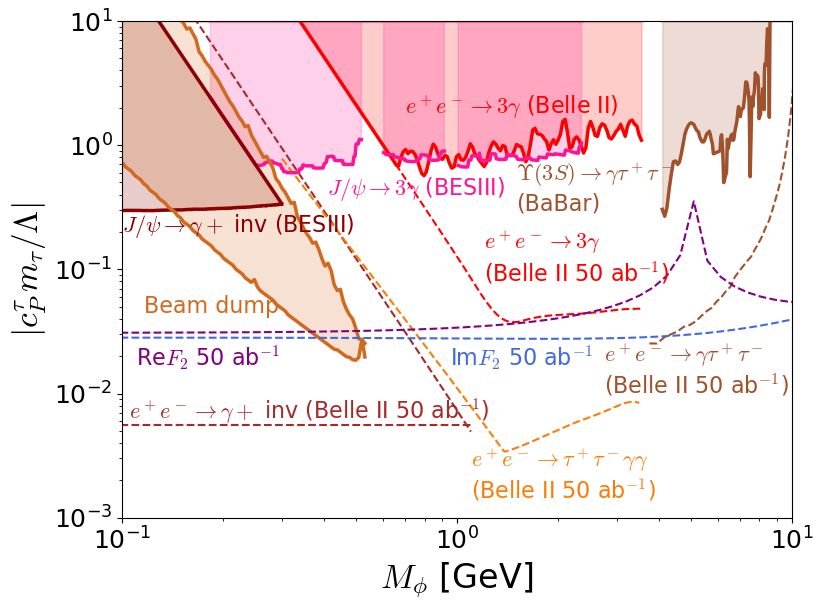}
    \includegraphics[width=0.45\linewidth]{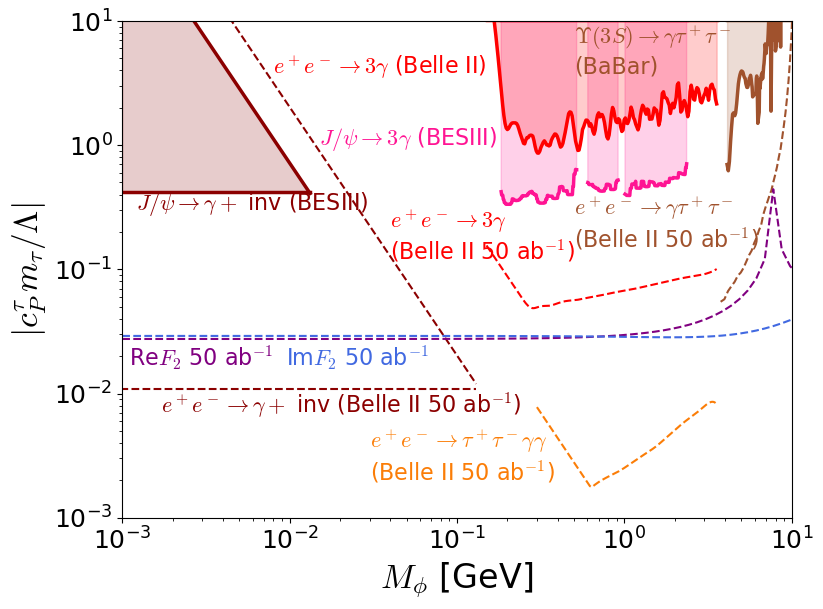}
    \includegraphics[width=0.45\linewidth]{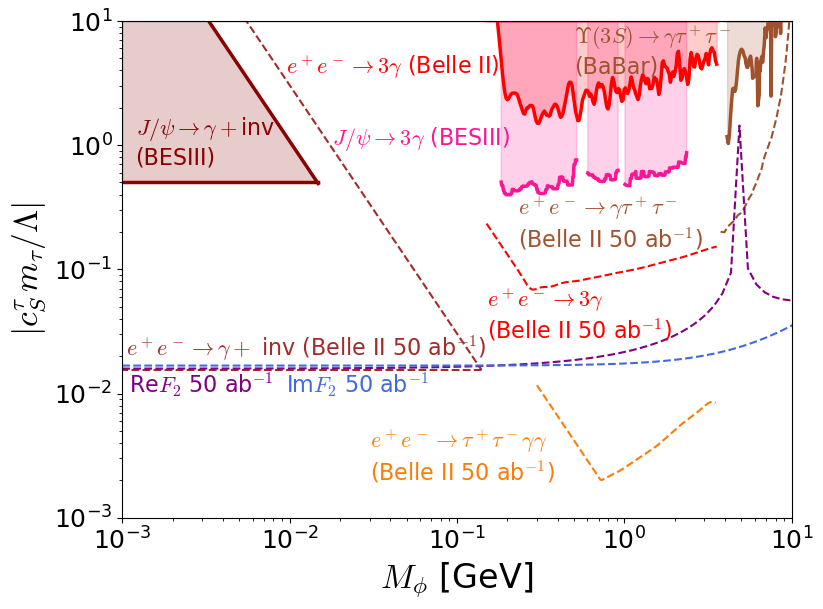}
    \caption{Experimental exclusion bounds on tauphilic ALP (\textit{top left}), pseudoscalar (\textit{top right}), and scalar (\textit{bottom}), as per Ref.~\cite{Alda:2024cxn} (to which we refer for a definition of the Lagrangian densities). The exclusion bounds from the imaginary part of the form factor $F_2$ are new; they are obtained assuming a sensitivity to the latter at a level of $10^{-6}$, as for the real part.}
    \label{fig:TauphilicALPs_ImPart}
\end{figure*}

In the previous sections we have provided the complete expressions for the contributions to $a_\tau$,  $d_\tau$, $F_2(q^2)$, and $F_3(q^2)$ as induced by the virtual exchange of a light NP state coupled to $\tau$ leptons.
As such, they can be employed in order to derive constraints on the parameters of specific models that couple mainly to the third generation of fermions.

Such an analysis was performed in Ref.~\cite{Alda:2024cxn}, where astrophysical and Belle II constraints on specific classes of tauphilic scalar and pseudoscalar particles were considered.\footnote{Further analyses concerning tauphilic ALPs are those exploring the phenomenological signatures of long-lived ALPs from $\tau$ decays~\cite{Ema:2025bww,Jiang:2025nie}.} In particular, it is important to notice that bounds from $a_\tau$, being indirect in nature, cover the whole mass range available for such a class of particles. Even if the cases in which such an observable poses the best constraints are limited (with the noteworthy exception of a pure scalar), it represents a solid benchmark against which to compare other direct searches. 
Furthermore, we stress that in the absence of a dedicated search in the channel $e^+ e^- \to \tau^+ \tau^- \gamma \gamma$, which has not been performed so far, the bounds obtained from the imaginary part of the form factor $F_2$ could allow one to place the best constraint on this kind of candidates for masses above $(0.1$--$1)\GeV$, depending on the scenario under consideration.
The case study of light tauphilic spin-$0$ particles shows that the interplay between direct probes and indirect constraints from $a_\tau$ and $d_\tau$ represents a valuable source of information for constraining NP scenarios, see Fig.~\ref{fig:TauphilicALPs_ImPart}.

The same observation holds for tauphilic light vector bosons, for which, however, a dedicated phenomenological analysis has not been performed to date. In order to at least partly cover this topic, we consider the phenomenological probes at Belle II of a new vector state coupled to the third generation of leptons in the SM.

\subsection{Testing a light tauphilic vector boson at Belle II}
\label{subsec:testing_light_tauphilic_VB}

\begin{figure*}[t]
    \centering    \includegraphics[width=0.45\linewidth]{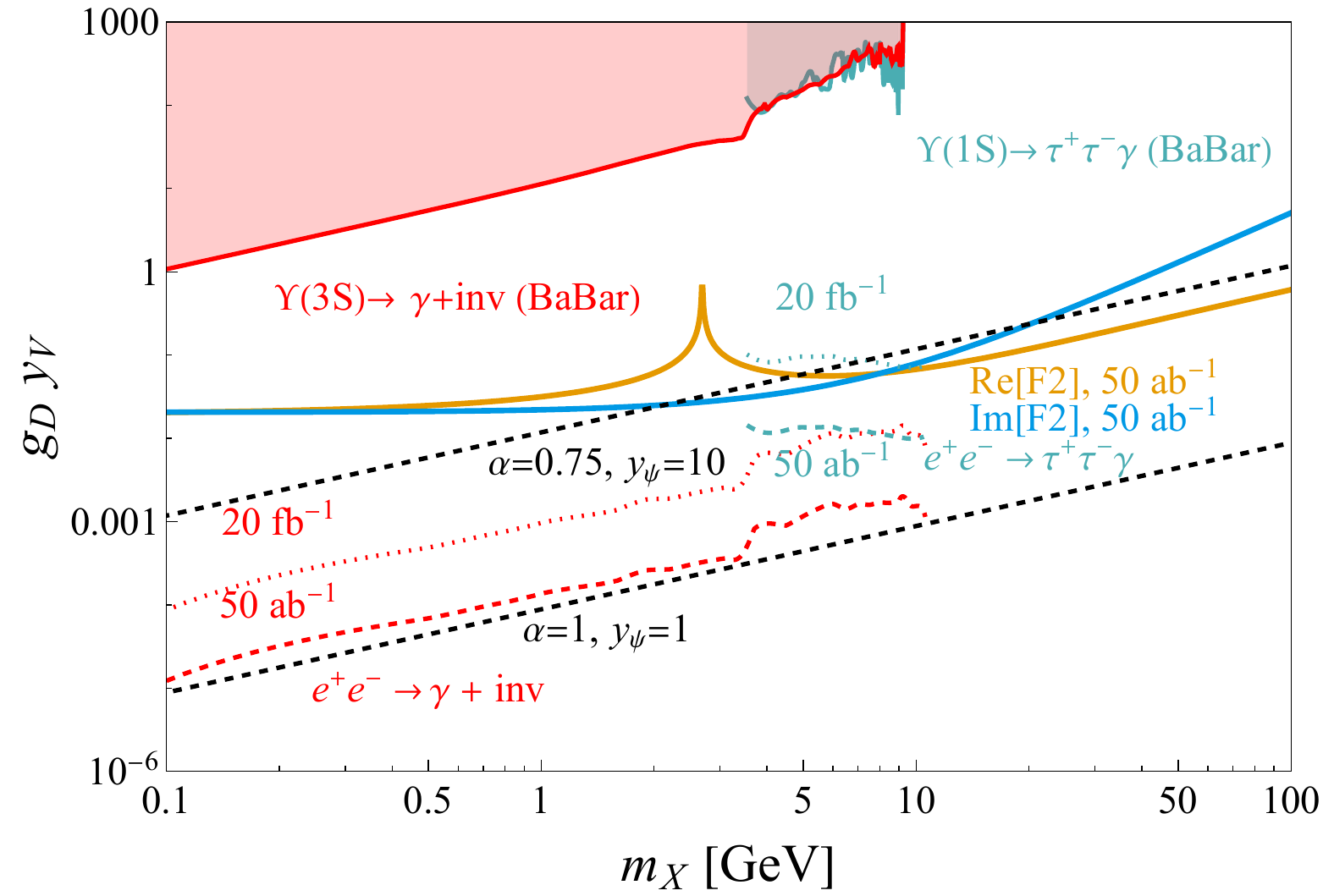}
    \includegraphics[width=0.45\linewidth]{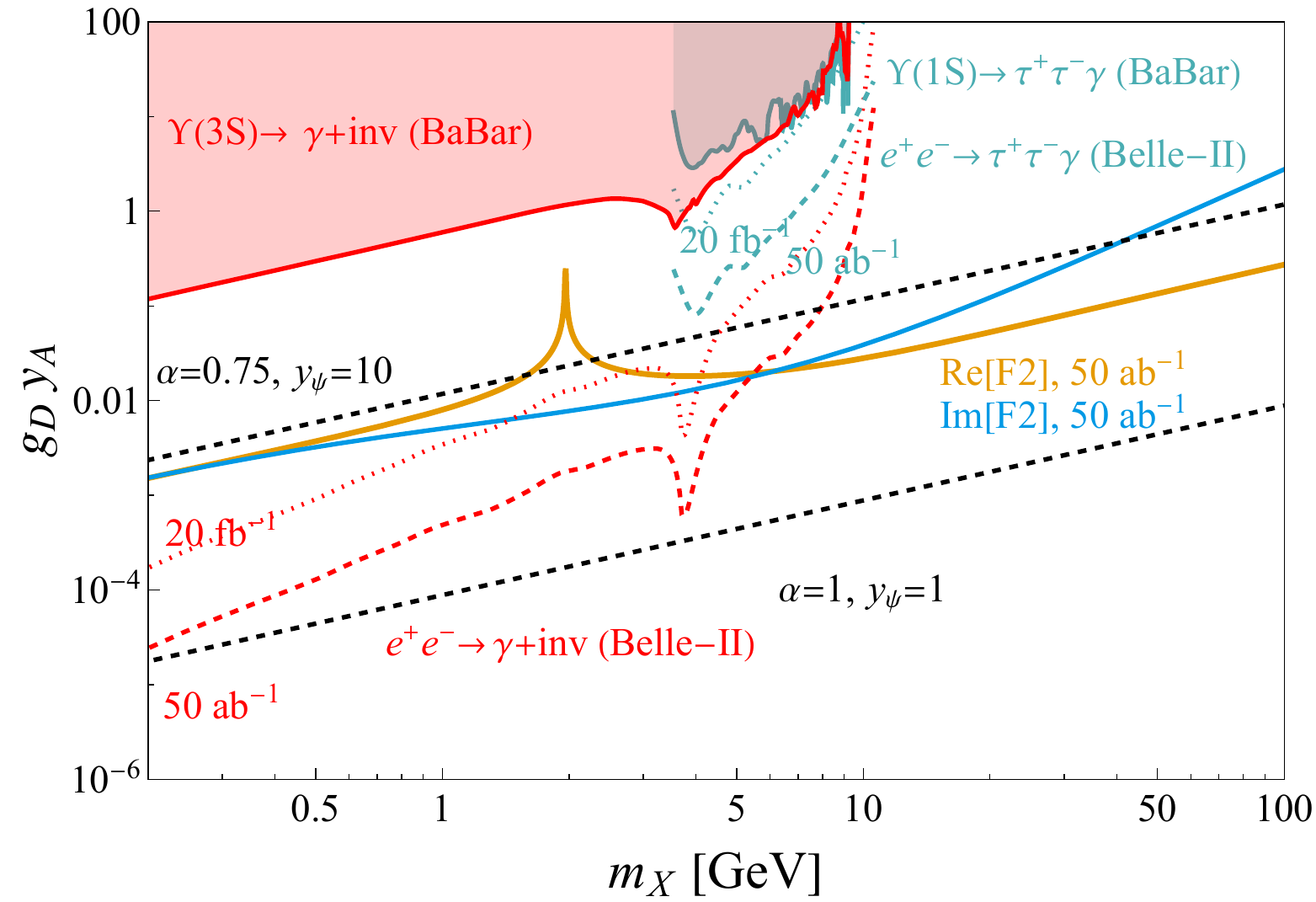}
    \caption{Exclusion bounds from Belle II on a purely tauphilic light vector boson. The two black dashed lines indicate potential regions of interest for such NP candidates, see App.~\ref{sec:Tauphilic_VBs} for further details. \textit{Left/Right:} Purely axial/vectorial couplings.}
    \label{fig:Tauphilic_Vectors_VA}
\end{figure*}

The starting point for our analysis, leading to the constraints in Fig.~\ref{fig:Tauphilic_Vectors_VA} and reported in detail in App.~\ref{sec:Tauphilic_VBs}, is the following Lagrangian
\begin{equation}
    \mathcal{L}^\text{int}_{\text{spin-1}} = - \, g_D\,\chi_L\, X_\mu \,\bar{\ell}_{3L}\, \gamma^\mu \ell_{3L}\,+ \, g_D\,\chi_R\, X_\mu \,\bar{\tau}_R\, \gamma^\mu \tau_R\,,
\end{equation}
which results in the following coupling to $\tau$ leptons:
\begin{equation}
    \mathcal{L}^{\text{int,}\tau}_{\text{spin-1}} \supset - \, g_D\, X_\mu \,\bar{\tau}\, \gamma^\mu \left(y_V + \,y_A \,\gamma_5\right) \tau\,,
\end{equation}
where we have defined $2\,y_V = \chi_R+\chi_L$ and $2\,y_A = \chi_R+\chi_L$. In the absence of further couplings to SM fermions, this Lagrangian leads to the appearance of anomalies induced by the exchange of virtual $\tau$ leptons and neutrinos in a fermionic triangle loop. The cancellation of the anomalies associated with such graphs requires the existence of new fermionic states in the UV, called \textit{anomalons}, whose charge assignment is dictated precisely by the anomaly cancellation condition. Integrating out of the theory these heavy fermionic fields then results in the generation of effective $\gamma \gamma X$ and $\gamma Z X$ vertices at low energies:
\begin{align}
\label{eq:XintLagAnoCou}
\mathcal{L}_\text{an} 
&\supset g_D \,e^2 C_{\gamma\gamma} \epsilon^{\mu\nu\alpha\beta}X_\mu A_\nu \partial_\alpha A_\beta \nonumber \\
&+ g_D \,g \,g' C_{\gamma Z} \epsilon^{\mu\nu\alpha\beta}(X_\mu A_\nu \partial_\alpha Z_\beta+X_\mu Z_\nu \partial_\alpha A_\beta)\,.
\end{align}
The requirement of anomaly cancellation directly relates the coefficients of such terms to the anomalies induced by the SM fermions alone: 
\begin{align}
C_{\gamma\gamma} &=  -\frac{y_A}{6\pi^2}\,, \nonumber \\
C_{\gamma Z} &=  -\frac{y_V}{24\pi^2}+ \frac{y_A}{24\pi^2}(c_w^2-3s_w^2)\,,
\end{align}
where $c_w=\cos \theta_w$ and $s_w=\sin \theta_w$ are the cosine and the sine of the weak mixing angle.
Despite the presence of a nonzero $X \gamma \gamma$ coupling, the decay of a $X$ boson to two on-shell photons is prohibited by the Landau--Yang theorem~\cite{Landau:1948kw,Yang:1950rg}. As a consequence, the $X$ boson decays exclusively to neutrinos below the $\tau^+\tau^-$ threshold; above it, also the $\tau^+\tau^-$ decay channel opens, with a comparable rate:
\begin{align}
\label{eq:decayrates}
\Gamma_{\tau\tau} &=g_D^2 \,\frac{m_X}{12\pi}\sqrt{1-\frac{4 m_\tau^2}{m_X^2}}\, \bigg[|y_A^\tau|^2 \left(1-\frac{4m_\tau^2}{m_X^2}\right) \nonumber \\
& \qquad \qquad \qquad \qquad \qquad  + |y^\tau_V|^2 \left(1+\frac{2m_\tau^2} {m_X^2}\right)\bigg]\,,
\nonumber \\
\Gamma_{\nu\nu}&= g_D^2 \frac{m_X}{48 \pi} |y_V-y_A|^2\,. 
\end{align}
Decay channels to other leptons are loop-level suppressed and proceed via the mixing of the $X$ vectorial component with a photon, $\frac{\Pi}{2} F_\gamma^{\mu\nu}F^X_{\mu\nu}$. This effect can be quantified, in the limit of a light vector boson, to be~\cite{Araki:2017wyg}
\begin{equation}
\Pi(m_X \to 0) \simeq \frac{e g_D}{12\pi^2} \log \frac{m_\tau^2}{\Lambda^2}\,.
\end{equation}
The phenomenology associated to tauphilic vector bosons is rich and interesting, and would deserve a thorough investigation, which is beyond the scope of the current paper. Nonetheless, we point out that Belle II offers unique opportunities to probe such scenarios, with a high degree of complementarity with respect to the indirect searches for light vector bosons via their impact on $a_\tau$, as discussed previously.

The most promising avenues for detecting or constraining a new tauphilic vector bosons are based on the exploitation of the longitudinal enhancement experienced by the axial $\tau \tau X$ coupling or by the anomalous coupling $\gamma \gamma X$, which are a consequence of the $\tau$ current not being individually conserved~\cite{Dror:2017nsg, Dror:2018wfl}.

\begin{figure}[t]
    \centering
    \includegraphics[width=1\linewidth]{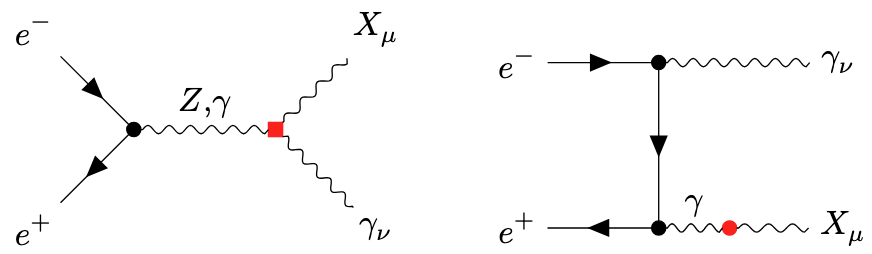}
    \caption{Diagrams contributing to the scattering process $e^-(p_1) \,  e^+(p_2) \rightarrow X_\mu(k_1) \, \gamma_\nu(k_2)$. Red squared dots represent NP insertions, while black dots denote the insertion of a SM vertex. The $u$-channel topology is not depicted but understood to be included.}
    \label{fig:MMTTDiag}
\end{figure}

The former coupling is responsible for the longitudinally enhanced emission of a light $X$ boson from a $\tau$ lepton in the processes  $e^+ e^- \to \tau^+ \tau^- + \text{inv}$ or $e^+ e^- \to \tau^+ \tau^- \ell^+\ell^-$:
\begin{equation}
\label{eq:ee_tautauX1}
\sigma_{ee\to \tau \tau X} \overset{m_X\to 0}{=} \frac{\alpha_{\text{em}}^2 g_D^2}{48\pi} |y_A|^2 \frac{m_\tau^2}{s}\frac{1}{m_X^2} \log \frac{s}{m_\tau^2}\,.
\end{equation}
The possibility to make use of such a process at $e^+ e^-$ colliders had already been proposed in Ref.~\cite{Dror:2017nsg}.
We argue, however, that experimental searches for such processes are difficult to perform and might not lead to significant bounds, if compared to other opportunities. Indeed, once the $X$ boson is produced in $e^+ e^- \to \tau^+ \tau^- X$, the detection strategies depend on whether it decays to neutrinos or to light leptons.
In the former process, experimental difficulties in reconstructing the invariant mass of a $\tau$ pair are expected to lead to a significant loss of statistical sensitivity. The second process is instead suppressed by the small couplings of the new tauphilic vector to leptons of the first and second generation, which are only induced at the loop level, see App.~\ref{sec:Tauphilic_VBs} for further details.
A peculiarity of tauphilic vector bosons, however, is that they necessarily generate also $\gamma \gamma X$ couplings, which lead as well to longitudinally enhanced production of $X$ bosons and can be probed both in $2\to 2$ processes and, potentially, in vector boson fusion processes at colliders.
The phenomenology associated to these new vertices can then be explored at Belle II by considering the processes $e^+ e^- \to \gamma + \text{inv}$ and $e^+ e^- \to \tau^+ \tau^- \gamma$. The leading contributions to the differential cross section for the process $e^+ e^- \to \gamma X$ are found to be:
\begin{align}
\frac{d\sigma_{ee\to \gamma X}}{d\cos \theta} &\simeq |y_A|^2 \frac{e^6 g_D^2}{128\pi^5}\frac{1}{m_X^2}\frac{1}{16}(3+\cos 2\theta) \nonumber \\
& \quad + |y_V|^2\frac{e^6g_D^2}{144\pi^5}\frac{1}{s}\frac{1}{16}\frac{1+\cos^2 \theta}{1-\cos^2 \theta}\,.
\end{align}
The contribution in the first line is generated by the $s$-channel exchange of a $\gamma^*$ in the presence of an anomalous $\gamma \gamma X$ vertex, which experiences the expected longitudinal $1/m_X^2$ enhancement. Interestingly enough, the anomalous coupling also plays a role in quarkonia decays, which can be tested at $B$ factories. The contribution in the second line instead provides the leading dependence on the vectorial $X$ boson coupling and is generated by the conversion of a photon to an $X$ boson in a $t$- or $u$- channel pair production graph, see Fig.~\ref{fig:MMTTDiag}.

In order to place sensitivity bounds on these processes, we consider a search for a bump in the photon energy $E_\gamma$. The latter is related to the mass of the $X$ boson and to the Belle II CM energy $\sqrt{s_B}=10.58\GeV$ via
\begin{equation}
    E_\gamma = \frac{\sqrt{s}}{2}-\frac{m_X^2}{2\sqrt{s}}\,.
    \label{eq:2final_energies}
\end{equation}
Similarly, the $X$ boson energy $E_X=\sqrt{s}/{2}+m_X^2/(2\sqrt{s})$ is fixed by the kinematics of the process and can be employed to compute the decay length of the $X$ boson to $\tau$ leptons. The boost factor then sets the fraction of $X$ boson candidates that decay after having traveled for a distance $L_\text{dec}$ in the detector
\begin{equation}
    p_X \approx 1 - e^{-L_\text{dec}\Gamma_X/\sqrt{\gamma_X^2-1}}\,.
\label{eq:prob_dec}
\end{equation}
In order to perform the proposed search, we have simulated the cross section for the SM background processes $e^+ e^- \to \gamma \, \bar{\nu} \, \nu$ and $e^+ e^- \to \gamma \, \tau^+ \tau^-$ by making use of {\sc\small MadGraph5\_aMC@NLO}~\cite{Alwall:2014hca}.

We then focus on photon energies $E_\gamma > 0.1\GeV$, which can be reconstructed by the Belle II calorimeters, and impose the angular acceptance cut $\theta_\text{min}< \theta< \theta_\text{max}$. These angles refer to the CM frame of the Belle II detector and correspond to $\cos \theta_{\text{min}} = 0.941$ and $\cos \theta_{\text{max}} =-0.821$~\cite{Araki:2017wyg}.
According to the Belle II technical design report, we take the photon energy resolution to be the one of the internal calorimeter~\cite{Belle-II:2010dht,Belle:1999bhb}:
\begin{equation}
    \frac{\sigma_{E}}{E}=\sqrt{\left(\frac{0.066\%}{E}\right)^2+\left(\frac{0.81\%}{E^{1/4}}\right)^2+\left(1.34\%\right)^2}\,.
    \label{eq:BelleII_res}
\end{equation}
We then explore the $X$ mass region $m_X=[0.2\GeV,2m_\tau]$ and $m_X=[2m_\tau, \sqrt{s_B}]$ and impose upper bounds on $g_D \,y_{A/V}$. To do so, we assume Poisson statistics of signal events in each bin and find the value of $g_D \,y_{A/V}$ corresponding to $S/\sqrt{B} = 2$, where $S$ is the number of events in each bin and $B$ the number of background events, after having imposed all the relevant cuts discussed before.
Our phenomenological analysis leads to the constraints in Fig.~\ref{fig:Tauphilic_Vectors_VA}, where we have included also the bounds obtained from quarkonia decays. For further details on the latter, on the procedure we followed in placing our bounds, and on the expected SM background we refer the reader to the detailed discussion in App.~\ref{sec:Tauphilic_VBs}.

\begin{figure*}
    \centering
    \includegraphics[width=0.45\linewidth]{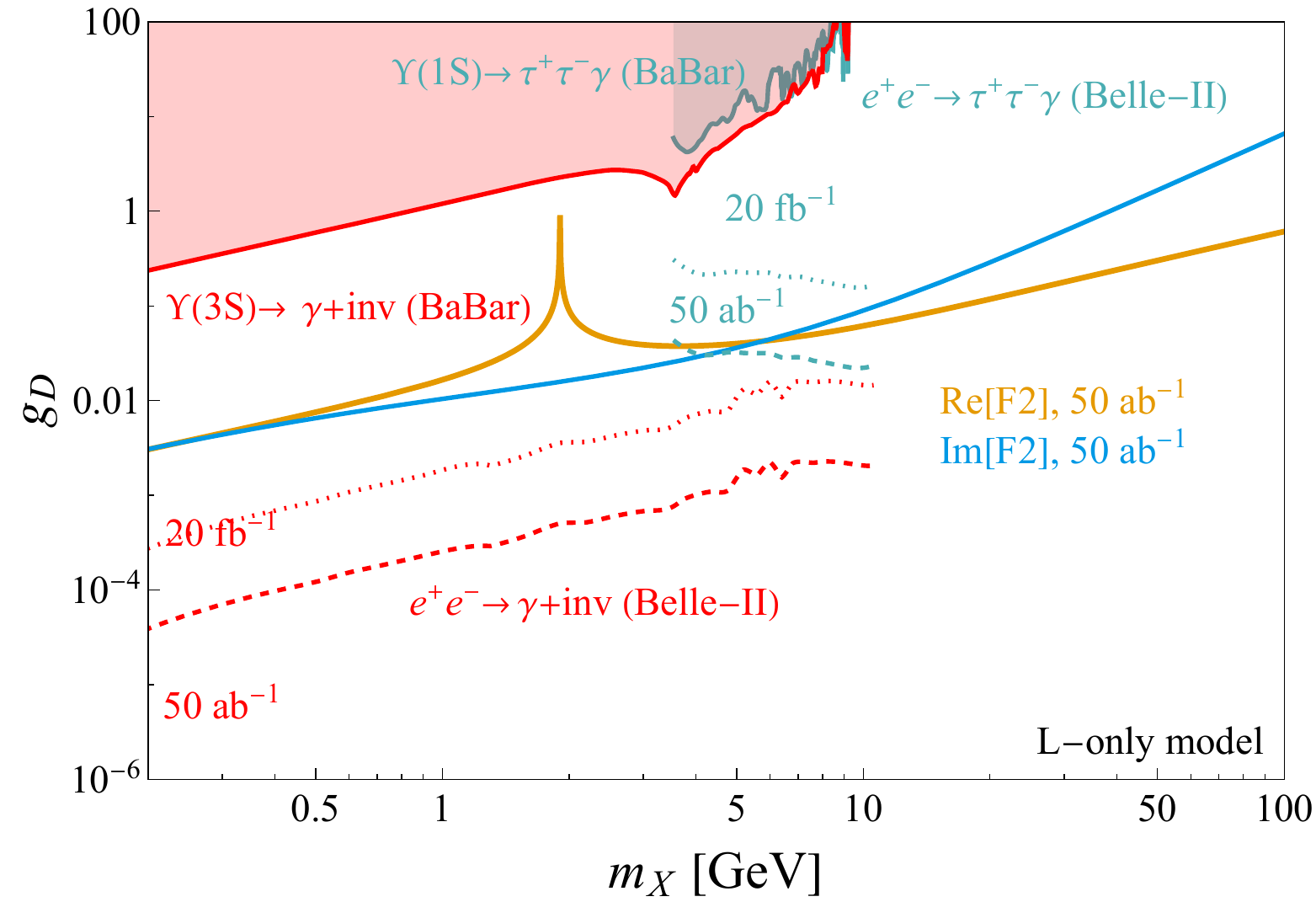}
    \includegraphics[width=0.45\linewidth]{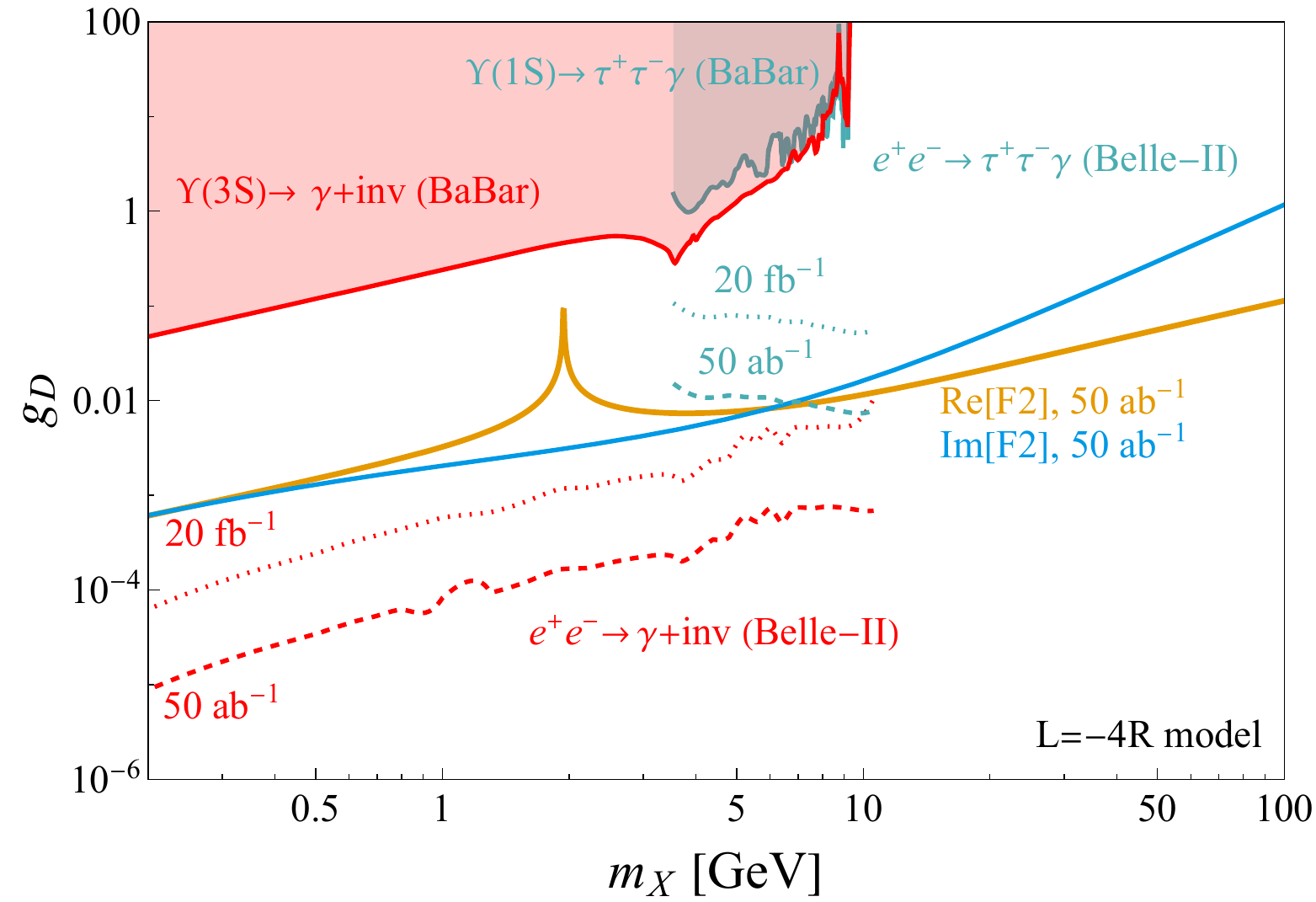}
    \caption{Exclusion bounds from Belle II experiments on the two classes of models proposed in Ref.~\cite{DiLuzio:2025qkc} to solve the observed discrepancy in $B\to K^* + E_\text{miss}$. \textit{Left:} $L$-only model. \textit{Right:} $L=-4R$ model.}
    \label{fig:UVCompletions}
\end{figure*}

\subsection{Testing a specific scenario}
\label{sec:specific_scenario}
The phenomenological analysis we have carried out so far can be applied to specific cases of interest. A particularly relevant one for our scenario is the one discussed in Ref.~\cite{DiLuzio:2025qkc}. There the authors advance the hypothesis of a new light vectorial state coupling mainly to the third generation of leptons to solve the observed discrepancy in the process $B\to K^* + E_\text{miss}$~\cite{Belle-II:2023esi}.
For a best-fit point of $m_X = 2.1\GeV$ and couplings at the level of $g_D \simeq 10^{-3}$, such a discrepancy can be accommodated without violating any experimental constraint, primarily $Z\to Z' \gamma$.
The authors also provide some specific UV completions for such phenomenologically viable models, among which, for instance, are the ``$L$-only model,'' with $\alpha_L=1, \alpha_R=0$ and best fit point of $(m_X, g_D) = (2.1\GeV, 0.018)$ and the ``$L=-4R$ model,'' with $\alpha_L = 4, \alpha_R \simeq -1$, and best fit point of $(m_X, g_D) = (2.1\GeV, 0.0042)$.
These models can be further phenomenologically tested via the experimental searches we have considered so far, with the advantage of having under complete control the full UV completion. This, in particular, implies that we can compute exactly the two-loop contribution from the effective three-boson vertex in Eq.~\eqref{eq:Rosenberg} by making use of the results in Ref.~\cite{Hoferichter:2025yih}.
We report our results in Fig.~\ref{fig:UVCompletions}. Interestingly enough, the two proposed models are not yet excluded by currently available data, but will be probed extensively by future upgrades of the Belle II experiment. In particular, a luminosity of $20\,\text{fb}^{-1}$ will be enough to either confirm or discard the proposal of Ref.~\cite{DiLuzio:2025qkc}. As a comment, we find it particularly compelling that possible solutions to the anomaly reported at Belle II can be actually probed at the same experimental facility.

\section{Conclusions and outlook}
\label{sec:conclusions}

In this paper we have investigated the possibility to probe light NP candidates through the impact they have on the anomalous electric and magnetic dipole moments of the $\tau$ lepton. In order to do so, we have exploited the techniques outlined in Refs.~\cite{Bernabeu:2007rr, Bernabeu:2008ii, Crivellin:2021spu, Gogniat:2025eom}, which show that information on $a_\tau$ and $d_\tau$ can be gathered by considering properly constructed polarization asymmetries in the cross section for the process $e^+ e^- \to \tau^+ \tau^-$. These can be employed to relate the experimentally measured form factors $F_{2,3}(s_B)$ to the quantities of interest, $a_\tau$ and $d_\tau$. Our main results~\cite{Hoferichter:2025ijh}, obtained for a wide variety of popular classes of EFTs for light NP candidates including light scalars, pseudoscalars, vectors and axial vectors, are reproduced in Fig.~\ref{fig:Money_Plots}, derived from the results described in Sec.~\ref{sec:light_NP} for the various scenarios.
Whereas for heavy NP candidates the only quantity of interest is the local contribution to $\Re F_2$ induced by the exchange of particles with mass $m_\text{NP}^2 \gg s_B$, the case of light new states is more interesting. Indeed, for sufficiently large collider energies, a new light propagating state also contributes to the generation of a nonzero imaginary part for the form factor $F_2$, $\Im F_2$. This in turn can be more easily constrained experimentally, as it does not require polarized initial-state beams, but rather only access to the polarization information on the decay products of the final-state $\tau$ leptons. This makes searches for light NP particles through this channel already viable with the present setup of the Belle II experiment.

An interesting feature of our results is that $\Re F_2$ shows the largest possible sensitivity to NP at large mediator mass energies, where it basically collapses back to the heavy NP case. On the other hand, the imaginary part displays a power-law suppression in the mass of the mediator, as a direct consequence of generalized unitarity arguments. Indeed, the one-loop imaginary part of any given diagram is proportional to the sum of the product of two tree-level cut amplitudes. As one of the two features the virtual exchange of a NP state, in the high mass limit the imaginary part will experience at least a $1/(s-m^2)\to 1/m^2$ suppression. At smaller mediator masses, the imaginary part of the form factor $F_2$ can even place better constraints than the real one, thus highlighting an interesting complementarity between the two search strategies depending on the mass of the NP candidate.

Another interesting observation concerns the near-threshold enhancement of the form factors for $s \simeq 4 m_\tau^2$. This is a common feature displayed, to different degrees, by most of the cases we have considered. Experimentally, such a low energy could be reached either by tuning the CM energy of an $e^+e^-$ collider to the $\tau^+\tau^-$ threshold, or by exploiting radiative-return techniques. The latter could be already considered at current $B$ factories,  however,  at the expense of the loss of a significant fraction of statistics. As shown in Fig.~\ref{fig:DiTau_threshold}, as long as the loss in luminosity is smaller than a factor of 100, searches above the $\tau^+\tau^-$ threshold could become competitive.

We furthermore identify several interesting directions for future analyses. First of all, both the generation of an imaginary part for $s> 4m_\tau^2$, the near-threshold enhancements, and logarithmic sensitivities to NP are not necessarily limited to propagating light states. They are indeed prompted by the two virtual fermions going on the mass shell and are expected also in the presence of specific four-fermion operators. It would be interesting to explore this possibility further, which we leave for future work. Second, the results we have obtained only include flavor-conserving couplings of NP states to $\tau$ leptons. This choice is well motivated in the framework of NP mainly coupled to the third generation of fermions, or in the presence of a strong hierarchy among the three leptonic families (e.g., if new scalars couple derivatively or via Higgs mixing to SM fermions). However, it would be interesting to consider more general scenarios, where also lepton-flavor-violating couplings exist, together with couplings to the other generations of leptons. In this case, one should also take into account nuisance contributions to the definition of the experimentally measured asymmetries generated by virtual corrections to the electron electromagnetic vertex or by one-loop box diagrams, where a light new state is exchanged between an initial-state electron and a final-state $\tau$.

Looking for light NP by investigating its impact on form factors in the process $e^+ e^- \to \tau^+ \tau^-$ allows one to place indirect constraints that extend up to any mass value of the virtually exchanged particle, thus making them particularly appealing. Nonetheless, these kind of indirect searches are only a fraction of the possible phenomenological analyses that can be performed to probe light NP candidates, which naturally include other indirect probes at the precision frontier and direct probes in other collider-based experiments. In this paper we have also highlighted the high degree of complementarity of such searches to map the available parameter space for light NP candidates. In this context, we have explored for the first time the phenomenology of light tauphilic vector bosons at Belle II, which fits well in the growing interest that tauphilic NP candidates have been receiving~\cite{Alda:2024cxn}.
In particular, we have proposed to make use of the anomalous $X\gamma \gamma$ coupling, which is unavoidably generated in such scenarios, to directly constrain tauphilic vector boson interactions in the processes $e^+ e^- \to \gamma \text{ + inv}$ and $e^+ e^- \to \gamma \,\tau^+ \tau^-$. These represent the most constraining processes in a wide region of the parameter space, being complemented in intermediate mass regions by indirect bounds from the form factors $F_{2,3}$. 
Interestingly enough, these searches have the potential to test some light NP models that have been proposed in order to provide a viable solution to the observed rate for the process $B \to K^{(*)} + E_\text{miss}$ at Belle II~\cite{DiLuzio:2025qkc}, so that a solution to an experimental anomaly at Belle II could actually be probed at the same experimental facility.

The phenomenological analysis we have performed is far from complete and simply focused on the possibility to probe light tauphilic vectors at Belle II. A thorough analysis would require considering as well flavor probes for such candidates, prompted by the anomalous coupling $X WW$ and the investigation of the effect that such candidates can have on astrophysical observables, such as the dynamics of core-collapse supernovae or neutron star mergers~\cite{Alda:2024cxn}.
An experimental feasibility study would also be required in order to assess the precise range of the longitudinally enhanced $X$-boson production in $e^+ e^- \to \tau^+ \tau^- X$ and of the vector boson fusion process $e^+ e^- \to e^+ e^- X$.
Finally, more refined analyses on the background for the process $e^+ e^- \to \gamma X^*$ would then need to be performed to further refine our estimates.
We leave these tasks for future investigations.

 \section*{Acknowledgments}

We would like to thank Jorge Alda
for sharing code for tauphilic light spin-0 states~\cite{Alda:2024cxn}, used for Fig.~\ref{fig:TauphilicALPs_ImPart}, and Nud\v{z}eim Selimovi{\'c} for useful discussions about collider searches.
Financial support by the Swiss National Science Foundation (Project No.\ TMCG-2\_213690) is gratefully acknowledged.

\appendix

\begin{widetext}

\section{Notation and conventions}
\label{app:conventions}

Throughout our computations we work in NDR.
The conventions we adopted match those employed in \texttt{Package-X}~\cite{Patel:2015tea, Patel:2016fam}, whose notation we abbreviate as follows
\begin{align}
\mathcal{B}_0^{\tau\tau\phi} &\equiv \texttt{DiscB}[m_\tau^2, m_\tau, M_\phi] \,, & \mathcal{B}_0^{\tau\tau V} &\equiv \texttt{DiscB}[m_\tau^2, m_\tau, M_V]\,, \nonumber\nonumber \\
\mathcal{B}_0^{q\tau\tau} &\equiv \texttt{DiscB}[q^2,m_\tau, m_\tau] \,,&
\mathcal{B}_0^{q\tau V} &\equiv \texttt{DiscB}[q^2,m_\tau, M_V] \,,
\end{align}
where
\begin{equation}
\texttt{DiscB}[q^2, m, M] = \frac{\lambda^{1/2}(q^2,m^2, M^2)}{q^2}\log \frac{m^2+M^2-q^2+\lambda^{1/2}(q^2,m^2, M^2) }{2mM}
\end{equation}
and $\lambda(a,b,c) = a^2+b^2+c^2-2ab-2ac-2bc$.
Furthermore, we define:
\begin{align}
\mathcal{C}_0^{\tau\tau\phi} &\equiv \texttt{ScalarC0}[m_\tau^2, m_\tau^2, q^2,m_\tau, M_\phi, m_\tau] \,,\nonumber \\
\mathcal{C}_0^{\tau\tau V} &\equiv \texttt{ScalarC0}[m_\tau^2, m_\tau^2, q^2,m_\tau, M_V, m_\tau] \,,\nonumber \\
\mathcal{C}_0^{\tau \gamma V}& \equiv\texttt{ScalarC0}[m_\tau^2, m_\tau^2, q^2, M_V, m_\tau,0]\,.
\end{align}
An explicit expression for $\texttt{ScalarC0}[p_1^2, p_2^2, q^2=(p_1-p_2)^2, m_2,m_1,m_0]$ in terms of polylogarithms is given in Ref.~\cite{Denner:1991kt}.

\section{Complete expressions for the  form factors}
\label{sec:Explicit_Expressions}

In the following we report complete analytical expressions for the form factors discussed in the main text.

\subsection{Light scalar bosons}

The one-loop contributions to the form factors $F_2$ and $F_3$ as mediated by the virtual exchange of a light spin-$0$ particle possessing the interactions discussed in Eq.~\eqref{eq:ALP_Lag} correspond to the diagrams in Figs.~\ref{fig:YukDiagrams} and \ref{fig:MixDiagrams}.

\subsubsection{Yukawa couplings only}

\begin{figure}[t]
    \centering
    \includegraphics[width=0.48\linewidth]{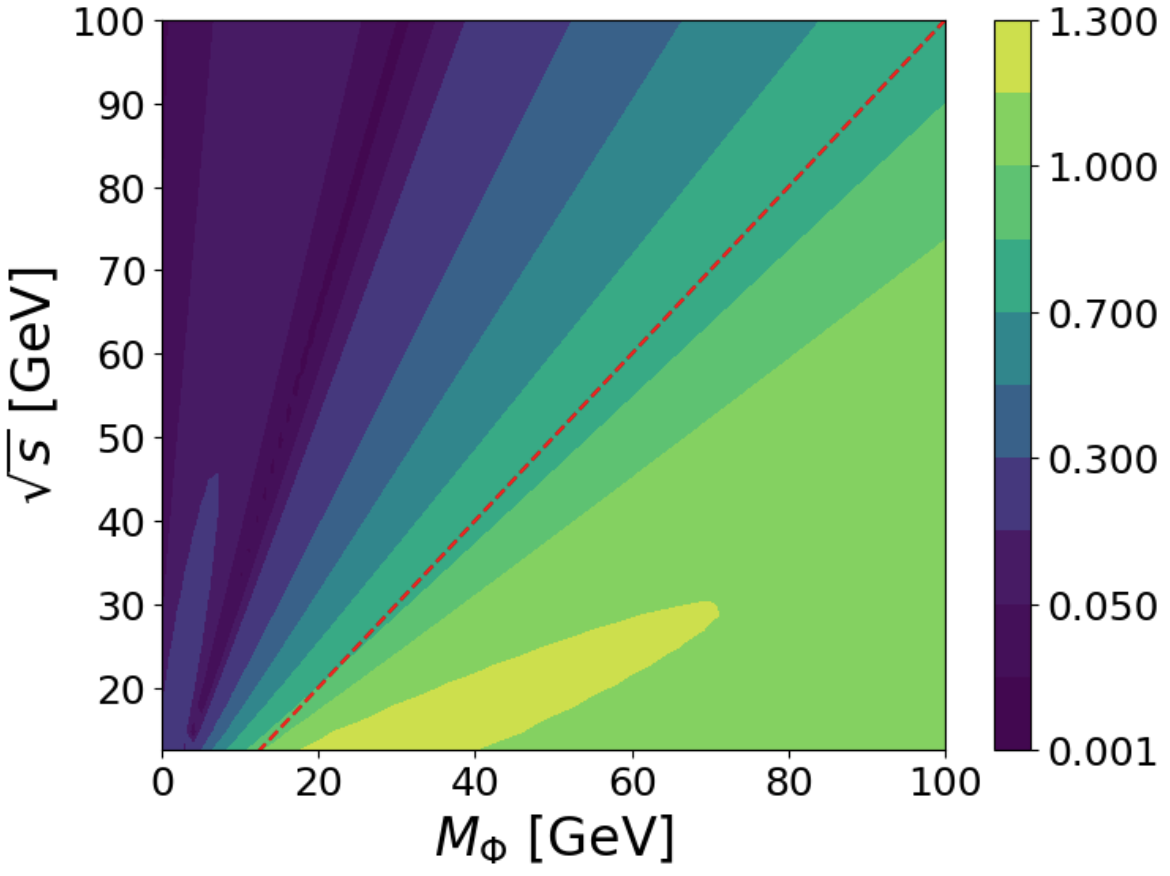}
    \includegraphics[width=0.48\linewidth]{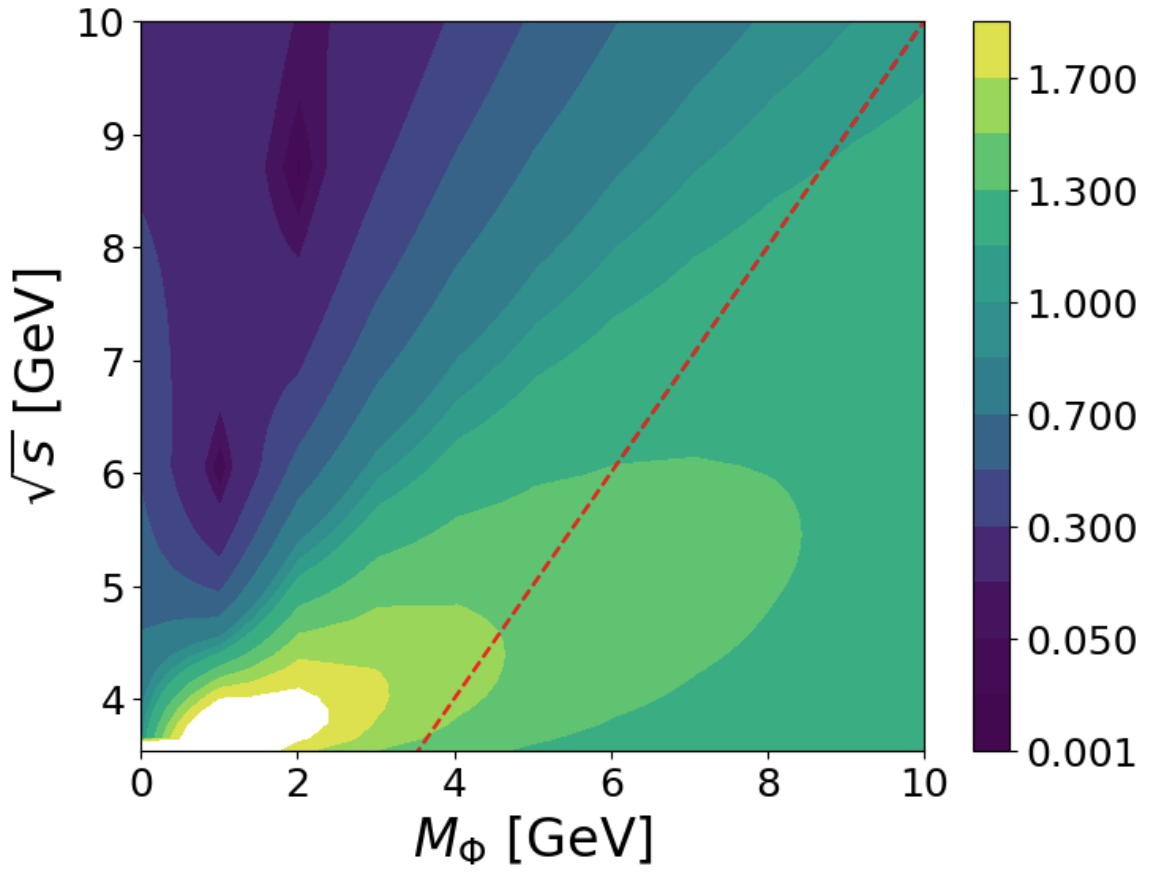}
    \caption{Momentum dependence of the ratio $|\Re F_2^\text{y}|/|a_\tau^\text{y}|$ for $c_P^\tau = c_S^\tau=1$. The diagonal line $\sqrt{s}= M_\phi$ serves as an indicative threshold separating the light NP scenario from heavy NP.
    }
    \label{fig:Energy_Comparison_CPVALP_tau_yuk}
\end{figure}

The complete expressions for the form factors $F_2$ and $F_3$ as induced by the virtual exchange of a light scalar possessing only Yukawa couplings to the $\tau$ lepton read:
\begin{align}
F_2^\text{y} &= -\frac{1}{16\pi^2 }\bigg(\frac{c_P^\tau m_\tau}{\Lambda}\bigg)^2 \frac{1}{m_\tau^2 (q^2-4m_\tau^2)^2}\Big[2m_\tau^2 M_\phi^2(10m_\tau^2-q^2) \mathcal{B}_0^{\tau\tau\phi} - 2m_\tau^4(4m_\tau^2+6M_\phi^2-q^2)\mathcal{B}_0^{q\tau\tau}\nonumber\\&\qquad  - M_\phi^2 (-8m_\tau^4 + 2m_\tau^2 q^2)- M_\phi^2(-2m_\tau^2 q^2 + M_\phi^2 q^2+ 8m_\tau^4-10m_\tau^2M_\phi^2)\log\frac{m_\tau^2}{M_\phi^2}  \nonumber \\
&  \qquad - 4m_\tau^4M_\phi^2(q^2-4m_\tau^2+3M_\phi^2) \mathcal{C}_0^{\tau\tau\phi}\Big]  \nonumber \\
&\qquad +\frac{1}{16\pi^2 m_\tau}\bigg(\frac{c_S^\tau m_\tau}{\Lambda}\bigg)^2 \frac{1}{m_\tau^2 (q^2-4m_\tau^2)^2}\Big[- M_\phi^2 (-8m_\tau^4 + 2m_\tau^2 q^2)+ 6m_\tau^4(4m_\tau^2-2M_\phi^2-q^2)\mathcal{B}_0^{q\tau\tau} \nonumber \\
& \qquad + ( 8 m_\tau^4 q^2- 2m_\tau^2 M_\phi^4q^2-32m_\tau^6 + 20 m_\tau^4 ) \mathcal{B}_0^{\tau\tau \phi} + M_\phi^2(6m_\tau^2 q^2 - M_\phi^2 q^2- 24m_\tau^4+10m_\tau^2M_\phi^2)\log\frac{m_\tau^2}{M_\phi^2}  \nonumber \\
& \qquad - 12m_\tau^4(q^2-4m_\tau^2+M_\phi^2) \mathcal{C}_0^{\tau\tau\phi}\Big] \,,\notag\\
F_3^\text{y} &= -\frac{1}{4\pi^2}\frac{c_S^\tau c_P^\tau m_\tau^2}{\Lambda^2} \frac{1}{q^2 (q^2-4m_\tau^2)} \Big[q^2M_\phi^2 \log \frac{m_\tau^2}{M_\phi^2} + 2 m_\tau^2 q^2 \, \mathcal{B}_0^{\tau\tau\phi} - 2m_\tau^2 q^4\mathcal{B}_0^{q\tau\tau}- 2 m_\tau^2 M_\phi^2 q^2 \mathcal{C}_0^{\tau\tau\phi} \Big]\,,
\end{align}
while the corresponding expressions for the AMM and EDM are found to be:
\begin{align}
a_\tau^{\text{y}} &= \frac{-1}{16\pi^2\Lambda^2}\bigg[\frac{c_P^{\tau\, 2}}{m_\tau^2(4m_\tau^2-M_\phi^2)}\bigg(4m_\tau^6 + 7 m_\tau^4 M_\phi^2 - 2 m_\tau^2 M_\phi^4 - \log \frac{m_\tau^2}{M_\phi^2}\left(4m_\tau^4 M_\phi^2 - 5 m_\tau^2 M_\phi^4 + M_\phi^6\right)\notag\\
&\qquad\qquad- 2 M_\phi^2 m_\tau^2(M_\phi^2-3m_\tau^2) \mathcal{B}_0^{\tau\tau\phi}\bigg) \nonumber \\
&\qquad -\frac{c_S^{\tau\,2}}{m_\tau^2}\bigg(3 m_\tau^4 - 2 m_\tau^2 M_\phi^2  + \log \frac{m_\tau^2}{M_\phi^2}(3m_\tau^2 M_\phi^2 - M_\phi^4) + 2m_\tau^2 (m_\tau^2-M_\phi^2)\mathcal{B}_0^{\tau\tau\phi}\bigg)\bigg]\,, \nonumber\\
d_\tau^{\text{y}} &= \frac{e}{16\pi^2 \Lambda^2}\frac{c_S^\tau c_P^\tau}{m_\tau(4m_\tau^2-M_\phi^2)} \left[8m_\tau^4 - 2 m_\tau^2 M_\phi^2 - 2 M_\phi^2(-4m_\tau^2+M_\phi^2)\log \frac{m_\tau}{M_\phi}+ 2 m_\tau^2  (-2m_\tau^2-M_\phi^2) \mathcal{B}_0^{\tau\tau\phi}\right]\,.
\end{align}
The resulting momentum dependence of $|\Re F_2^\text{y}|/|a_\tau^\text{y}|$ is illustrated in Fig.~\ref{fig:Energy_Comparison_CPVALP_tau_yuk}, converging towards the EFT limit when $M_\phi^2\gg s$.

\subsubsection{Including nonrenormalizable couplings}

The complete expressions for the form factors $F_2$ and $F_3$ as induced by the virtual exchange of a light scalar featuring both Yukawa-like couplings to $\tau$ leptons and nonrenormalizable couplings to photons read:
\begin{align}
F_2^\text{m}&= \frac{m_\tau^2}{2\pi^2 \, q^2\,(q^2-4m_\tau^2)^2}\frac{\tilde{c}_{\gamma\gamma}c_P^\tau}{\Lambda^2} \frac{\alpha_\text{em}}{4\pi}\bigg[q^2(q^2-4m_\tau^2)^2 \left(\frac{1}{\epsilon} + \log \frac{\mu^2}{m_\tau^2}\right) \nonumber \\
& \qquad  + q^2(q^2-4m_\tau^2)(4m_\tau^2-M_\phi^2) + (q^2-M_\phi^2)^2(q^2+2m_\tau^2)\log \frac{m_\tau^2}{M_\phi^2-q^2} \nonumber\\
& \qquad -2m_\tau^2(2q^4-2m_\tau^2q^2- M_\phi^2q^2- 2m_\tau^2M_\phi^2)\mathcal{B}_0^{\tau\tau\phi} +4m_\tau^2(q^2-m_\tau^2)(q^2-M_\phi^2)^2 \mathcal{C}_0^{\tau\tau\phi} \bigg]\nonumber\\
&  +\frac{m_\tau^2}{2\pi^2 \, q^2\,(q^2-4m_\tau^2)^2}\frac{c_{\gamma\gamma}c_S^\tau}{\Lambda^2} \frac{\alpha_\text{em}}{4\pi}\bigg[q^2(q^2-4m_\tau^2)^2 \left(\frac{1}{\epsilon} + \log \frac{\mu^2}{m_\tau^2}\right) \nonumber \\
& \qquad + q^2(q^2-4m_\tau^2)(2q^2-12m_\tau^2+M_\phi^2) - 2m_\tau^2(-6 q^2 m_\tau^2+M_\phi^2 q^2 + 2 m_\tau^2 M_\phi^2)\mathcal{B}_0^{\tau\tau\phi} \nonumber \\
& \qquad + (q^2-M_\phi^2)(q^4-10q^2m_\tau^2+q^2M_\phi^2+ 2 m_\tau^2 M_\phi^2)\log \frac{m_\tau^2}{M_\phi^2-q^2}\nonumber \\
&\qquad - 4m_\tau^2(q^2-m_\tau^2)(3q^2m_\tau^2-q^2M_\phi^2+q^2m_\tau^2 )\mathcal{C}_0^{\tau\tau\phi} \bigg]\,,\notag\\
F_3^{\text{m}}&= \frac{1}{4 \pi^2 m_\tau^2 q^2(q^2-4m_\tau^2)}\frac{\tilde{c}_{\gamma\gamma} c_S^{\tau} m_\tau}{\Lambda^2} \frac{\alpha_\text{em}}{4\pi} \bigg[2m_\tau^2 q^2 (4m_\tau^2-q^2)\left(\frac{1}{\epsilon} + \log \frac{\mu^2}{m_\tau^2}\right) \nonumber \\
&\qquad +(-2q^4m_\tau^2 + 4q^2m_\tau^2M_\phi^2 - q^2M_\phi^4 + 2m_\tau^2M_\phi^2)\log \frac{m_\tau^2}{M_\phi^2} -  2m_\tau^2 (q^2-M_\phi^2)^2 \log \frac{M_\phi^2}{M_\phi^2-q^2}  \nonumber \\
&\qquad + 2m_\tau^2 (2m_\tau^2 q^2- M_\phi^2 q^2 + 2 m_\tau^2 M_\phi^2) \mathcal{B}_0^{\tau\tau\phi}- 4m_\tau^2(q^2-M_\phi^2)^2 \mathcal{C}_0^{\tau\tau\phi} \bigg] \nonumber \\
& \qquad  -\frac{1}{4 \pi^2 m_\tau^2 q^2(q^2-4m_\tau^2)}\frac{c_{\gamma\gamma} c_P^{\tau} m_\tau}{\Lambda^2} \frac{\alpha_\text{em}}{4\pi}\bigg[2m_\tau^2 q^2(q^2-4m_\tau^2)\left(2+\frac{1}{\epsilon} + \log \frac{\mu^2}{m_\tau^2}\right) \nonumber \\
& \qquad +(2m_\tau^2q^4-q^2M_\phi^4+ 2m_\tau^2M_\phi^4)\log \frac{m_\tau^2}{M_\phi^2} + 2 m_\tau^2 (q^4-M_\phi^4) \log \frac{M_\phi^2}{M_\phi^2-q^2}  \nonumber \\
& \qquad + (-2m_\tau^2M_\phi^2 q^2- 2m_\tau^4q^2+4m_\tau^4M_\phi^2) \mathcal{B}_0^{\tau\tau\phi}+ 4m_\tau^4(q^4-M_\phi^4) \mathcal{C}_0^{\tau\tau\phi} \bigg]\,,
\end{align}
while the expressions for the AMM and EDM of the $\tau$ lepton become:
\begin{align}
\label{atau_dtau_m}
a_\tau^{\text{m}}&= \frac{1}{\Lambda^2 }\frac{\alpha_\text{em}}{(4\pi)^3}\bigg[\frac{4}{3}\frac{\tilde{c}_{\gamma\gamma}\, c_P^\tau }{m_\tau^2}\bigg(6m_\tau^4 \Big[
\frac{1}{\epsilon} + \log \frac{\mu^2}{m_\tau^2}-1\Big] +  M_\phi^4 \log \frac{m_\tau^2}{M_\phi^2}+ 2m_\tau^2 (2m_\tau^2+ M_\phi^2)\mathcal{B}_0^{\tau\tau\phi} + 2 m_\tau^2 M_\phi^2 \bigg)\nonumber \\
&+ \frac{c_{\gamma\gamma}\,c_S^\tau}{m_\tau^2}\bigg(8m_\tau^4\Big[\frac{1}{\epsilon} + \log \frac{\mu^2}{M_\phi^2}+3 \Big] - \frac{8}{3}  (6m_\tau^4-6M_\phi^2m_\tau^2+M_\phi^4)\log\frac{m_\tau^2}{M_\phi^2} -\frac{8}{3}m_\tau^2 M_\phi^2 +\frac{8}{3} m_\tau^2 (M_\phi^2- 4m_\tau^2) \mathcal{B}_0^{\tau\tau\phi}\bigg)\bigg]\,,\notag\\
d_\tau^{\text{m}} &= \frac{e}{2\Lambda^2} \frac{\alpha_\text{em}}{(4\pi)^3}\bigg[\frac{c_{\gamma\gamma} \,c_P^{\tau}}{m_\tau^3} \bigg(8 m_\tau^4 \Big[\frac{1}{\epsilon} + \log \frac{\mu^2}{M_\phi^2}+ 2\Big] + \frac{4}{3}M_\phi^4\log \frac{m_\tau^2}{M_\phi^2} + \frac{8}{3}m_\tau^2 M_\phi^2 + \frac{8}{3}m_\tau^2 (M_\phi^2+2m_\tau^2) \mathcal{B}_0^{\tau\tau\phi}\bigg) \nonumber \\
& - \frac{\tilde{c}_{\gamma\gamma} \,c_S^{\tau}}{m_\tau^3} \bigg(8 m_\tau^4 \Big[\frac{1}{\epsilon} + \log \frac{\mu^2}{M_\phi^2}\Big]- \frac{4}{3}M_\phi^2(M_\phi^2-6m_\tau^2)\log \frac{m_\tau^2}{M_\phi^2} - \frac{8}{3}m_\tau^2 M_\phi^2 - \frac{8}{3}m_\tau^2(M_\phi^2-4m_\tau^2) \mathcal{B}_0^{\tau\tau\phi}\bigg)\bigg]\,.
\end{align}
The previous expressions feature a UV divergence, signaled by the pole $1/\epsilon$. This is a consequence of the fact that we are working in a setup that involves nonrenormalizable dimension-5 operators. Such a divergence is to be reabsorbed by including appropriate counterterms in the ALP EFT, as detailed in the main body of the paper, see Eq.~\eqref{eq:UVpole_Ren} and the related discussion. Since the finite value of such counterterms depends on the specific underlying UV completion, the finite contributions we have computed are subject to $\mathcal{O}(1)$ corrections and hence have to be interpreted with caution.
The universal logarithmic dependence on the NP scale $\Lambda$ can be obtained by identifying $1/\epsilon + \log \mu^2 \to \log \Lambda^2$.

\subsection{Light vector bosons}

The one-loop contributions to the form factor $F_2$ as mediated by the virtual exchange of a light vector boson interacting with SM fields as defined in Eq.~\eqref{eq:XBos} correspond to the diagrams in Figs.~\ref{fig:MinCouDiagrams} and~\ref{fig:CSdiagram}.

\subsubsection{Minimal couplings only}

\begin{figure}[t]
    \centering
    \includegraphics[width=0.48\linewidth]{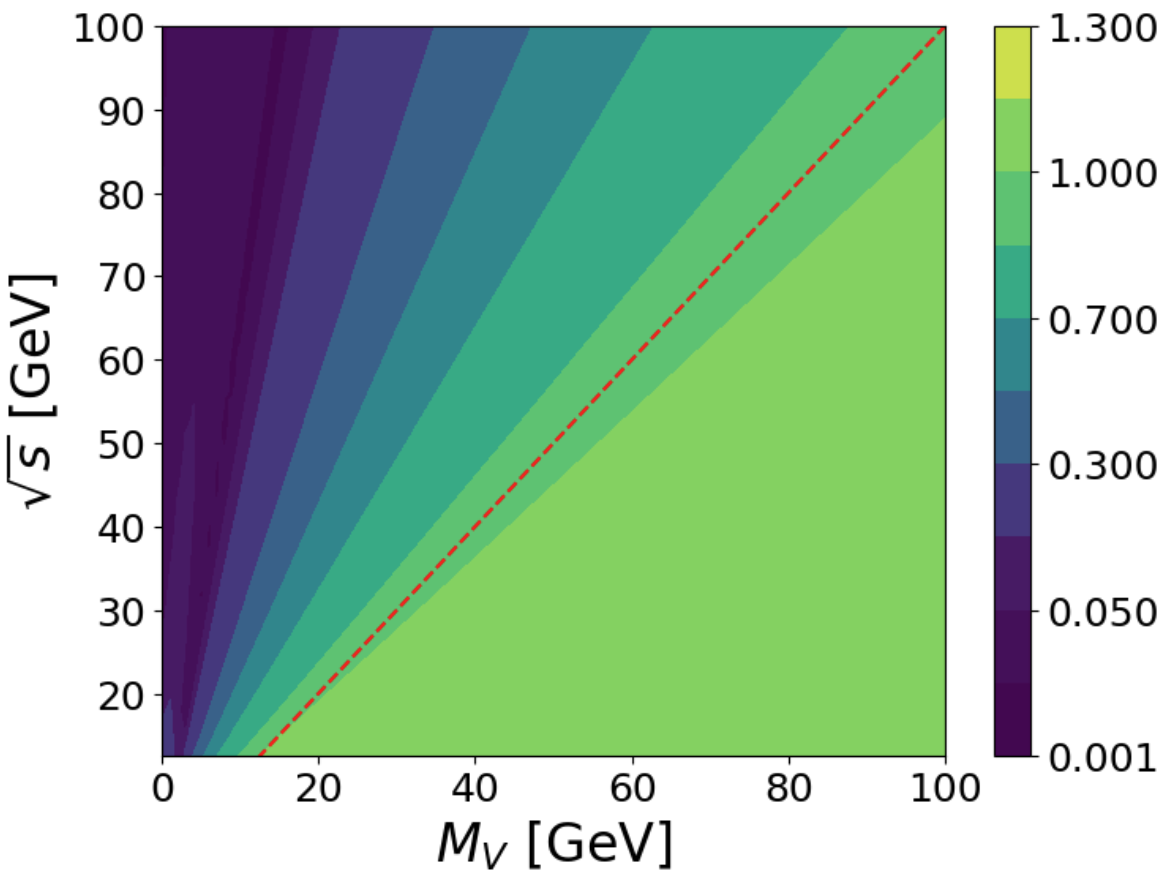}
    \includegraphics[width=0.48\linewidth]{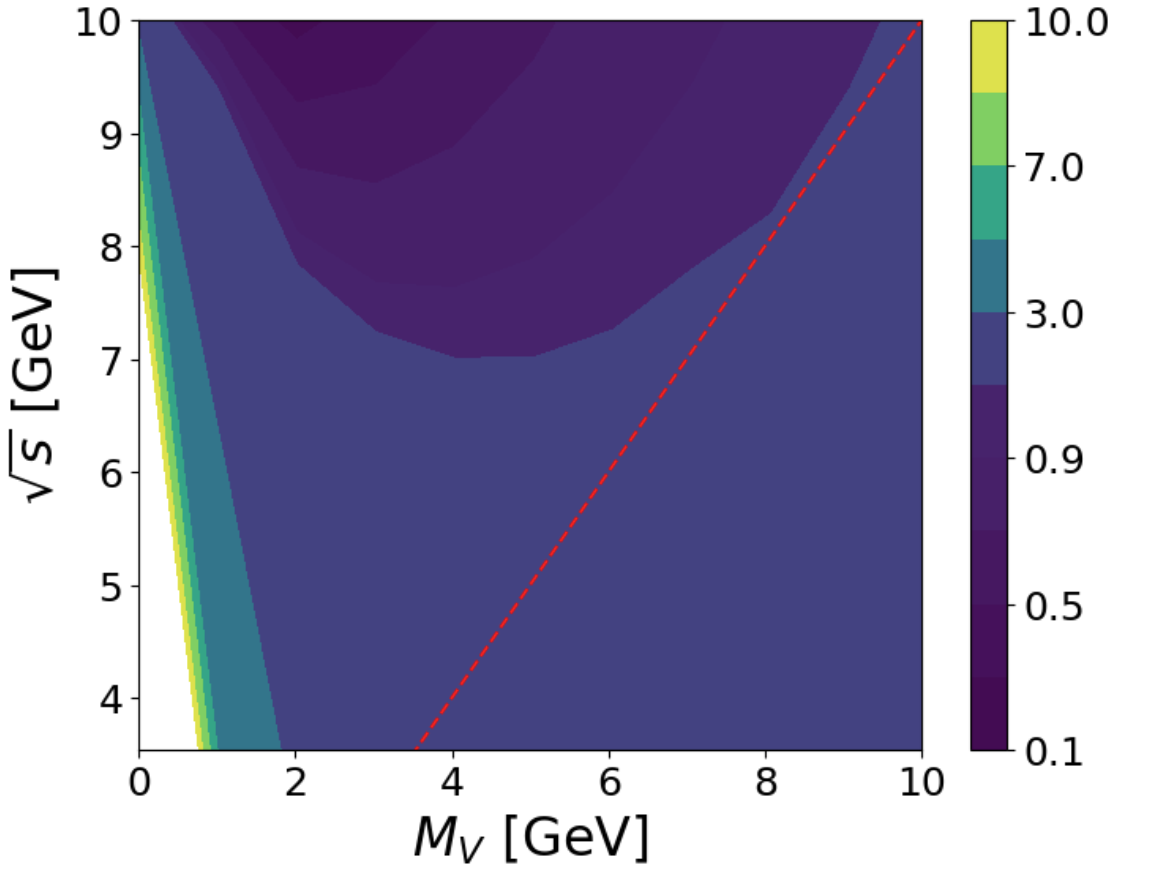}
    \caption{Momentum dependence of $|\Re F_2^\text{mc}|/|a_\tau^\text{mc}|$ for $y_A^\tau =1, \, y_V^\tau=0$. The diagonal line $\sqrt{s}= M_V$ serves as an indicative threshold separating the light NP scenario from heavy NP.
    }
    \label{fig:Energy_Comparison_MinCou}
\end{figure}

The expression for the form factor reads
\begin{align}
\label{eq:F2_MinCou2}
F^\text{mc}_2(q^2) &=  \frac{g_D^2 |y_A|^2}{8 \pi^2}\frac{1}{m_\tau^2 M_V^2 (q^2- 4 m_\tau^2)^2} \bigg\{2 m_\tau^2 M_V^2 \mathcal{B}_0^{\tau\tau V} \Big(20 m_\tau^4 + M_V^2 q^2- 2 m_\tau^2(5 M_V^2 + 4 q^2)\Big)\nonumber \\
& \qquad + 2 m_\tau^4 \mathcal{B}_0^{q\tau V} \Big(8 m_\tau^4 + 6 M_V^4 + 9 M_V^2 q^2 - 2 m_\tau^2(q^2+12M_V^2)\Big)\notag\\
&\qquad+ M_V^2 \bigg[ \log \frac{m_\tau^2}{M_V^2}\Big(16m_\tau^6+ 4 m_\tau^2(7M_V^2-q^2)+ q^2 M_V^4-10m_\tau^2 M_V^2 (M_V^2 + q^2)\Big)\nonumber \\
&\quad\qquad+4\mathcal{C}_0^{\tau\tau V} \Big(24 m_\tau^4 + 3 M_V^4 + 6M_V^2 q^2 + 2 q^4- 2 M_V^2(9M_V^2+ 7 q^2)\Big)-2m_\tau^2(2m_\tau^2 +M_V^2) (4m_\tau^2-q^2)
\bigg] \bigg\}\nonumber \\
&+\frac{g_D^2 |c_V|^2}{8 \pi^2}\frac{1}{m_\tau^2 M_V^2 (q^2- 4 m_\tau^2)^2}\bigg\{ 2 m_\tau^4 \big(q^2+6 M_V^2-4m_\tau^2 \big)\mathcal{B}_0^{q\tau \tau} 
-2 \mathcal{B}_0^{\tau \tau V}\Big(8 m_\tau^6 + m_\tau^2 M_V^2 q^2 - 2 m_\tau^4(q^2+ 5 M_V^2)\Big)\nonumber \\
& \qquad + M_V^2 \bigg[2 m_\tau^2 q^2 - 8 m_\tau^4 + \log\frac{m_\tau^2}{M_V^2}\Big(16m_\tau^4 + q^2 M_V^2 - 2 m_\tau^2(5M_V^2+ 2q^2)\Big) \notag\\
&\qquad\quad+ 4 m_\tau^4 \mathcal{C}_0^{\tau\tau V} \big(2q^2+ 3M_V^2-8m_\tau^2\big)\bigg]\bigg\}\,,
\end{align}
while for the AMM we find
\begin{align}
\label{eq:atauMinCou2}
a^{\text{mc}}_\tau &= - \frac{g_D^2 |y_A|^2}{m_\tau^4 8\pi^2M_V^2} \bigg[2m_\tau^6-5m_\tau^4M_V^2 + 2 m_\tau^2 M_V^4 + 2m_\tau^2M_V^2\big( M_V^2-2m_\tau^2\big) \mathcal{B}_0^{\tau\tau V} + \big(2 m_\tau^4 M_V^2 - 4 m_\tau^2 M_V^4 + M_V^6\big)\log\frac{m_\tau^2}{M_V^2} \bigg]\nonumber \\
& + \frac{g_D^2 |c_V|^2}{m_\tau^4 8\pi^2(4m_\tau^2- M_V^2)} \bigg[2m_\tau^2 \big(2m_\tau^4-4m_\tau^2 M_V^2+M_V^4\big) \mathcal{B}_0^{\tau\tau V}  \notag\\
&\qquad+ (4m_\tau^2-M_V^2)\, \bigg(m_\tau^4 - 2 M_V^2 m_\tau^2 + (2M_V^2m_\tau^2 - M_V^4)\log\frac{m_\tau^2}{M_V^2}\bigg) \bigg]\,.
\end{align}
The resulting momentum dependence of $|\Re F_2^\text{mc}|/|a_\tau^\text{mc}|$ is illustrated in Fig.~\ref{fig:Energy_Comparison_MinCou}, converging towards the EFT limit when $M_\phi^2\gg s$.

\subsubsection{Including Chern--Simons couplings}

Finally, including Chern--Simons couplings we obtain
\begin{align}
\label{eq:F2AnoCou2}
F^\text{an}_2(q^2) &= -\frac{C_A \,c_A}{16\, m_\tau^2 \,M_V^2 \,\pi^2}\frac{e^2 g_D^2}{q^2(q^2-4m_\tau^2)^2} \bigg[-2\,q^2\, m_\tau^2\,(q^2-4m_\tau^2) (-8\,q^2\,m_\tau^2+ 16 m_\tau^4+ 8 m_\tau^2 M_V^2 -M_V^4) \nonumber \\
&  \qquad +2m_\tau^2 \Big(q^4(M_V^4-2m_\tau^2 M_V^2 - 16 m_\tau^4) + q^2 (16m_\tau^6 + 36 m_\tau^4 M_V^2 - 8 m_\tau^2 M_V^4)+ 4m_\tau^4 M_V^4 - 16 m_\tau^6 M_V^2\Big)\mathcal{B}_0^{\tau \tau V}\notag\\
& \qquad +\log\frac{M_V^2}{M_V^2-q^2}\Big(8\,q^6\, m_\tau^4 + q^4(16m_\tau^6 - 20 m_\tau^4 M_V^2) + q^2\, (24m_\tau^4 M_V^2 - 64 m_\tau^6 M_V^2)
+48 m_\tau^6 M_V^4 - 12 m_\tau^4 M_V^6 \Big)\notag\\
&\qquad+\log\frac{M_\tau^2}{M_V^2}\Big(8q^6 m_\tau^4+ q^4(M_V^6- 4m_\tau^2 m_V^4-20 m_\tau^4 m_V^2+ 16 m_\tau^6)
+q^2 (-64 m_\tau^6 m_V^2+ 56 m_\tau^4 m_V^4- 8 m_\tau^2 m_V^6)\nonumber \\
& \qquad + 4m_\tau^4 M_V^6 - 16 m_\tau^6 M_V^4\Big)+ 8q^2m_\tau^4 (q^2-4m_\tau^2)^2\log \frac{\Lambda^2}{m_\tau^2}  +4m_\tau^4 \Big(q^6(8m_\tau^2-3M_V^2)\nonumber \\
&\qquad + q^4 (2M_V^4+ 2M_V^2 m_\tau^2-8m_\tau^4)- q^2 (M_V^6+8m_\tau^2M_V^4-32m_\tau^4M_V^2)
+8 M_V^4 m_\tau^4 - 2 m_\tau^2 M_V^6\Big) \mathcal{C}_0^{\tau \gamma V}\bigg]\,,
\end{align}
and
\begin{align}
\label{eq:atau_AnoCou2}
a_\tau^{\text{an}}&= -\frac{C_A\,c_A}{24 m_\tau^4 M_V^2 \pi^2}\,e^2g_D^2 \bigg[12 m_\tau^6 + 11 m_\tau^4 M_V^2 - 2 m_\tau^2 M_V^2+ 12 m_\tau^6 \log \frac{\Lambda^2}{m_\tau^2}  \notag\\
&  \qquad -M_V^2(12m_\tau^4-7m_\tau^2 M_V^2 + M_V^4)\log \frac{m_\tau^2}{M_V^2}+2m_\tau^2(4m_\tau^4-5m_\tau^2M_V^2 + M_V^4)\,\mathcal{B}_0^{\tau \tau V}\bigg]\,.
\end{align}
The previous expressions feature a UV divergence, signaled by a pole $1/\epsilon$. This is a consequence of the fact that we are working in a setup that involves nonrenormalizable Chern--Simons operators, see comment below Eq.~\eqref{atau_dtau_m} and the discussion following Eq.~\eqref{eq:F2an_thr} in the main text. In the expression for $a_\tau^{\text{an}}$, 
the universal logarithmic dependence on the NP scale $\Lambda$ has been identified via $1/\epsilon + \log \mu^2 \to \log \Lambda^2$.

\section{Searches for a tauphilic vector boson at Belle II}
\label{sec:Tauphilic_VBs}

\subsection{EFT of a light tauphilic vector boson}

\begin{figure}[t]
    \centering
    \includegraphics[width=0.75\linewidth]{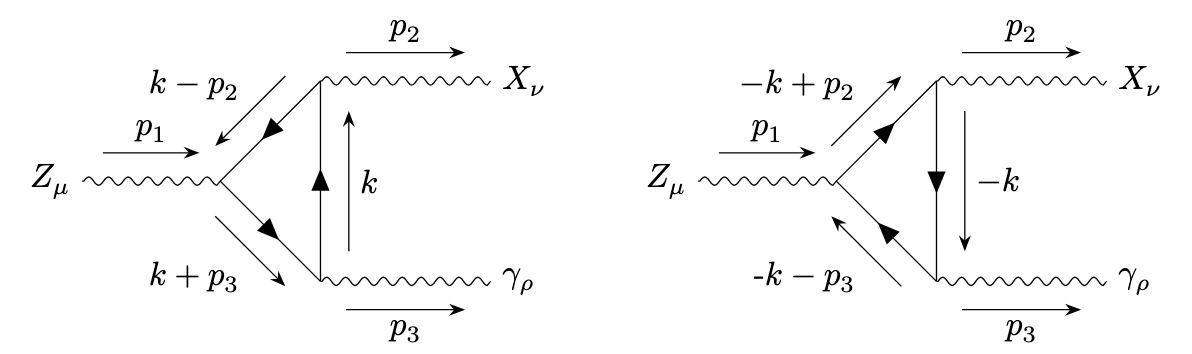}
    \caption{Triangle diagrams contributing to the anomalous three-boson vertices. For the sake of concreteness, we illustrate the diagrams for the assignment $Z_\mu$, $X_\nu$, and $\gamma_\rho$, matching the discussion in the main text. Generalizations of such results are immediate.}
    \label{fig:Triangles}
\end{figure}

In this section we will introduce the EFT setup in which we will work to discuss the phenomenology of light tauphilic vector bosons.
The starting point in our discussion consists of defining the tree-level couplings of a new vector state to the third generation of leptons in the SM:
\begin{equation}
\label{eq:XintLagGauge2}
    \mathcal{L}^\text{int}_{\text{spin-1}} = - \, g_D\,\chi_L\, X_\mu \,\bar{\ell}_{3L}\, \gamma^\mu \ell_{3L}\,+ \, g_D\,\chi_R\, X_\mu \,\bar{\tau}_R\, \gamma^\mu \tau_R\,,
\end{equation}
which results in the following coupling to $\tau$ leptons:
\begin{equation}
\label{eq:XintLagMinCou2}
    \mathcal{L}^{\text{int,}\tau}_{\text{spin-1}} \supset - \, g_D\, X_\mu \,\bar{\tau}\, \gamma^\mu \left(y_V + \,y_A \,\gamma_5\right) \tau\,,
\end{equation}
where we have conveniently defined $2\,y_V = \chi_R+\chi_L$ and $2\,y_A = \chi_R+\chi_L$.
In the absence of further couplings to SM fermions, this Lagrangian leads to the appearance of anomalies induced by the exchange of virtual $\tau$ leptons and neutrinos in a fermionic triangle loop, see Fig.~\ref{fig:Triangles} for an example.
The resulting amplitude can be parameterized as follows (Rosenberg parameterization~\cite{Rosenberg:1962pp}):
\begin{align}
\label{eq:Rosenberg}
g_i g_j g_k\,\Gamma_{\mu\nu\rho}^{ijk}(k_1, k_2; w, z)&= g_i g_j g_k \,t^{ijk} \Big[A_1 \,\epsilon_{\mu\nu\rho\sigma}\,k_2^\sigma+ A_2 \,\epsilon_{\mu\nu\rho\sigma}\,k_1^\sigma  \nonumber \\
&\qquad + (B_1 k_{2\nu} + B_2 k_{1\nu})\, \epsilon_{\mu\rho\alpha\beta}\,k_2^\alpha \,k_1^\beta+  (B_3 k_{2\rho} + B_4 k_{1\rho})\, \epsilon_{\mu\nu\alpha\beta}\,k_2^\alpha \,k_1^\beta\Big]\,,
\end{align}
where the $g_i$ are the gauge couplings associated to the gauge vector species $i$ and $t^{ijk}$ represents the value of the anomaly, which depends exclusively on the charge assignment of the leptons in the loop.
The coefficients $B_i(k_1, k_2)$ are implicitly defined via the following integrals~\cite{Anastasopoulos:2006cz}:
\begin{align}
B_1(k_1, k_2) &= - B_4(k_2, k_1) = - \frac{i}{8\pi^2} \int_0^1 d\alpha \int_0^{1-\alpha} d\beta\, \frac{2 \alpha \beta }{ \alpha k_2^2 + \beta k_1^2 - (\alpha k_2 - \beta k_1)^2 - m_f^2}\,, \notag\\
B_2(k_1, k_2) &= - B_3(k_2, k_1) = - \frac{i}{8\pi^2} \int_0^1 d\alpha \int_0^{1-\alpha} d\beta\, \frac{2 \beta\,(1-\beta) }{ \alpha k_2^2 + \beta k_1^2 - (\alpha k_2 - \beta k_1)^2 - m_f^2}\,.
\end{align}
The coefficients $A_1(k_1, k_2)$ and $A_2(k_1, k_2)$ depend explicitly on the regularization scheme chosen for the fermionic loop via the constants $w$ and $z$. They can be expressed in terms of the other $B_i$ by noticing that
\begin{align}
k_1^\nu \,\Gamma_{\mu\nu\rho}^{ijk} &= t^{ijk} (A_1+ B_1 \,k_1 k_2 + B_2\,k_1^2)\, \epsilon_{\mu\rho\alpha\beta}\,k_2^\alpha k_1^\beta \nonumber \,,\\
k_2^\rho \,\Gamma_{\mu\nu\rho}^{ijk} &= t^{ijk}\,(A_2+ B_3 \,k_2^2 + B_4\,k_1 k_2) \, \epsilon_{\mu\nu\alpha\beta}\,k_2^\alpha k_1^\beta \nonumber \,,\\
(k_1+ k_2)^\mu \,\Gamma_{\mu\nu\rho}^{ijk} &=t^{ijk}\,(A_2 - A_1) \, \epsilon_{\nu\rho\alpha\beta}\,k_2^\alpha k_1^\beta\,,
\end{align}
has to be matched on the Ward identities for the three currents entering the vertices of the triangle diagram~\cite{Michaels:2020fzj}:\footnote{Here we are considering a case in which the vector boson associated to the index $\nu$ is purely vectorial with unit charge, while the $X$ and $Z$ vector bosons partaking in the amplitude can have both a vectorial $y_V^{X/Z}$ and an axial component $y_A^{X/Z}$. $Z$ need not necessarily be the SM $Z$ boson.} 
\begin{align}
k_1^\nu \,\Gamma_{\mu\nu\rho}^{ijk} &=\frac{1}{4\pi^2} \left[(w-1)(y_V^X y_A^Z + y_A^X y_V^Z) - 4 m_f^2\, y_V^X\,  y_A^Z\, C_0(m_f^2)\right]\, \epsilon_{\mu\rho\alpha\beta}\,k_2^\alpha k_1^\beta \nonumber \,,\\
k_2^\rho \,\Gamma_{\mu\nu\rho}^{ijk} &=\frac{1}{4\pi^2} \left[(1+z)(y_V^X y_A^Z + y_A^X y_V^Z)\right]\,\epsilon_{\mu\nu\alpha\beta}\,k_2^\alpha k_1^\beta \nonumber \,,\\
(k_1+ k_2)^\mu \,\Gamma_{\mu\nu\rho}^{ijk} &= -\frac{1}{4\pi^2} \left[(w-z)(y_V^X y_A^Z + y_A^x y_V^Z) + 4 m_f^2\, y_A^X\,  y_V^Z\, C_0(m_f^2)\right]\,\epsilon_{\nu\rho\alpha\beta}\,k_2^\alpha k_1^\beta \,,
\end{align}
where $C_0(m_f)= C_0(0, M_Z^2, m_{X}^2 ,m_f, m_f, m_f)$ is the scalar $C_0$ Passarino--Veltman function.
As was mentioned before, $A_1$ and $A_2$ depend on the  regularization scheme, which has to be chosen in the same way for all the fermion fields appearing in the theory~\cite{Michaels:2020fzj}.
The symmetric regularization scheme, corresponding to the so-called \textit{consistent anomaly} case, can be found by choosing $w=-z=1/3$. Alternatively, it is possible to work in the so-called \textit{covariant anomaly} regularization scheme, where all the anomalous divergences are associated to a single current, by setting $w=-z=1$.
In order to take care of such anomalies, one then has to postulate the existence of new fermionic degrees of freedom beyond the SM, the so-called \textit{anomalons}.  Integrating them out of the theory leads to results that are particularly simple in the \textit{consistent} case, where it is easy to see that the regularized heavy-fermion loop gives rise to effective dimension-4 Wess--Zumino terms~\cite{Dror:2017nsg, Michaels:2020fzj,DiLuzio:2022ziu}:
\begin{align}
\label{eq:XintLagAnoCou2}
\mathcal{L}_\text{an} &= g_D \, (g')^2\, C_{BB} \,\epsilon^{\mu\nu\alpha\beta}X_\mu B_\nu \partial_\alpha B_\beta +g_D \, g^2\, C_{WW}\,\epsilon^{\mu\nu\alpha\beta}X_\mu \left(W^I_\nu \partial_\alpha W^I_\beta + \frac{g}{3}\, \epsilon^{IJK} \,  W_\nu^I W_\alpha^J W_\beta^K\right)\nonumber \\
&\supset g_D \,e^2 C_{\gamma\gamma} \epsilon^{\mu\nu\alpha\beta}X_\mu A_\nu \partial_\alpha A_\beta+ g_D \,g \,g' C_{\gamma Z} \epsilon^{\mu\nu\alpha\beta}(X_\mu A_\nu \partial_\alpha Z_\beta+X_\mu Z_\nu \partial_\alpha A_\beta)\,.
\end{align}
In our setup in Eq.~\eqref{eq:XintLagMinCou2} for instance, provided that the mass eigenstates $\tau$ are also $U(1)'$ gauge eigenstates, one would find:\footnote{The anomaly coefficient reads $\mathcal{A}_{\alpha, \beta, \gamma} = \text{Tr}[T_\alpha, \{T_\beta, T_\gamma\}]|_R-\text{Tr}[T_\alpha, \{T_\beta, T_\gamma\}]|_L$, where the $T_\alpha$ are the generators of the gauge group whose vector boson is attached to the vertex $\alpha$.}
\begin{equation}
C_{BB} = - \frac{1}{24\pi^2} (y_V+ 3y_A)\,,\qquad C_{WW} =  \frac{1}{24\pi^2}(y_V-y_A)\,.
\end{equation}

Clearly physical observables are independent of the specific framework one is using to perform computations, and physical results are only those that stem from the sum of the one-loop SM amplitude and the ones involving the (possibly indirect) effect of heavy anomalons. 
In particular, it is important to notice that the presence of heavy anomalons manifests itself at low energies only provided that there is a consistent mass gap between the SM fermions and the heavy NP ones; if no mass gap existed, anomaly cancellation would leave no physical effect on any observable. In this sense, what is called an anomalous effect should be rather regarded as the nondecoupling phenomenology related to the existence of heavy massive states that are responsible for the overall UV consistency of the theory.

The Lagrangian terms in Eqs.~\eqref{eq:XintLagMinCou2} and \eqref{eq:XintLagAnoCou2} contain all the relevant information to explore the low-energy phenomenology of a new vector boson that couples exclusively to the third generation of leptons. This is the objective we are going to pursue in the remainder of this appendix. Before devoting ourselves to such a task, however, some comments are in order.
First of all, the interactions in Eqs.~\eqref{eq:XintLagMinCou2} and \eqref{eq:XintLagAnoCou2} completely determine the decay rate of a light vector boson. 
Below the $\tau^+\tau^-$ threshold, the decay to two photons is inhibited by the Landau--Yang theorem and the only available decay channel is the one to $\tau$ neutrinos, see Eq.~\eqref{eq:decayrates}.
Above the $\tau^+\tau^-$ threshold, the $\tau^+\tau^-$ channel opens up with a comparable rate, as detailed in Eq.~\eqref{eq:decayrates}.
Couplings to the first and second generation of leptons are induced at the one-loop level via $\gamma$--$X$ or $Z$--$X$ mixing. As far as the mixing with the electromagnetic field is concerned, $\frac{\Pi}{2} F_\gamma^{\mu\nu}F^X_{\mu\nu}$, these effects amount to~\cite{Araki:2017wyg}:
\begin{equation}
\label{eq:mixing_1L}
\Pi = \frac{e\,g_D}{36\pi^2} \left[-5-12 \frac{m_\tau^2}{m_X^2} + 6 \left(1+ \frac{2m_\tau^2}{m_X^2}\right)f_\Pi\left(\frac{m_X^2}{4m_\tau^2}\right)-3 \log \frac{\Lambda^2}{m_\tau^2}\right]\,,
\end{equation}
where
\begin{equation}
f_\Pi(x) = \begin{cases}\frac{\sqrt{1-x^2}}{2}\log\frac{1+\sqrt{1-x^2}}{1-\sqrt{1-x^2}}\quad &\text{ if } \quad x\leq1\,,\\
\sqrt{x^2-1} \arctan\frac{1}{\sqrt{x^2-1}}\quad &\text{ if } \quad x>1\,.
\end{cases}
\end{equation}
Similarly, one can obtain the mixing with the $Z$ boson, even if it plays a subdominant role at low energies. While these mixing effects exist and play an important phenomenological role, as we will discuss in the following, they can be neglected as far as the decay of a light vector boson is concerned, leading to loop-suppressed branching ratios to electrons and muons.

Second, and partially related to this, the phenomenology related to light tauphilic vector bosons crucially depends on the axial or vectorial nature of its couplings to $\tau$ leptons. Indeed, in the presence of a nonvanishing axial coupling, a nonzero contribution to the anomalous $X\gamma \gamma$ coupling exists,
\begin{equation}
C_{\gamma \gamma} =C_A = C_{BB} + C_{WW} = -\frac{y_A}{6\pi^2}\ne 0\,.
\end{equation}
This is not the case in the absence of an axial coupling, and the most relevant anomalous coupling at low energies is $XZ\gamma$, whose anomalous coupling reads
\begin{equation}
C_{\gamma Z} = c_w^2 C_{WW}-s_w^2 C_{BB} = \frac{y_V}{24\pi^2}-\frac{y_A}{24\pi^2}(c_w^2-3 s_w^2)\,.
\end{equation}
 The contributions associated to this vertex are, however, expected to be suppressed by inverse powers of the $Z$ boson mass at low energies, and hence to play a subdominant role in the low-energy phenomenology of light tauphilic vectors.
On the other hand, as mentioned before, in the case of light vectors with $y_V \ne 0$, one has a relevant contribution to the $X$--$\gamma$ mixing, which results in a rich phenomenology at low energies. The mixing with the $Z$ boson as induced by both vectorial and axial couplings can be neglected in a first approximation as one is interested in energy scales $q^2 \ll M_Z^2$.

In the following sections we will explore the possibility to probe this kind of scenario at low energies, with a particular focus on the experimental searches that can be carried out at $B$ factories such as Belle II. For the sake of simplicity, we organize the material in two subsections: probes from $a_\tau$ and collider probes.

\subsection{\texorpdfstring{$\boldsymbol{a_\tau}$}{atau} as a probe of light tauphilic vector bosons}

We can directly employ the interactions in Eqs.~\eqref{eq:XintLagMinCou2} and~\eqref{eq:XintLagAnoCou2} to compute the impact of light tauphilic vector bosons on $a_\tau$ by making use of the formulae we have given in the previous sections. Neglecting the subdominant contribution due to the exchange of a virtual $Z$ boson, it amounts to the sum of Eqs.~\eqref{eq:atauMinCou2} and \eqref{eq:atau_AnoCou2}, which, in the limit of a light vector boson ($m_X \ll m_\tau$), reduce to
\begin{equation}
a_\tau(m_\tau \gg m_X) = - \frac{g_D^2}{4 \pi^2}\,|y_A|^2\left[\frac{m_\tau^2}{m_X^2}- \frac{5}{2} + \log \frac{m_\tau^2}{m_X^2}\right]+ \frac{g_D^2}{8\pi^2}\,|y_V|^2\,- \frac{C_{\gamma\gamma}\,y_A}{2 \pi^2}\frac{m_\tau^2}{m_X^2} e^2 g_D^2 \left(1+  \log \frac{\Lambda^2}{m_\tau^2}\right)\,.
\end{equation}
From this expression it is apparent that $a_\tau$ receives a longitudinal enhancement as long as it possesses an axial coupling. Moreover, as the anomalous contribution is formally of one-loop order, it generally plays a subdominant role in setting the overall contribution to $a_\tau$, the only exception being given by accidental cancellations between the axial and the vectorial contributions stemming from the minimal couplings of $X$ to $\tau$ leptons or by large anomaly values~\cite{Anastasopoulos:2022ywj}.

In principle, the full two-loop contribution due to the effective three-gauge boson vertex should be taken into account, rather than only the mass-independent terms related to the anomaly. This can be done along the lines of Ref.~\cite{Hoferichter:2025yih} once the full particle content of the theory in the UV is known, see Sec.~\ref{sec:specific_scenario}.
However, we will argue here that the dominant contributions are actually captured by the anomalous couplings of a new vector boson to photons, governed by the combination $C_A = C_{BB}+ C_{WW}$. The impact of dimension-4 Wess--Zumino terms on $a_\mu$ has been studied in detail in Ref.~\cite{Anastasopoulos:2022ywj}. There it was shown that the leading contribution is induced by the anomalous contributions of both SM particles and heavy anomalons, with subleading contributions induced by fermion-mass dependent effects stemming from the fermionic triangle loop. These effects are encoded in the coefficients $B_i$, which have mass dimension $-2$; since the only mass available in a triangle loop is given either by external massive vectors, or by internal fermions, the impact of a $\tau$ loop cannot exceed the one from an electron or a muon loop. Moreover, in the limit of a large energy $s$ injected in the triangle loop, the coefficients $B_i$ decouple at least as fast as $1/s$, thus preventing any possible $s/m_f^2$ enhancement from the mass-dependent terms. As these are seen to be subdominant with respect to the effects induced by the anomaly, we can safely estimate the largest contribution to the form factor $F_2(q^2)$ by simply focusing on the anomalous contributions to the amplitude. 

The total contribution amounts to the sum of Eqs.~\eqref{eq:F2_MinCou2} and \eqref{eq:F2AnoCou2}.
Again, the contributions proportional to $y_A$ are the only ones that experience a longitudinal enhancement for low vector masses, as can be appreciated from Figs.~\ref{fig:VB_anom} and \ref{fig:Energy_Comparison_MinCou}.

\subsection{Collider probes of light tauphilic vector bosons}

In this section we will discuss the main collider probes of light tauphilic vector bosons at $B$ factories. 
These can be thought of as belonging to two classes: processes that test directly the coupling to leptons, and processes proceeding through an anomalous $X\gamma \gamma$ or an $X Z \gamma$ effective vertex.

\subsubsection{\texorpdfstring{$e^+ e^- \to \tau^+ \tau^- X$}{eetotautau} processes}

The first class includes those processes where the $X$ is directly radiated off a final-state $\tau$, such as $e^+ e^- \to \tau^+ \tau^- X$, with the $X$ further decaying to either visible or invisible states. 
In the limit $m_X^2\ll m_\tau^2, s$ the corresponding longitudinally enhanced differential cross section reads
\begin{equation}
\frac{d\sigma_{L}}{dx_1 dx_2} = \frac{e^4 g_D^2}{768 \pi^3}|y_A|^2 \frac{m_\tau^2}{m_X^2}\frac{1}{s} \frac{(-2+x_1+x_2)^2}{(1-x_1)(1-x_2)}\,,
\end{equation}
where $s = (p_1 + p_2)^2$ represents the momentum exchanged in the scattering process and the $x_i$ variables are defined as 
\begin{equation}
x_i = \frac{2Q\cdot k_i}{Q^2}\,.
\end{equation}
Here, the $k_i$ are the four-momenta of the final-state particles, with the following assignments, valid for all the diagrams shown above: $\tau^-(k_1)$, $\tau^+(k_2)$, and  $X(k_3)$.
The total cross section can be obtained by integrating the previous expression as follows:
\begin{equation}
\label{eq:Integration}
\sigma_{L} = \int^{x_1^{\text{max}}}_{x_1^{\text{min}}}  dx_1 \int_{x_2^{\text{min}}}^{x_2^{\text{max}}}  dx_2 \,\frac{d^2\sigma_{L}}{dx_1dx_2} \,,
\end{equation}
where 
\begin{align}
\label{eq:x2Boundaries}
    x_2^{\text{max}/\text{min}} &= \frac{2 - 2 r - 3 x_1 + r x_1 + x_1^2 + 4 \epsilon  - 2 x_1 \epsilon  \pm
   \sqrt{x_1^2 - 4 \epsilon} \, \sqrt{
     1 - 2 r + r^2 - 2 x_1 + 2 r x_1 + x_1^2 - 4 r \epsilon}}{2 (1 - x_1 + \epsilon)}\,,
\end{align}
and 
\begin{equation}
\label{eq:x1Boundaries}
    x_1^{\text{max}} = 1- r -2 \sqrt{\epsilon} \, \sqrt{r}\,,\qquad  
    x_1^{\text{min}} = 2 \sqrt{\epsilon}\,,
\end{equation}
where $r = m_X^2/s$ and $\epsilon = m_\tau^2/s$.
In the limit $m_X^2\ll m_\tau^2\ll s$, we obtain the result in Eq.~\eqref{eq:ee_tautauX1}.

Once the cross section for the process $e^+ e^- \to \tau^+ \tau^- X$ is either computed or simulated by making use of a Monte-Carlo code, a bump hunt for the resonant production of an $X$ boson can in principle be performed, with the details depending on whether it decays visibly or invisibly.

The former case proceeds through the one-loop conversion of an on-shell $X$ to an energetic photon that eventually decays to an electron pair or a muon pair, provided that the $X$ has a vectorial coupling.
It is then possible to perform a bump hunt in the invariant mass of the produced lepton pair to reconstruct the mass of the decaying $X$. If the $X$ only features axial couplings, such a possibility is further suppressed by powers of $q^2/M_Z^2$. 
Given the suppression expected in this case, we do not delve further in the extraction of a bound from this kind of process. 

In case the $X$ decays to invisible states, instead, the search can in principle be performed, but faces serious issues. Indeed, as the $X$ decays invisibly, the bump hunt has to be performed by making use of the information from the invariant mass of the remaining visible states, i.e., that of the $\tau$ pair. However, reconstructing the invariant mass of two $\tau$ leptons is difficult due to the presence of neutrinos in their decay products and has so far been performed only under the assumptions that they are produced back-to-back, carrying all the energy available in the CM. In general, this is not our case, in which the $X$ can in principle be radiated with a portion of the total available energy, $2m_\tau < E_X < \sqrt{s_B}$.
If, on the other hand, the $X$ is emitted as a soft particle, with little kinematical influence on the distribution of the final-state decay products, the search is in principle still viable, but has to confront a large QED background from the process $e^+ e^- \to \tau^+ \tau^-$.

\subsubsection{\texorpdfstring{$e^+ e^- \to \gamma\, X$}{eetoXgamma} processes}

This class of processes includes both resonant quarkonium decays and searches in anomalous processes such as $e^+ e^- \to \gamma \,X$, with the $X$ further decaying to either a visible state (a $\tau^+\tau^-$ pair) or to an invisible state (a neutrino pair), see Fig.~\ref{fig:MMTTDiag2}.
As far as the former are concerned, the 
Breit--Wigner approximation can be employed and one can write the resonant production $e^+e^-\to V \to \gamma X$ cross section as
\begin{equation}
\sigma_{\rm R} = \frac{12\pi\Gamma_V^2}{(s-M_V^2)^2 + M_V^2\Gamma_V^2}\,{\mathcal{B}}(V \to e^+e^-)
    {\mathcal{B}}(V \to \gamma X)\,.
    \label{eq:sigmaR_quarkonia}
\end{equation}
In the previous expression, $M_V$ and $\Gamma_V$ are the mass and decay width of the quarkonium state $V$, while $f_V$ represents its decay constant~\cite{Merlo:2019anv,DiLuzio:2024jip}. The  experimental value for the branching ratio $\mathcal{B}(V\to e^+e^-)$ can be found in Refs.~\cite{CLEO:2005ejw,BaBar:2008xay,Belle:2022cce,ParticleDataGroup:2024cfk}.
The leading contribution to the decay of a quarkonium state $V$ to $X\,\gamma$ is instead found to be
\begin{equation}
\label{eq:BR_quarkonia}
\Gamma(V\to \gamma \, X)\big|_L = Q_V^2\frac{(e^2g_D)^2}{48\pi^4}\frac{f_V^2}{m_X^2}M_V\,\bigg|\sum_f 2A^f_1-B^f_2\bigg|^2+Q_V^2\frac{(e \,g\,g_D)^2}{48\pi^4(4c_w)^2}\frac{f_V^2} 
{m_X^2}\frac{M_V^3}{M_Z^2}\bigg|\sum_f {2A'_1}^{f}-{B'_2}^f\bigg|^2\,,
\end{equation}
where $Q_V$ is the electric charge of the quarks making up the quarkonium state, and where the primed quantities are computed with the kinematics and the anomaly requirements associated with an anomalous vector boson rather than an anomalous axial boson.

Requiring no evidence for a signature of an $X$ boson decaying to a pair of $\tau$ leptons or to invisible states, we can then set limits on $g_D \,y_A$ and $g_D \,y_V$, with the latter being suppressed with respect to the former by an unavoidable factor of $M_V^2/{M_Z^2}$ due to the vectorial nature of quarkonia resonances.

An important point to be stressed regarding these searches is that Eq.~\eqref{eq:BR_quarkonia} can only be used for resonant searches, namely, when the parent meson can be reconstructed by the kinematics of the process, as it would be the case, for instance, when identifying the $\Upsilon(1S)$ through
$\Upsilon(2S)\to \Upsilon(1S) \pi^+ \pi^-$. Contrarily, if the experiment is performed at the energy $\sqrt{s} = M_V$, but the quarkonium state cannot be kinematically identified, the search results in being sensitive to both nonresonant and resonant [Eq.~\eqref{eq:sigmaR_quarkonia}] cross sections~\cite{Merlo:2019anv}. 
While the $\Upsilon (1S)$, $\Upsilon (2S)$, and $\Upsilon (3S)$ have a relatively narrow width and can be kinematically reconstructed at BaBar and Belle II, the $\Upsilon(4S)$ resonance has a width that is much larger than the energy spread of the beam. As a consequence, in this latter case, nonresonant contributions  dominate over the resonant ones.

\begin{figure}[t]
    \centering
    \includegraphics[width=1\linewidth]{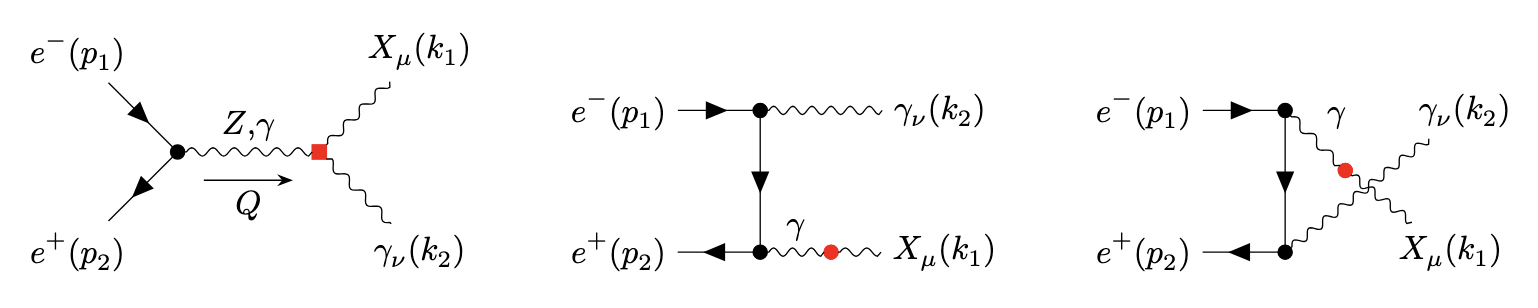}
    \caption{Diagrams contributing to the scattering process $e^-(p_1) \,  e^+(p_2) \rightarrow X(k_1) \, \gamma(k_2)$. Red squared dots represent NP insertions, while black dots denote the insertion of a SM vertex.}
    \label{fig:MMTTDiag2}
\end{figure}

The nonresonant cross sections for the process $e^+ e^- \to \gamma\, X$ receive two kinds of contributions: a longitudinally enhanced $s$-channel contribution stemming from the insertion of an effective, anomalous vertex, and $t$- and $u$-channel contributions that originate from the virtual conversion of a photon to an $X$ boson mediated by a fermion loop, see Fig.~\ref{fig:MMTTDiag}. We obtained the full cross section as induced by these three diagrams, by properly taking into account the effect of the full one-loop anomalous vertex in the Rosenberg parameterization. As these are quite lengthy expressions, we do not report them here; yet, in order to qualitatively understand their behavior, we report and comment on the leading contribution from each class of diagrams.

First of all, the $s$-channel amplitude displays the well-understood longitudinal enhancement for small vector boson masses and reads: 
\begin{align}
\frac{d\sigma}{d\cos\theta}(e^+ e^- \to \gamma^* \to  \gamma X) &= |y_A^\tau|^2\frac{(e^3\,g_D)^2}{384\,\pi \,m_X^2}\frac{3}{16}(3+ \cos 2\theta) \bigg|\sum_f 2A^f_1-B^f_2\bigg|^2\,, \notag\\
\frac{d\sigma}{d\cos \theta}(e^+ e^- \to Z^* \to  \gamma X) &= |y_V^\tau|^2\frac{(e^3\,g_D)^2}{384\,\pi \,m_X^2 c_w^4 s_w^4}\frac{s^2}{M_Z^4} \frac{3}{16}(3+ \cos 2\theta)\bigg|\sum_f {2A'_1}^f-{B'_2}^f\bigg|^2\,,
\end{align}
where the primed quantities are computed with the kinematics and the anomaly requirements associated with an anomalous vector boson rather than an anomalous axial boson. As far as these leading contributions are concerned, we find agreement with the results in Ref.~\cite{Kribs:2022gri}.
In the limit $m_X^2\ll s$, the sum of the pure $t$- and $u$-channel contributions, together with their interference, are instead found to be
\begin{equation}
\frac{d\sigma}{d\cos\theta}(e^+ e^- \to \gamma^* \to  \gamma X) = \frac{e^4}{16\pi^2}\frac{|\Pi|^2}{s}\frac{1+\cos^2 \theta}{1-\cos^2 \theta}\,,
\end{equation}
where $\Pi$ is given by the expression in Eq. \eqref{eq:mixing_1L}.
The interference between the $s$- and the $t$- or $u$-channel amplitudes is zero in the case of a virtual $\gamma$ being exchanged in the amplitude. This fact can be easily understood, as the anomalous $X\gamma \gamma $ coupling is $P$-odd, while standard QED interactions are $P$-even. The interference is different from zero in the case of a virtual $Z$ boson exchange, but we do not report here the corresponding expression, as it does not display any longitudinal enhancement. 

Having at hand explicit expressions for the $e^+ e^- \to X \,\gamma$ cross section, we can easily identify the leading contributions in the case of either a purely vectorial or a purely axial spin-$1$ state.
In the case of an anomalous axial-vector boson, the leading contribution is generated by the $s$-channel exchange of a photon, with subdominant effects being given by the virtual exchange of a $Z$ boson in the $s$-channel (suppressed by both a factor $s/M_Z^2$ and by an accidentally small vectorial couplings to leptons, $1- 4s_w^2\simeq 0.08$). A mixing between the axial component of the $Z$ boson can give rise to $t$- and $u$-channel exchanges, which, however, suffer from the same kind of suppression, and hence lead to subleading contributions. The situation is drastically different for the case of an anomalous vector having vectorial couplings to SM fields. In this case, there can be a direct mixing with the photon, and hence a relevant contribution from the $t$- and $u$-channels; the $Z$-mediated $s$-channel contribution proceeds in this case via the axial coupling of fermions, and is hence less suppressed with respect to the axial $X$ case.

Having discussed the main features of the NP signal, we turn now to the discussion of the procedure we have followed to place experimental bounds on both quarkonia decays and on nonresonant processes at Belle II.

\paragraph{Quarkonia decays}

\begin{table}[t]
    \centering
    \renewcommand{\arraystretch}{1.3}
    \begin{tabular}{lcl}\toprule
      Process & Type &  Experiment \\
      \colrule
      {$\Upsilon(1S) \to \gamma \tau^+ \tau^ -$} & {Resonant}  & BaBar~\cite{BaBar:2012sau}, Belle~\cite{Belle:2021rcl}\\
    {$\Upsilon(1S) \to \gamma + \text{inv}$} & {Resonant}  & BaBar~\cite{BaBar:2010eww}, Belle~\cite{Belle:2021rcl}\\
        $\Upsilon(3S) \to \gamma \tau^+\tau^-$ & Mixed  & BaBar~\cite{BaBar:2009oxm} \\\botrule
        \renewcommand{\arraystretch}{1.0}
    \end{tabular}
    \caption{Visible and invisible quarkonia decays relevant for the discussion in the main text.}
    \label{tab:quarkonia_processes}
\end{table}

As far as quarkonia decays are concerned, we employ the processes reported in Table~\ref{tab:quarkonia_processes}. In particular, we can directly translate the upper bounds from Refs.~\cite{BaBar:2010eww, BaBar:2012sau, Belle:2021rcl} to our framework.
These were originally intended to test light Higgs bosons employing $24.91\,\text{fb}^{-1}$ of data regarding the tagged process $\Upsilon(2S)\to \Upsilon(1S) \pi^+ \pi^-$ at Belle~\cite{Belle:2021rcl} and $14.4\,\text{fb}^{-1}$ of data regarding the tagged process $\Upsilon(2S)\to \Upsilon(1S) \pi^+ \pi^-$ at BaBar~\cite{BaBar:2010eww, BaBar:2012sau}. In these processes the $\Upsilon(1S)$ was then tagged and we can perform a resonant search on its decay products.

In particular, as the photon and the $X$ bosons are monochromatic, one can perform a bump hunt in the energy of the emitted photon, which is related to the mass of the $X$ boson via the simple relation
\begin{equation}
    E_\gamma = \frac{m_\Upsilon}{2}-\frac{m_X^2}{2m_\Upsilon}\,.
\end{equation}
The corresponding bounds, accordingly recast, are reported in Fig.~\ref{fig:Tauphilic_Vectors_VA}.

\paragraph{Nonresonant searches}

Nonresonant searches at Belle II, even if not yet performed, are expected to bring about significant improvements to the bounds discussed in the previous section due to the large luminosity that is expected at such a facility.
In order to place these sensitivity bounds, we consider, as for quarkonia decays, a search for a bump in the photon energy $E_\gamma$. The procedure we have followed is the one discussed in 
Sec.~\ref{subsec:testing_light_tauphilic_VB}.

The bounds obtained for the two scenarios (purely axial and purely vectorial vector bosons) are reported in Fig.~\ref{fig:Tauphilic_Vectors_VA}.
The behavior of such bounds at low energies can be better appreciated by considering the shape of the SM background process $e^+ e^- \to \gamma\, \bar{\nu} \,\nu$~\cite{Araki:2017wyg, Ma:1978zm}:
\begin{align}
\label{eq:SMbackgroundneutrinos}
\frac{d\sigma_{\text{SM}}}{dE_\gamma} &= \frac{\alpha_\text{em}G_F^2}{3\pi^2}\left(1-\frac{2E_\gamma}{\sqrt{s}}\right)(g_R^2+g_L^2)E_\gamma\left[\left(1-\frac{\sqrt{s}}{E_\gamma}+ \frac{s}{2E_\gamma^2}\right)\log \frac{(1+\cos \theta_\text{max})(1-\cos \theta_\text{min})}{(1-\cos \theta_\text{max})(1+\cos \theta_\text{min})}+ \cos \theta_\text{min}- \cos \theta_\text{max}\right]\,,
\end{align}
where
\begin{equation}
g_L = \begin{cases} -1/2 + s_w^2 \quad &\text{for } \nu_\mu, \nu_\tau\\
-1/2 + s_w^2 + 1\quad &\text{for } \nu_e
\end{cases}\,,\qquad 
g_R = s_w^2\,.
\end{equation}
Indeed, the quantity in Eq.~\eqref{eq:SMbackgroundneutrinos},  is directly proportional to $m_X^2 = s\,(1- 2 E_\gamma/\sqrt{s})$.
This explains the further enhancement at low energies of our bounds with respect to the standard linear dependence.
In our analysis we have only considered the irreducible SM background effects. However, other sources of background at Belle II exist and consist of processes of the type $e^+ e^- \to \gamma + X_\text{SM}$, where by $X_\text{SM}$ we denote any SM state that might escape detection. $X_\text{SM}$ can then consist of undetected $e^+ e^-$ pairs, or $n \,\gamma$ states that are not experimentally reconstructed, or hadronic states. In order to properly take these effects into account and maximize the signal yield over the background a complete Monte-Carlo analysis has to be performed, taking into account the specifics of the Belle II experimental setup. In particular, the background event with $X_\text{SM}=\gamma$ can represent a significant obstacle to such an analysis, as the signal photon possesses the same kinematics as the background ones in some regions of the parameter space, up to corrections that can be estimated of being of size $\mathcal{O}(m_X^2/s)$~\cite{Essig:2013vha, BaBar:2017tiz}.
We postpone such an analysis to a dedicated paper, and for the time being we consider only the irreducible SM background, thus assuming perfect reconstruction of the signal events over the background.

The two black dashed lines in Fig.~\ref{fig:Tauphilic_Vectors_VA} indicate potential regions of interest for this kind of candidate. Assuming that both the light vector boson and the anomalons acquire mass from a Higgs-like mechanism involving a SM singlet field with vacuum expectation value $\bar{v}$, one has
\begin{equation}
m_X \simeq \alpha \frac{g_D \bar{v}}{2}\,,\qquad m_\psi \simeq \frac{y_\psi\bar{v}}{\sqrt{2}}\,,
\end{equation}
where $\alpha$ is a factor parameterizing the impact of a possible mass mixing between the $X$ boson and the $Z$. Since the diagonalization of a $2 \times 2$ matrix reduces the smallest eigenvalue and increases the largest, we expect $\alpha < 1$. Requiring the lightest anomalons to have a mass larger than $1.5\TeV$,
one then finds that the mass of the $X$ boson and the coupling $g_D$ are related by the following inequality:
\begin{equation}
\frac{\sqrt{2}y_\psi m_X}{\alpha g_D} < 1.5\TeV\,,
\end{equation}
which we imposed for two plausible limiting regimes in Fig.~\ref{fig:Tauphilic_Vectors_VA}. 
These lines are to be intended as purely illustrative of a region in the parameter space that might be interesting to explore, but one has to keep in mind that the above bounds can be easily avoided by specific UV completions. For instance, the lightest anomalons might be neutral particles, thus avoiding the constraint on long-lived electromagnetically charged particles. Yet, it is interesting to notice that heavy anomalons interacting with a light gauge boson can be probed at Belle II, in complementarity to collider searches, and with a similar discovery potential.

\end{widetext}

\bibliography{tau}

\end{document}